\documentclass[aps,prx,reprint,groupedaddress]{revtex4-2} % You should use BibTeX and apsrev.bst for references
\usepackage{graphicx}% Include figure files
\usepackage{dcolumn}% Align table columns on decimal point
\usepackage{bm}
\usepackage{graphicx}
\usepackage[caption=false]{subfig}
\usepackage[colorlinks=true, citecolor=blue, linkcolor=blue, urlcolor=blue]{hyperref}
\usepackage{pinlabel} % pdfLatex can't load eps somehow. use pdf for figures
\usepackage{amsmath}
\usepackage{dcolumn}% Align table columns on decimal point

\usepackage{amssymb}
\usepackage{natbib}
\usepackage{mathrsfs}
\usepackage[T1]{fontenc}

\newcommand{\phiz}{\phi_{\textup{zpf}}}

\newcommand{\nz}{n_{\textup{zpf}}}
\newcommand{\beq}{\begin{equation}}
\newcommand{\eeq}{\end{equation}}
\newcommand{\beqr}{\begin{eqnarray}}
\newcommand{\eeqr}{\end{eqnarray}}

\newcommand{\ah}{\hat{a}}
\newcommand{\ahd}{\hat{a}^\dagger}
\newcommand{\RN}[1]{%
  \textup{\uppercase\expandafter{\romannumeral#1}}%
}

\usepackage{color}

\begin{document}

%\preprint{}

\title{ZZ freedom in two qubit gates}

\author{Xuexin Xu}
\affiliation{Institute for Quantum Information,
RWTH Aachen University, D-52056 Aachen, Germany}
\affiliation{Peter Gr\"unberg Institute, Forschungszentrum J\"ulich, J\"ulich 52428, Germany}
\affiliation{J\"ulich-Aachen Research Alliance (JARA), Fundamentals of Future Information Technologies, J\"ulich 52428, Germany}
\email{x.xu@fz-juelich.de}
\author{M.H. Ansari}
\affiliation{Institute for Quantum Information,
RWTH Aachen University, D-52056 Aachen, Germany}
\affiliation{Peter Gr\"unberg Institute, Forschungszentrum J\"ulich, J\"ulich 52428, Germany}
\affiliation{J\"ulich-Aachen Research Alliance (JARA), Fundamentals of Future Information Technologies, J\"ulich 52428, Germany}


%\date{\today}

\begin{abstract}
Superconducting qubits on a circuit exhibit an always-on state-dependent phase error. This error is due to sub-MHz parasitic interaction that repels computational levels from non-computational ones. We study a general theory to evaluate the `static' repulsion between seemingly idle qubits as well as the `dynamical' repulsion between entangled qubits under microwave driving gate. By combining qubits of either the same or opposite anharmonicity signs we find the characteristics of static and dynamical ZZ freedoms. The latter universally eliminate the parasitic repulsion, leading us to mitigate high fidelity gate operation.  Our theory introduces new opportunities for making perfect entangled and unentangled states which is extremely useful for quantum technology.

\end{abstract}


\maketitle


\section{Introduction}  \vspace{-0.1in}

High-performance quantum processors require improvements in  the fidelity of quantum logic gates. Such improvements provide opportunities for near term quantum systems to demonstrate multi-partite entanglement \citep{ayanzadeh2020reinforcement,Brydges_2019} and full-fledged quantum error correction for  surpassing classical computer power  \citep{foxen2020demonstrating,Arute:2019aa}. These milestones can be achieved with the advent of long coherence qubits that can quickly go from perfect isolated state to  strongly interacting entanglement and vice versa, using  programmable electronics \citep{kjaergaard2020superconducting,martinis2020quantum,Bialcza12Fast}.  In  the last decade {many} advances have taken place to suppress degrading interactions among qubits as well as between a qubit and environmental noise and quasiparticles \citep{gustavsson2016suppressing,ansari2015rate,ansari2013effect,serniak2018hot,ansari2011noise,bal2015dynamics}.  However state-of-the-art quantum systems are yet far from being perfect \cite{krantz2019quantum}. 

In today's quantum processors single qubit rotation is fast and precise, however two-qubit entanglement  is yet to achieve a high contrast on/off operation with logical error rates below error correction threshold \citep{Caldwell2018para,McKay2016universal,Walter2017rapid,blais2020circuit}. A pair of idle qubits initialized at either $|00\rangle$  or $|11\rangle$ accumulate the phase error $\exp(i \zeta t/4)$ after time $t$, while  $|01\rangle$ and  $|10\rangle$ do it differently with the error being $\exp(- i\zeta t/4)$.  Experiments show $\zeta$ is a sub-MHz coupling strength. In theory this extra force is generated due to level repulsion between computational levels, e.g. $E_{11}$, and non-computational, e.g. $E_{02}$ or $E_{20}$ \cite{ku2020suppression}. Within  computational subspace this repulsion is a ZZ interaction, with Z being $\sigma_z$ Pauli operator, and is always internally present between any pair of qubits on the circuit and is called \emph{static} ZZ interaction. The effect of static ZZ interaction goes beyond accumulating idle phase error. Externally driven qubits by 2-qubit gates produces additional ZZ component on top of the static one that  keeps two-qubit gates from achieving high fidelity entanglement \citep{mundada2019suppression, mckay2019three,PhysRevApplied.14.024042}.  Therefore freeing qubits from the unwanted ZZ interaction is highly demanded for increasing gate fidelity, which is the purpose of this paper.  

In the first part we discuss Hamiltonian circuit analysis for demonstrating static ZZ freedom. This  is better to take place first in the circuit model before fabricating the actual circuit as it requires delicate parameter tuning. We keep this part at general and self-contained as possible and indicate how all circuit parameters can impact achieving the freedom. Some of the main game players are qubit-qubit direct coupling and qubit anharmonicity values and signs.  We examine several circuits with the same-sign anharmonicity, such as 2 transmon devices. Some of our samples are set to be similar to those tested in \citep{zhao2020high,li2019tunable,malekakhlagh2020first}. We show new possibilities for zeroing repulsion between two transmons, for example by controlling their capacitive direct coupling. We develop our theory on circuits with opposite sign anharmonicity, such as in \cite{ku2020suppression}, and show under what conditions the exact ZZ freedom becomes possible.

In the second part we discuss a new strategy for zeroing unwanted ZZ interaction, the \emph{dynamical} ZZ freedom. This is applicable for circuits with built-in ZZ interaction, for instance in large processors with many qubits eliminating all  static ZZ interactions seems not to be possible. We propose to apply microwave pulses on qubits in a way similar to cross resonance (CR) gates \cite{Rigetti2010CR}. The microwave pulse produces additional ZZ component on top of the static part and we show that for certain circuit parameters this is possible to have the two parts canceling each other, resulting in zero total ZZ interaction. This allows to make perfect entanglement as well as unentanglement in absence of parasitic interaction. Interestingly dynamical ZZ freedom can be universally achieved in qubits with any anharmonicity sign.

  %_____ SECTION  II _______
   \vspace{-0.27in}

\section{General Model} \label{sec general model}   
\begin{figure}[t]
\begin{center}
\includegraphics[width=0.45\textwidth]{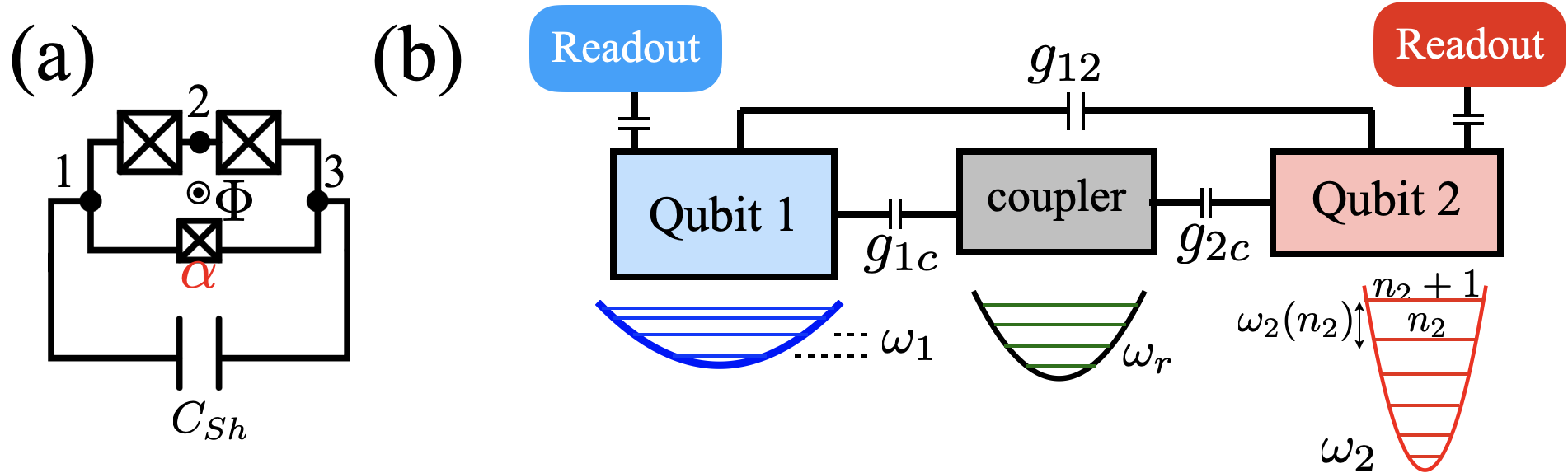}
 \includegraphics[width=0.43\textwidth]{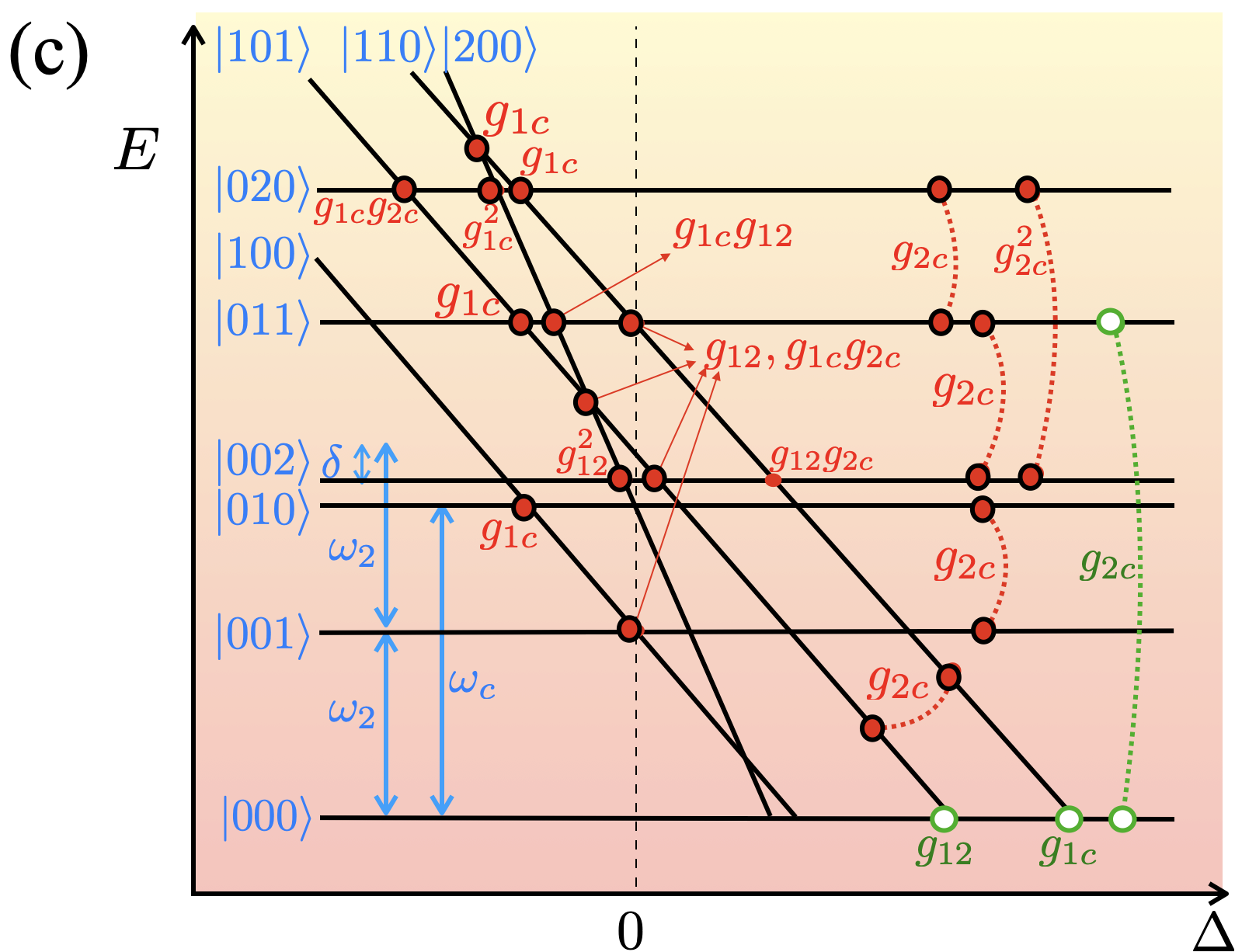}
 \vspace{-0.15in}
\caption{(a) A CSFQ, (b) A CSFQ-Transmon circuit coupled capacitively by $g_{12}$ and via a coupler, each qubit is measured, (c) Energy diagram of a circuit with two qubits Q1 and Q2 and a resonator coupler C. Qubits have similar anharmonicity $\delta$ and frequency detuning $\Delta$, i.e.  $\omega_1=\omega_2-\Delta$.  The plot shows how different noninteracting eigenstates $|n_{Q1}, n_{C}, n_{Q2}\rangle$ are coupled to one another by headed lines based on different interaction terms in the Hamiltonian (\ref{eq. ham}). Red head lines indicate level repulsion due to co-rotating terms in the interactions. White head lines indicate level repulsions by counter-rotating terms.} \vspace{-0.27in}
\label{fig. diag}
\end{center}
\end{figure}

\vspace{-0.1in}
Physical qubits have more than two energy levels and can be found in two classes (anharmonicity species): positive or negative anharmonicity. In qubits with positive (negative)  anharmonicity higher levels are farther (closer) apart. A negative anharmonic qubit is transmon and an example of positively anharmonic qubit is a capacitively shunted flux qubit (CSFQ) \cite{kjaergaard2020superconducting}. In a single Josephson junction (JJ) transmon a  rather large capacitance shunts the charging energy to warrant less sensitivity to charge fluctuations. A CSFQ, as depicted in Fig. \ref{fig. diag}(a), replaces the JJ with a closed loop of three junctions, i.e. two in-series JJ parallel to a smaller JJ (smaller in critical current and capacitance) \cite{Steffen2010high}. 

A single JJ transmon  can be classically characterized  by the flux $\Phi$ and its canonical conjugate $Q=2e n$ being the electric charge of $n$ Cooper pairs tunnelling across the JJ, $e$ being electron charge.  The device has a periodic potential energy, i.e. $ -E_J \cos{(2\pi \Phi/ \Phi_0)}$ with the Josephson energy $E_J\equiv I_0 \Phi_0/2\pi$ proportional to the critical current $I_0$ and  the flux quantum $\Phi_0 =h/2e$, $h$ being Planck constant.  Variables can be simplified to $\phi\equiv 2\pi \Phi/ \Phi_0$ and its conjugate $n \hbar $ whose quantization can take place in the Fock space of quantum states $| m\rangle $ with $m=0,1,2,\cdots$ and annihilation and creation operators $\ah  = \sum_m \sqrt{m} |m\rangle \langle m +1 |$ and  $\ahd =\sum_m \sqrt{m+1} |m+1\rangle \langle m |$. A transmon canonical operators are defined $\hat{\phi}=\phiz (\ah +\ahd)$ and $\hat{n}= i \nz (\ah -\ahd)$  with the zero-point fluctuations $\phiz$ and $\nz$, satisfying the minimum uncertainty $\phiz  \nz=1/2$. The fluctuations can be determined by the characteristic impedance $Z_c$,  i.e. $\phiz = \sqrt{\hbar Z_c/2 }$.  A transmon's small anharmonicity allows to approximate the periodic potential energy in the vicinity of minimum with a 4th degree polynomial  making a Duffing oscillator, i.e. $H=\omega \hat{a}^\dagger \hat{a} + (\delta/12) (\hat{a} + \hat{a}^\dagger)^4$ with $\delta$ being anharmonicity. Higher order corrections are  known \cite{Didier2018Analytical}.

The CSFQ depicted in Fig. \ref{fig. diag}(a) has two in-series identical JJ's in a closed loop with a smaller JJ that has smaller critical current $\alpha  I_0$ and capacitance $\alpha C_J$ with $\alpha<1$. The loop carries negligible kinetic inductance and an external magnetic flux $\Phi_{\textup{ext}}$ in the loop which can tune the potential energy  \cite{Steffen2010high}. A CSFQ potential energy is $-  E_{J} [\cos(\phi_1-\phi_2) +\cos(\phi_2-\phi_3) - \alpha  \cos (2\pi f -  \phi_1+\phi_3)]$ with $f$ being $\Phi_{\textup{ext}}/\Phi_0 $. Defining the two variables $\phi=\phi_1- \phi_3$ and $\phi'=\phi_1-2\phi_2 + \phi_3$ and working out corresponding charging energies, one can show $\phi'$ mode has negligible influence on the potential and can be discarded, see Appendix \ref{app.CSFQ}. This simplifies the potential to $  - E_{J} [2\cos\left(\phi/2\right)  + \alpha \cos\left(2\pi f -  \phi \right)]$. Under the condition $\alpha<1/2$  this potential can be used as a qubit because each period conveys one minimum, not more. The potential is the most symmetric at the so-called  `Sweet Spot' (SS) where $f=1/2$. A less symmetric potential at $f=1/2+\delta f$ with $\delta f \ll 1/2-\alpha $ has a minimum at $\phi_0=-2 \pi \alpha (\delta f)/(1/2-\alpha)$ and in the vicinity of the minimum with phase differece $\Delta \phi = \phi-\phi_0$ it can be approximated to 
 $ \left(1 -2 \alpha \right) E_{J}    \Delta \phi ^2/4   + \left( \alpha -1/8 \right)  E_{J} \Delta \phi ^4/4!   +  2 \pi (\delta f) \alpha E_{J}( \Delta \phi  - \Delta \phi ^3/6) $. This potential has the simple form of a Duffing oscillator at SS, i.e. $\delta f=0$,  with the periodic frequency 
$\hbar \omega= \sqrt{8 E_J E_C (1/2-\alpha)}$, anharmonicity $\delta=4E_C (\alpha-1/8)/(1-2\alpha)$, and zero point phase fluctuations $\phiz=[4E_C/E_J(1-2\alpha)]^{1/4}$. In contrast to a transmon,  a CSFQ with  $1/8<\alpha<1/2$ has positive anharmonicity; see Appendix \ref{app.CSFQ} for quantization.

\paragraph*{A hybrid two-qubit circuit:} We consider qubits Q1 and Q2  coupled via a coupler C, such as a bus resonator. Each one of the two qubits can have positive or negative anharmonicity. In the circuit depicted in Fig. \ref{fig. diag}(b) we give an example of Q1 being transmon and Q2 {being} a CSFQ, however we keep this section general without referring to the qubit species.   Q1 and Q2 may have small capacitive coupling $g_{12}$ on top of indirect coupling via the coupler.  The circuit Hamiltonian is
\beq 
H = \sum_{i, n_i} \omega_i(n_i) |n_i+1\rangle \langle n_i+1| + \sum_{j (\neq i)}{g}_{ij}(\ah_i+\ahd_i)(\ah_j+\ahd_j) 
\label{eq. ham}  
\eeq
with $i=1,2$ for qubits and $c$ for the coupler. The energy levels in the bare basis associated to free Hamiltonian are $E_{n_{1}}$ and $E_{n_2}$ for qubits  and $E_{n_c}$ for the coupler.  $\omega_i(n_i)$ is the difference between $E_{n_i+1}$ and $E_{n_i}$, therefore qubit frequencies are $\omega_{1/2}(0)$ or simply $\omega_{1/2}$. The coupler frequency $\omega_{c}$ is far detuned from qubits in order to warrant no backaction from the coupler on the qubit driving when external gates are applied.  

Let us further analyse the Hamiltonian (\ref{eq. ham}).  Consider a fixed frequency Q2 from which Q1 is detuned by $\Delta$, i.e. $\omega_1=\omega_2-\Delta$ and similar anharmonicity $\delta$. Given that $|n_1,n_{c},n_2\rangle$ is an eigenstate for noninteracting Hamiltonian, interaction with coupling strength $(g_{1c})^k$ provides transition to  $ |n_1\pm k,n_{c}\mp k,n_2\rangle$, with the strength  $(g_{2c})^k$ to $|n_1,n_{c}\mp k,n_2\pm k\rangle$, with  the  strength  $(g_{12})^k$ to $ |n_1\pm k,n_{c},n_2\mp k\rangle$, with  the  strength  $g_{1c}g_{2c}$ to $|n_1\pm 1,n_{c}\mp2,n_2\pm1\rangle$,  and so on.  Figure \ref{fig. diag}(c)  shows these transitions marked by  co-rotating terms  such as $\ah_i \ahd_j $ by red heads. The  counter-rotating terms in intersaction such as $\ah_i \ah_j$ are indicated by white heads.  Each coupling is labelled by the interaction term in the Hamiltonian that is responsible for it. One can see in Fig. \ref{fig. diag}(c)  that the transition between  $|200\rangle$ and  $|101\rangle$ takes place either by direct coupling $g_{12}$ or via the intermediate state $| 110\rangle$. In absence of direct coupling the  avoided crossing between the two levels has been found $\sqrt{2} g_{1c}g_{2c}/\Delta_{2} (\Delta_{1}+\delta)$ in Ref. \cite{Reed:2012ab}. The seemingly non-interacting levels may stay so, or may interact via $n$ photons for $n>2$ or under external gates. 

\paragraph*{Effective circuit Hamiltonian:} To reduce quantum computation errors all circuit elements should be in the dispersive regime, i.e. with coupling strengths $g_{ij}$ much weaker than frequency detuning  $\Delta_{ij} \equiv \omega_i-\omega_j$  \cite{koch2007charge}. In this limit the Schrieffer-Wolff (SW) transformation \cite{BRAVYI20112793} simplifies Eq. (\ref{eq. ham}) to this qubit-qubit Hamiltonian:
\beqr
H&=& \sum_{n_{q}} \bar{\omega}_q({n_q}) |n_q+1\rangle \langle n_q+1| + \sum_{n_1,n_2} \sqrt{(n_1+1)(n_2+1)} \nonumber \\  &\times&J_{n_1,n_2}   \left(| n_1, n_2+1\rangle \langle n_1+1,n_2| +h.c. \right),
\label{eq.effH} 
\eeqr
with   dressed frequency $\bar{\omega}_q({n_q})$ being the difference between $\bar{E}_{n_q+1}$ and $\bar{E}_{n_q}$ in dressed basis associated to interacting Hamiltonian. The qubit-qubit coupling strength within two photons limit is
\beq
J_{n_1n_2}\equiv g_{12} -\frac{g_{1c}g_{2c}}{2} \sum_{q=1,2}\left[ \frac{1}{{\Delta_{q}(n_{q})}}+\frac{1}{{\Sigma_{q}(n_{qc})}}\right],
\label{eq.J}  \vspace{-0.1in}
\eeq 
with  $\Delta_{q}({n_q}) \equiv \omega_{c}-\omega_q(n_q)$ and $\Sigma_q(n_q) \equiv  \omega_{c} + \omega_q(n_q)$. One can find the qubit dressed frequency $\bar{\omega}_q({n_q}) = \omega_q({n_q}) -g_{qc}^{2}(n_q+1)/\Delta_{q}({n_q})$ and its dressed anharmonicity  $\bar{\delta}_q=\delta_q[1-2g_{qc}^2/\Delta_{q}(\Delta_{q}-\delta_q)]$.

Let us briefly discuss the impact of measurement of qubits. Readout measurement usually takes place by means of weakly coupling a qubit to a resonator with the interaction $H_{qR}=g_{qR}(\hat{a}_q+\hat{a}_q^\dagger) (\hat{a}_R+\hat{a}^\dagger_R)$.  The process of eliminating the readout resonator can take place before or after eliminating the coupler. The difference returns some small leftovers in dressed qubit frequency $\textup{\dj} \omega$ and  anharmonicity $\textup{\dj} \delta$, which  are
 \vspace{-0.05in}
\beqr 
&& \frac{\textup{\dj} \tilde{\omega}_2}{g^6}=-\frac{\textup{\dj} \tilde{\omega}_1}{g^6}= \frac{(\Delta+2\Delta_{2})^2(\Delta^2+\Delta\Delta_{2}+\Delta_{2}^2)}{2\Delta \Delta_{2}^4 (\Delta+\Delta_{2})^4},\ \  \\   
&& \frac{\textup{\dj} \tilde{\delta}_{1/2}}{2 g^6}=-  \frac{(\Delta_{2}-\delta_{1/2})^3+(\delta_{1/2}+\Delta_{2})\Delta_{2}^2}{(\Delta_{2}^3 \Delta\mp\delta_{1/2})(\Delta_{2}-\delta_{1/2})^4} \pm 
\frac{2}{\Delta\Delta_{2}^4}, \ \ \ \ 
\label{eq.order}
\eeqr
with universal coupling $g$ and two qubit detuning $\Delta=\omega_2-\omega_1$, and $\delta_{1},\delta_2,\Delta\ll\Delta_{2}$.

The effective Hamiltonian approach is limited to dispersive regime as shown in Ref. \cite{ansari2019superconducting}, therefore qubits with small frequency detuning must be treated nonperturbatively.  Moreover external driving Hamiltonian should be treated  with methods other than SW such as the method described in Ref. \cite{Cederbaum_1989,magesan2018effective} to express nonperturbative strong driving impacts.

\begin{figure}
    \centering
  \includegraphics[width=0.24\textwidth]{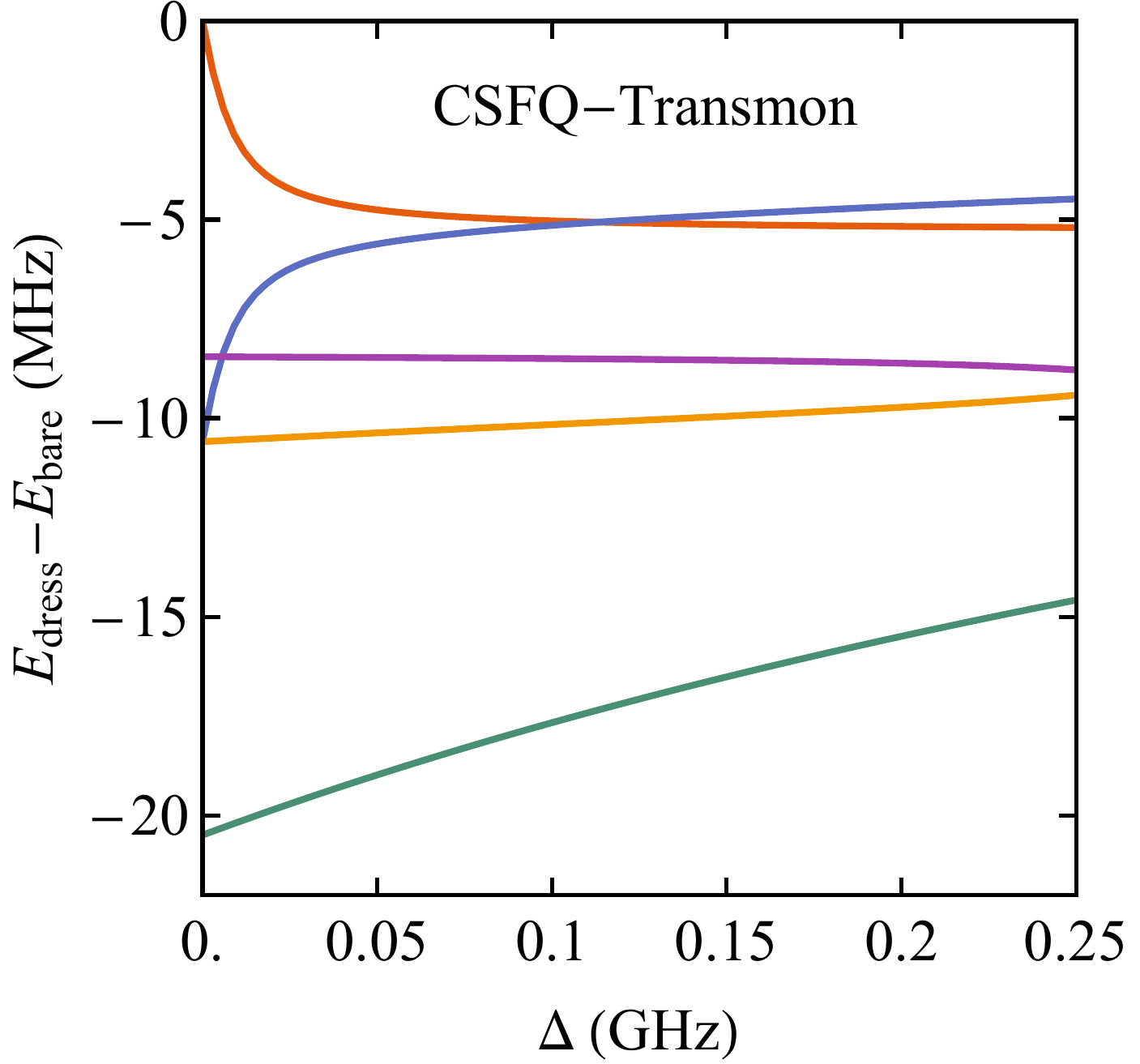}
   \put(-119,110){(a)}
  \includegraphics[width=0.24\textwidth]{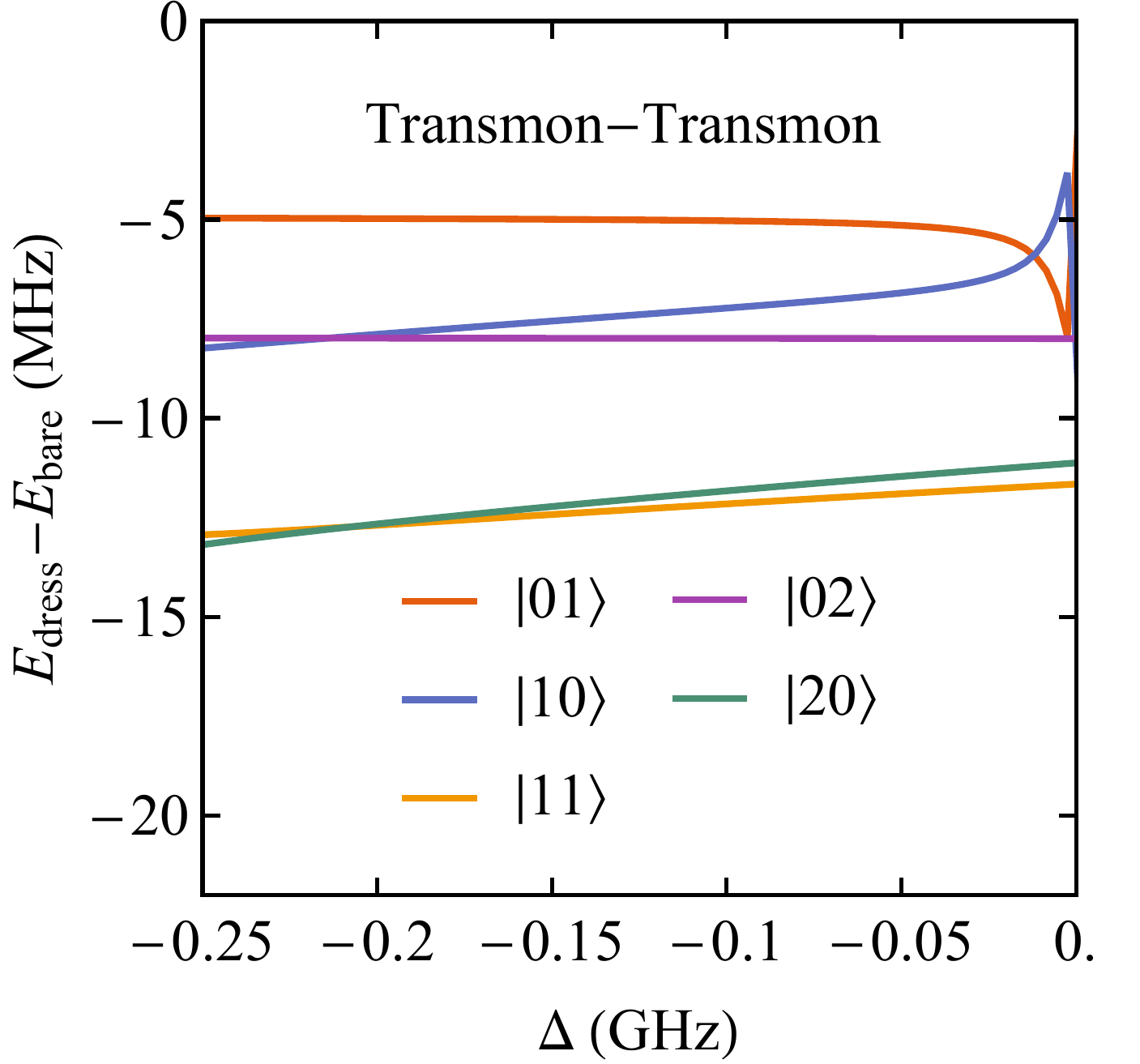}
   \put(-119,109){(b)}
    \caption{Difference between interacting and noninteracting energy levels obtained from Hamiltonian (\ref{eq. ham}) in (a) a CSFQ-transmon and (b) a transmon-transmon device. Circuit parameters are similar to those used in Fig. \ref{figZZ}(b) and Fig. \ref{figZZ}(d).}
    \label{fig:dress}
\end{figure}

\section{Static ZZ freedom}
\label{sec. statZZfree}

The qubit-qubit Hamiltonian (\ref{eq.effH}) has been written for any number of qubit energy levels. Further block diagonalization separates the computational subspace from higher excitations. Within the dispersive regime the computational form of qubit-qubit Hamiltonian turns out to have the following operator structure:
\beqr \nonumber
H_{\textup{eff}}&=&\tilde{\omega}_1\left|10\right>\left<10\right|+\tilde{\omega}_2\left|01\right>\left<01\right|+(\tilde{\omega}_1+\tilde{\omega}_2+\zeta)\left|11\right>\left<11\right|,\\
&=&-\frac{\tilde{\omega}_1+\zeta/2}{2} {\rm ZI}-\frac{\tilde{\omega}_2+\zeta/2}{2} {\rm IZ}+ \frac{\zeta}{4}{\rm ZZ}.
\label{eq.qqH}
\eeqr   
 
 Defining $E_{n_1n_2}$ the energy level with qubit 1 and 2 at levels $n_1$ and $n_2$,  one can find a general definition for the only interaction term present in Eq. (\ref{eq.qqH}) based on energy levels:
\beq 
\zeta= E_{11}-E_{10}-E_{01}+E_{00}, 
\label{eq.ZZoriginal}
\eeq

This interaction term is called \emph{static} ZZ interaction since it is always present even when qubits are idle.  Let us emphasize that  Eq. (\ref{eq.ZZoriginal}) is an original definition of static ZZ interaction that can be achieved only by taking the operator form of Eq. (\ref{eq.qqH}) into account.  Perturbation theory can evaluate the following $\zeta$ and dressed frequencies  within the dispersive regime:
\beqr
&& \zeta={2J_{10}^2}/({\bar\Delta-\bar{\delta}_1})-{2J_{01}^2}/({\bar\Delta + \bar{\delta}_2}), \label{eq.zeta}\\
&& \tilde{\omega}_1=\bar{\omega}_1-J_{00}^2/\bar\Delta, \ \ \ \text{and} \ \ \tilde{\omega}_2=\bar{\omega}_2+J_{00}^2/\bar\Delta, 
\label{eq.dw}
\eeqr
with $\bar\Delta\equiv \bar{\omega}_2-\bar{\omega}_1$.  There are a number of divergences in Eq. (\ref{eq.zeta}), however since the original definition Eq. (\ref{eq.ZZoriginal}) is divergence-free, the divergences are the consequence of our perturbative block-diagonalization, therefore results in the vicinity of these divergences are inaccurate.

The dressed frequency shifts depend on the coupling strength $J_{00}$, which causes $0 \to 1$ transition in one qubit and $1 \to 0$ transition in the other one. As shown in Eq. (\ref{eq.dw}) $J_{00}/\bar{\Delta}$ is subtracted from the frequency of a qubit and the same amount is added to the other one. Therefore based in the definition Eq. (\ref{eq.ZZoriginal}) $J_{00}$  does not contribute to the static ZZ coupling strength. However the contribution of higher excitations is different; $J_{01}$ and $J_{10}$ couple $0 \leftrightarrow 1$ transition in one qubit and $2 \leftrightarrow 1$ transition in the other one. The repulsive interaction between $E_{11}$ and the non-computational levels $E_{02}$ and $E_{20}$ can cancel one another, making ZZ interaction zero,  if $E_{11}$ can be in between and near the two non-computational levels. We find the first few eigenvalues of the Hamiltonian (\ref{eq. ham}) in presence and absence of interaction, and in Fig. \ref{fig:dress}  show the energy dispersion ($E_{\rm dress}-E_{\rm bare}$) of the two set to clarify how interaction affects energies. Equation (\ref{eq.ZZoriginal}) still holds valid to the energy dispersion since static ZZ is zero in absence of interactions, so one can evaluate $\zeta$ from Fig. \ref{fig:dress}. 
Usually the frequency shifting due to the interaction is much smaller than the energy gap, so this will not change the relative positions of energy levels. In a CSFQ-transmon pair with coupler frequency being far detuned from both qubits within the limit $|\Delta|<|\delta_{1/2}|$, $E_{11}$ falls in between $E_{02}$ and $E_{20}$ making it possible to have repulsion-free $E_{11}$ at a certain detuning frequency. In a transmon-transmon pair with coupler frequency being far detuned from both qubits,  $E_{11}$ is in one side of both non-computational levels $E_{02}$ and $E_{20}$, which makes the two repulsions to sum and not cancelled, possibly  except at very large detuning frequency which falls out of the domain of interest for quantum computation.

A pair of idle qubits that interact unwantedly by ZZ interaction accumulate state-dependent phase error. Evolving the idle quantum states $|00\rangle$ and  $|11\rangle$ after time $t$ results in the phase $\exp(+i\zeta t/4)$, while the states $|01\rangle$ and  $|10\rangle$ return a different phase $\exp(- i\zeta t/4)$. Therefore all qubits across a circuit accumulate such two-qubit-state-dependent phase error. 

Let us search for the possibility of vanishing phase error for interacting qubits. This can take place by finding ways to eliminate level repulsion between computational and non-computational levels while computational levels can still interact. Using Eq. (\ref{eq.zeta}) one can find the following condition for eliminating $\zeta$:  
 \beq
 \bar{\Delta}=\frac{\bar{\delta}_1+\bar{\delta}_2 \gamma^2}{1- \gamma^2}
 \label{eq.ZZfeq}
 \eeq
with $\gamma \equiv J_{10}/J_{01}$. Ignore counter-rotating terms one can find 
\beq
\gamma= \frac{1-\delta_1/ (2\Delta_{2} +\Delta)}{1-\delta_2/(2 \Delta_{2} +\Delta)} \frac{1-\delta_2/\Delta_{2}}{1-\delta_1/(\Delta_{2}+\Delta)}.
\label{eq.Jratio}
\eeq

The condition of the static ZZ elimination as shown in  Eq. (\ref{eq.ZZfeq})  makes it possible to investigate such a  possibility for certain circuit parameters.  Below we consider two types of circuits, a transmon-transmon pair coupled via a coupler, and a CSFQ-transmon pair coupled via a coupler.

\begin{figure}[t]
     \includegraphics[width=0.43\textwidth]{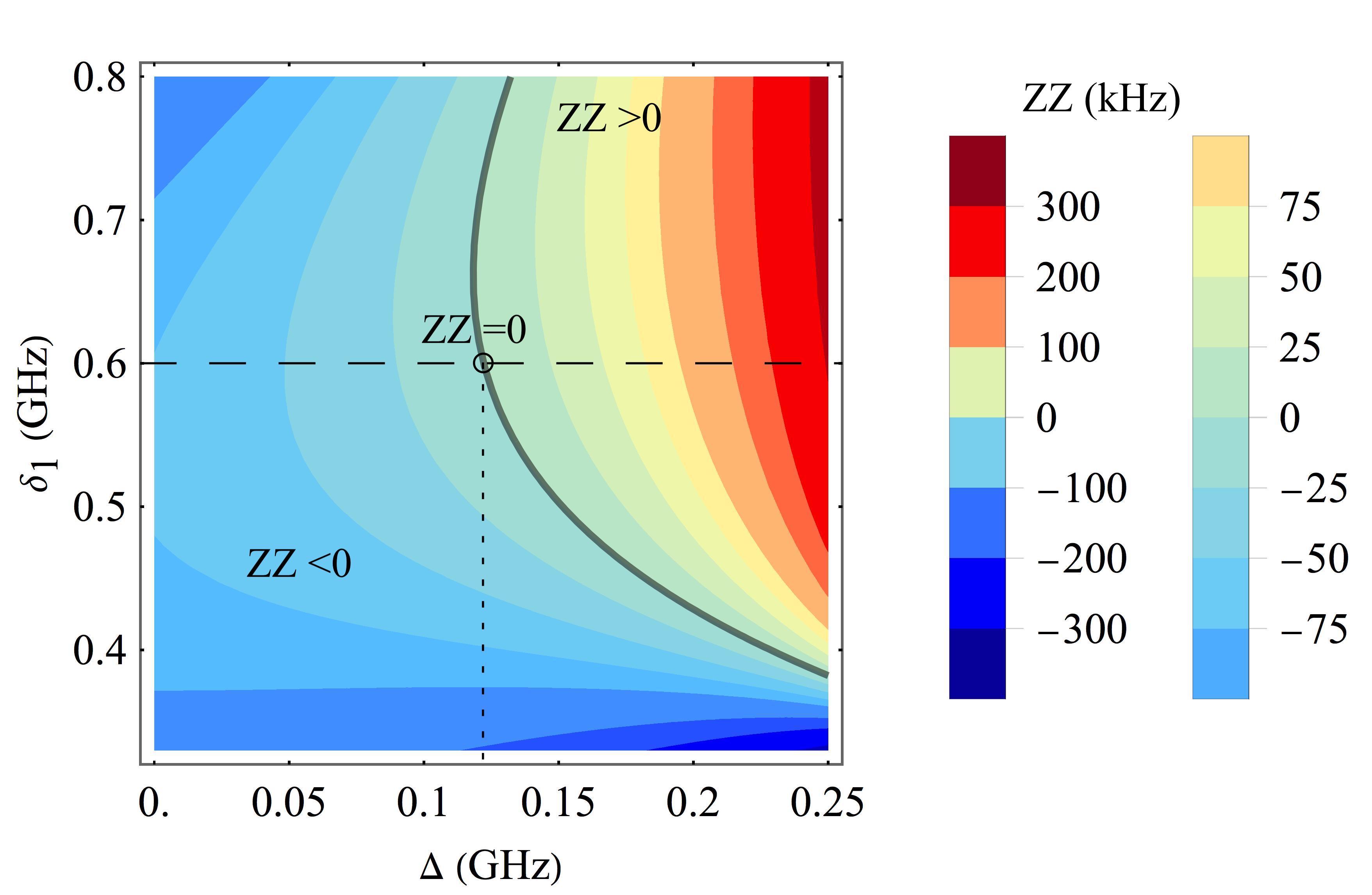} 
    \put(-227,133){(a)}\\
    \includegraphics[width=0.23\textwidth]{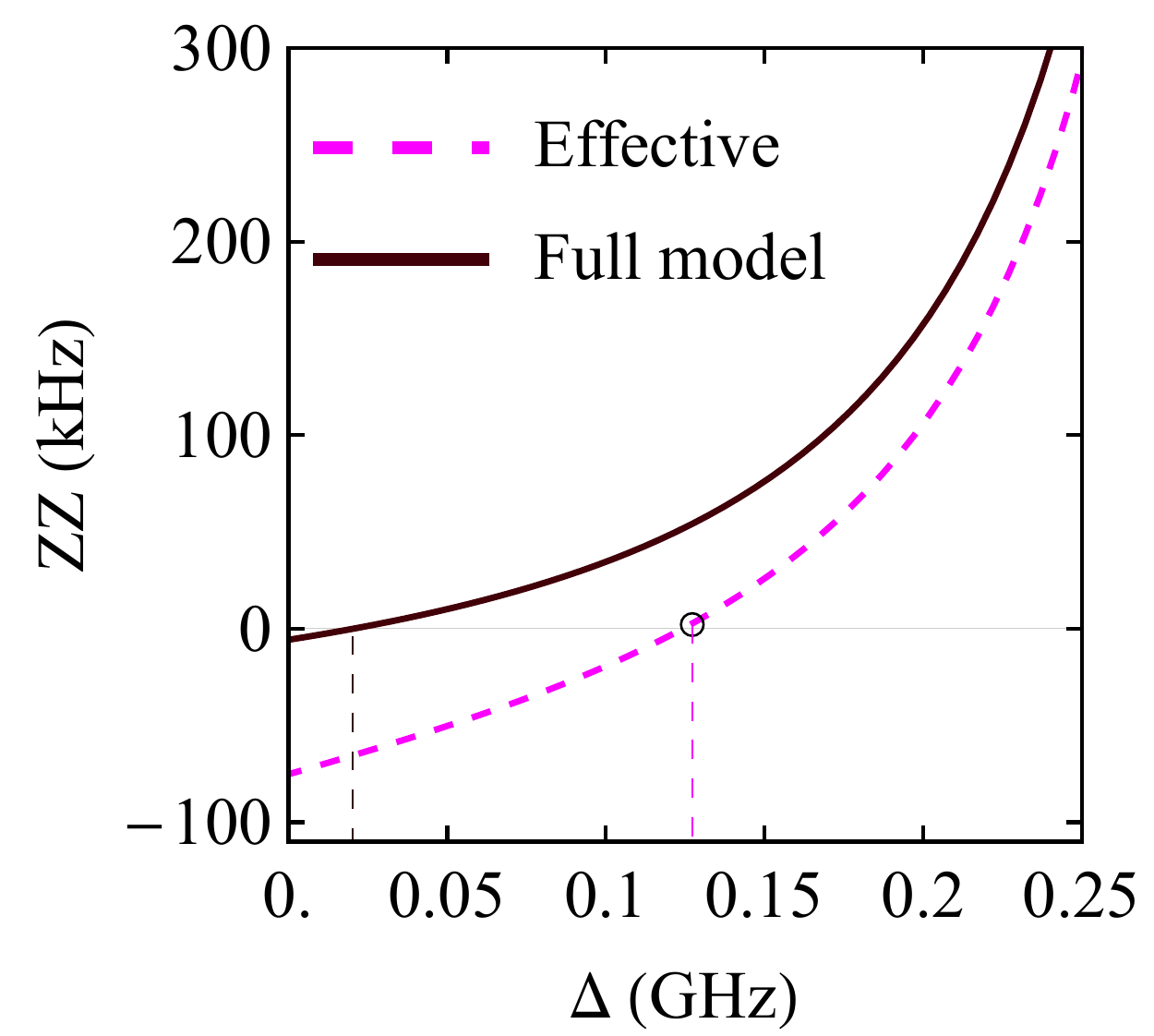}
    \includegraphics[width=0.23\textwidth]{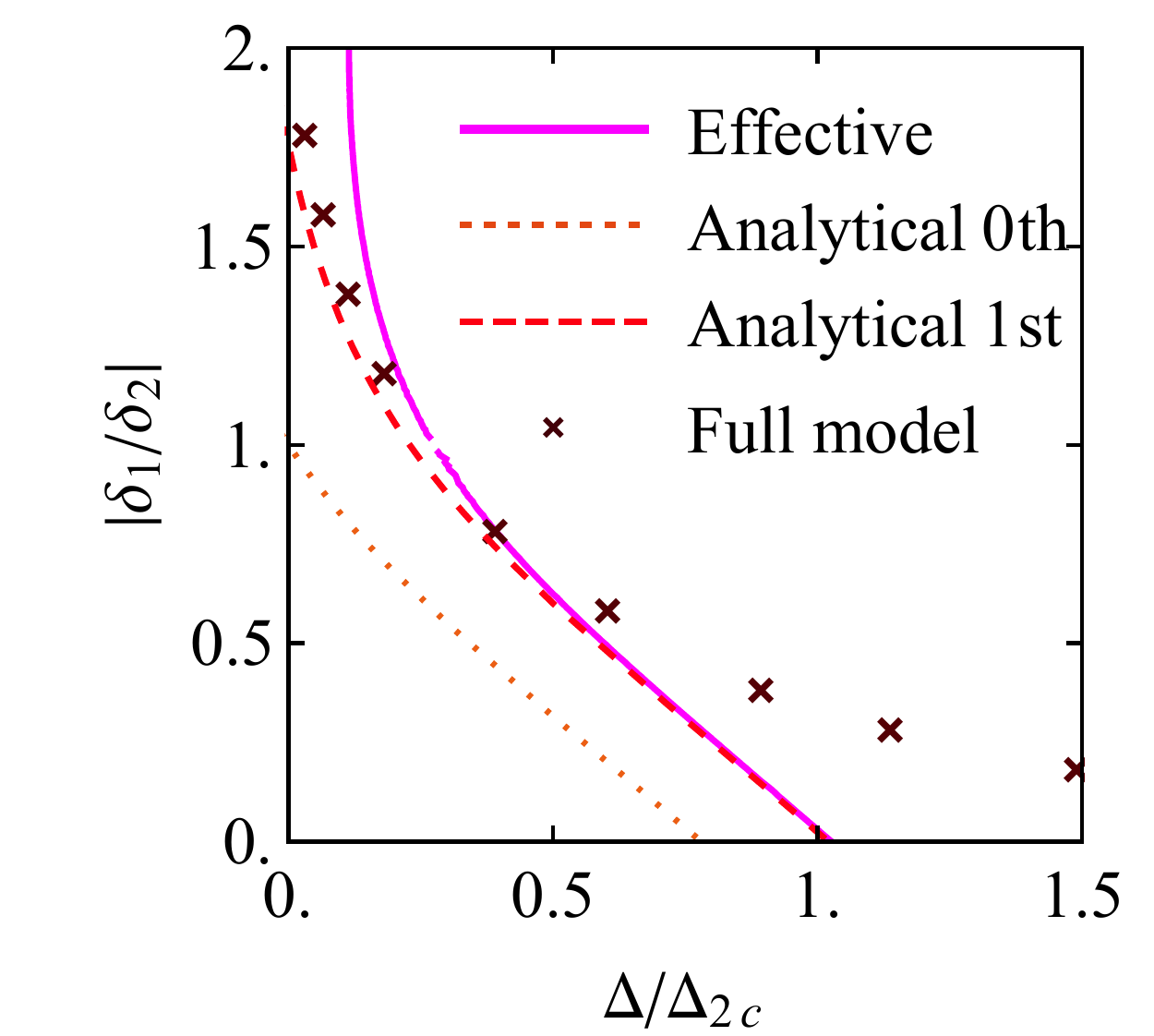} \put(-236,99){(b)}
    \put(-108,99){(c)}\\
     \includegraphics[width=0.23\textwidth]{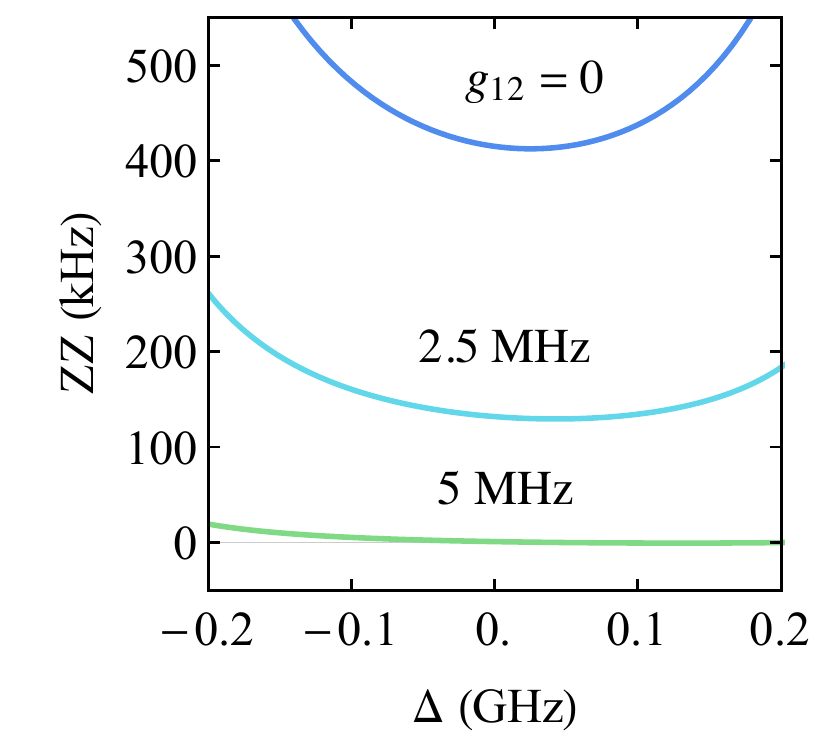}
    \includegraphics[width=0.23\textwidth]{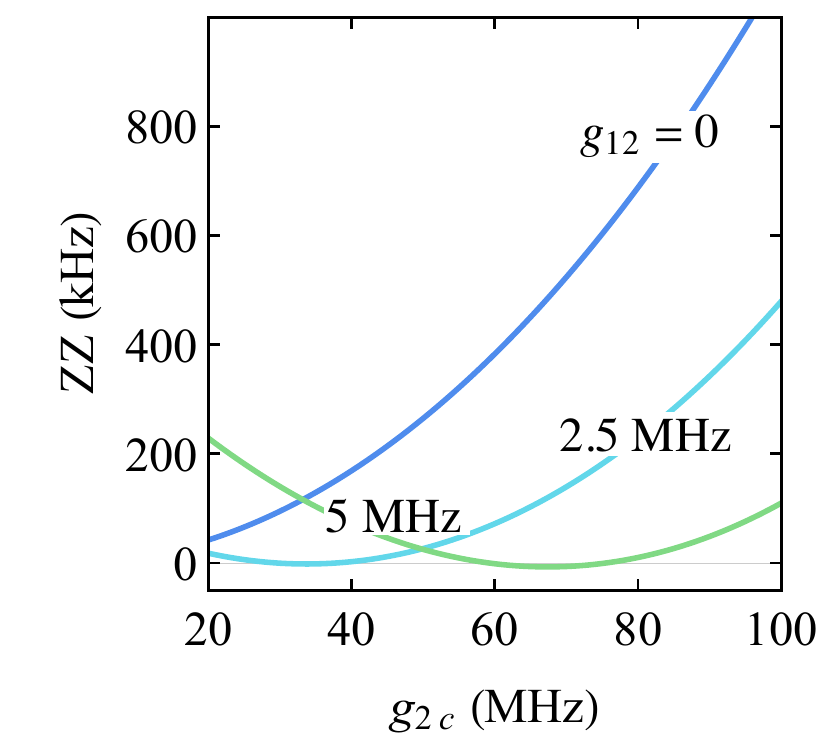}
    \put(-236,99){(d)}
    \put(-102,99){(e)}
    \caption{Static ZZ interaction in a CSFQ-transmon device (a)-(c) and in a transmon-transmon device (d) and (e). Q1 has the anharmoncity $\delta_1$ and is detuned from a transmon Q2 by $\Delta$. (a) Static ZZ interaction dependence on CSFQ anharmonicity and qubit qubit detuning. The two qubits are uncoupled directly, but they are indirectly coupled via a bus resonator with frequency $\omega_{c}=6.492$ GHz and coupling strengths $g_{1c}=g_{2c}=80$ MHz with $\omega_2=5.292$ GHz, $\delta_2=-0.33$ GHz. The effective ZZ interaction is positive (negative) in red (blue) areas and it vanishes on the solid lines. (b) Static ZZ interaction with $\delta_1=0.6$ GHz using the effective (dashed pink) and full model (solid brown) individually. (c) Static ZZ freedom criteria via the effective (solid), analytical zeroth-order (dotted), first order (dashed) and full Hamiltonian model (cross). Static ZZ interaction in a transmon-tranmon device with $\omega_2=4.914$ GHz, $\omega_c=6.31$ GHz, $\delta_1=\delta_2=-0.33$ GHz and direct coupling $g_{12}=0, 2.5, 5$ MHz as a function of (d) qubit-qubit detuning with $g_1=98$ MHz, $g_2= 83$ MHz, and (e) transmon-resonator coupling strength $g_{2c}$ with $g_{1c}=98$ MHz and $\Delta=-0.1$ GHz.}
    \label{figZZ}
\end{figure}

\paragraph*{CSFQ-Transmon pair: } A recent experiment has revealed that  combining  a positive and a negative anharmonic qubit can make  ZZ freedom \cite{ku2020suppression}. This freedom can take place in  qubits with non-zero $J$ interaction, which allows for entangling them with two-qubit gates.  Here we search for circuit characteristics that allow ZZ free qubits.  Let us consider  Q2 being fixed frequency  transmon is coupled to a CSFQ with $\Delta$ detuned frequency, i.e.   $\omega_1=\omega_2-\Delta$.  We consider a large Hamiltonian (\ref{eq.effH}) matrix with levels from $E_{00}$ up to $E_{04}$ and $E_{40}$ and block diagonalize to the computational subspace. This determines repulsion between computational and non-computational levels more accurately. The result has been plotted in Fig. \ref{figZZ}(a), showing ZZ coupling strength in colors over a wide domain of CSFQ anharmonicity $\delta_1$ and the frequency detuning $\Delta$.  ZZ interaction will have both negative and positive signs with a  nontrivial borderline  (solid black) between the two regions where the static ZZ coupling is zero. The marked circle indicates the  parameters of the CSFQ/Transmon circuit experimented in Ref. \cite{ku2020suppression} and showed zero static ZZ and one can see the marked circle is on the zero ZZ borderline.  

Figure \ref{figZZ}(b) compares two different approaches to determine the static ZZ for CSFQ/Transmon pair at different detuning frequency: dashed line obtained from SW effective Hamiltonian as found in Eq. (\ref{eq.zeta}), solid line obtains ZZ strength from numerical analysis of diagonalizing full Hamiltonian (\ref{eq. ham}). Comparing the two methods reveals that: both methods show consistent ZZ freedom and that perturbation theory is more accurate in large $\Delta$ domain.    Figure \ref{figZZ}(c) shows the parameters at which the static ZZ is zero. The parameters used here are the normalized frequency detuning $\Delta$ by transmon-coupler detuning $\Delta_{2}$ and the magnitude of anharmonicity ratio. Crossed points show exact results from diagonalizing of the full Hamiltonian, and the solid line is obtained from Eq. (\ref{eq.ZZfeq}). The two dashed lines are analytical solutions of  Eq. (\ref{eq.ZZfeq}) in the zeroth and first order in $|\delta_2/\Delta_{2}|$.  For obtaining these analytical solutions we define the qubits anharmonicities by  $\delta_1=  k \delta$,  $\delta_2=-\delta$ with  $k, \delta>0$. We consider no direct coupling between qubits and the universal qubit-coupler coupling strength $g$, and $\Delta=b\Delta_{2}$. Substituting these parameters in Eq.  (\ref{eq.Jratio}) evaluates  $J_{10}/J_{01}= (1+a)(1+b)(ak - b -2)/ (2+b+a) (ak -b -1)$, with $a\equiv \delta/\Delta_{2}$. Eq. (\ref{eq.ZZfeq}) can be simplified in the absence of $a$ to  $k=(2+b-3b^2-2b^3)/(2+5b+b^2)$, i.e. zeroth order solution.  Adding the first order of $a$ increases the solution precision. One can see for CSFQ anharmonicity being greater than transmon anharmonicity the analytical approximation is trustable.

 \paragraph*{Transmon-Transmon pair:}  In such a pair both anharmonicities are negative. Let us first based on the perturbation theory make some estimation about the possibility of ZZ freedom. A trivial possibility is when   $J_{01}\approx J_{10}\approx0$, more precisely direct coupling $g_{12}$ cancels out the indirect couplings in Eq. (\ref{eq.J}).  This freedom has been realized recently in several experiments \citep{goerz2017charting,zhao2020high,li2019tunable,malekakhlagh2020first}.  However non-interacting qubits cannot be entangled at such an operating point and are not useful for quantum computation. Searching for  ZZ freedom in the presence of $J$ coupling, results in such a possibility with qubit-coupler coupling  $g_{qc}$ being equal or larger than the qubit-resonator frequency detuning $\Delta_{q}$, which goes beyond the dispersive regime and perturbatively invalid. Exact numerical result with full Hamiltonian does not show any possibility for ZZ freedom in transmon-transmon pairs within the quantum computational domain of parameters. Other approaches e.g. two transmons coupled via a tunable coupler can only vary ZZ interaction above zero \citep{krinner2020demonstration, collodo2020implementation, xu2020highfidelity}.
 
Although ZZ freedom cannot be achieved for interacting pair of transmons, however one can achieve its suppression by tuning circuit parameters. For this purpose let us consider two transmons with almost the same anharmonicity and different frequencies. We numerically simulate the circuit and extract static ZZ by numerical diagonalization of the full Hamiltonian model. Figure \ref{figZZ}(d) shows static ZZ coupling strength decreases by lowering the magnitude of detuning frequency $\Delta$. This has been studied for three direct couplings $g_{12}$. At $g_{12}=5$ MHz the qubits are $J$-interaction-free.  Interestingly the weaker $g_{12}$ cases shows that suppression of  ZZ interaction can be achieved within a rather large domain of detuning frequency  $\Delta$, making it possible to design a transmon-transmon circuit with suppressed ZZ interaction within the dispersive regime $\Delta \gg J$. Figure \ref{figZZ}(e) shows how ZZ coupling changes by tuning transmon-resonator coupling $g_{2c}$.  Although the zero ZZ points belong to $J$-interaction-free transmons, however in their vicinity one can see a large class of transmons that not only interact but also their unwanted ZZ interaction is suppressed.

% ------- section II CR----

\vspace{-0.1in}
\section{Two-qubit gate: Cross Resonance}
\label{sec. CR}

Let us consider a two-qubit gate that rotates a qubit depending on the state of another qubit. Here we consider the cross-resonance (CR) gate --- a microwave pulse with the oscillation frequency equal to the frequency of Q2, which is applied on Q1. For this gate Q2 is target and Q1 is control qubit. The CR driving Hamiltonian is,
\vspace{-0.1in}
\beq
H_{\rm CR}=\Omega\cos(\omega_ d  t) \sum_{n_1}\left( \left|n_1\right\rangle \left\langle n_1+1\right|+h.c. \right),
\label{eq.CRHamil}\vspace{-0.06in}
\eeq 
with $\Omega$ being the amplitude  and $\omega_d=\tilde{\omega}_2+\zeta/2$ being the frequency of driving.  We let  Eq. (\ref{eq.CRHamil})  to co-rotate with the free Hamiltonian Eq. (\ref{eq.effH}) by transforming it by $R=\sum_n\exp(-i\omega_d t \hat{n})\left|n\right\rangle \left\langle n\right|$,  $H_{\rm CR} \to R^{\dagger}(H_{\rm CR}) R-i R^{\dagger}R$. Rotating wave approximation (RWA) helps to simplify fast oscillating terms making  the Hamiltonian time-independent . 

Driving qubits externally by weak amplitude pulses will not harm perturbative block diagonalization scheme  $\Omega \ll \Delta, g$. Since this domain of amplitude is too narrow we also compare perturbative transformation with nonperturbative least action transformation. We need a unitary transformation $T$ that can be constructed under a very weak constraint motivated by the idea that the only action that the transformation should perform is to bring $H$ into block diagonal form and does nothing otherwise\cite{magesan2018effective}. Such a block diagonalization technique has been previously worked out in chemistry in Ref. \cite{Cederbaum_1989}, namely the `Least Action' (LA)  block diagonalization. This transformation can be constructed after determining  the eigenvectors of  the Hamiltonian $H$.    The matrix of all eigenvectors S, and a block-diagonal part of it $S_{\rm BD}$, help to construct the transformation matrix $T$ in the following way:
\begin{equation}
T=S\ S_{\rm BD}^{\dagger}\ \left(S_{\rm BD} S_{\rm BD}^\dagger \right)^{-1/2},
\label{eq. leastaction}
\end{equation}

Block diagonalizing the CR driven multilevel interacting qubits by transforming their Hamiltonian  into $T (H+H_{\rm CR}) T^\dagger$ leaves us with the following qubit-qubit Hamiltonian in the computational subspace:  
\beqr \nonumber
H&=& \alpha_{\rm ZI} \frac{\rm ZI}{2} + \alpha_{\rm IX} \frac{  \rm IX}{2} + \alpha_{\rm ZX} \frac{ \rm ZX}{2}+\alpha_{\rm ZZ} \frac{\rm ZZ}{4} \\ & & \  + \alpha_{\rm ZY}\frac{\rm ZY}{2}+\alpha_ {\rm IY}\frac{\rm IY}{2}.
\label{eq.HCR}
\eeqr

\begin{figure}[h]
\begin{center}
    \includegraphics[width=0.24\textwidth]{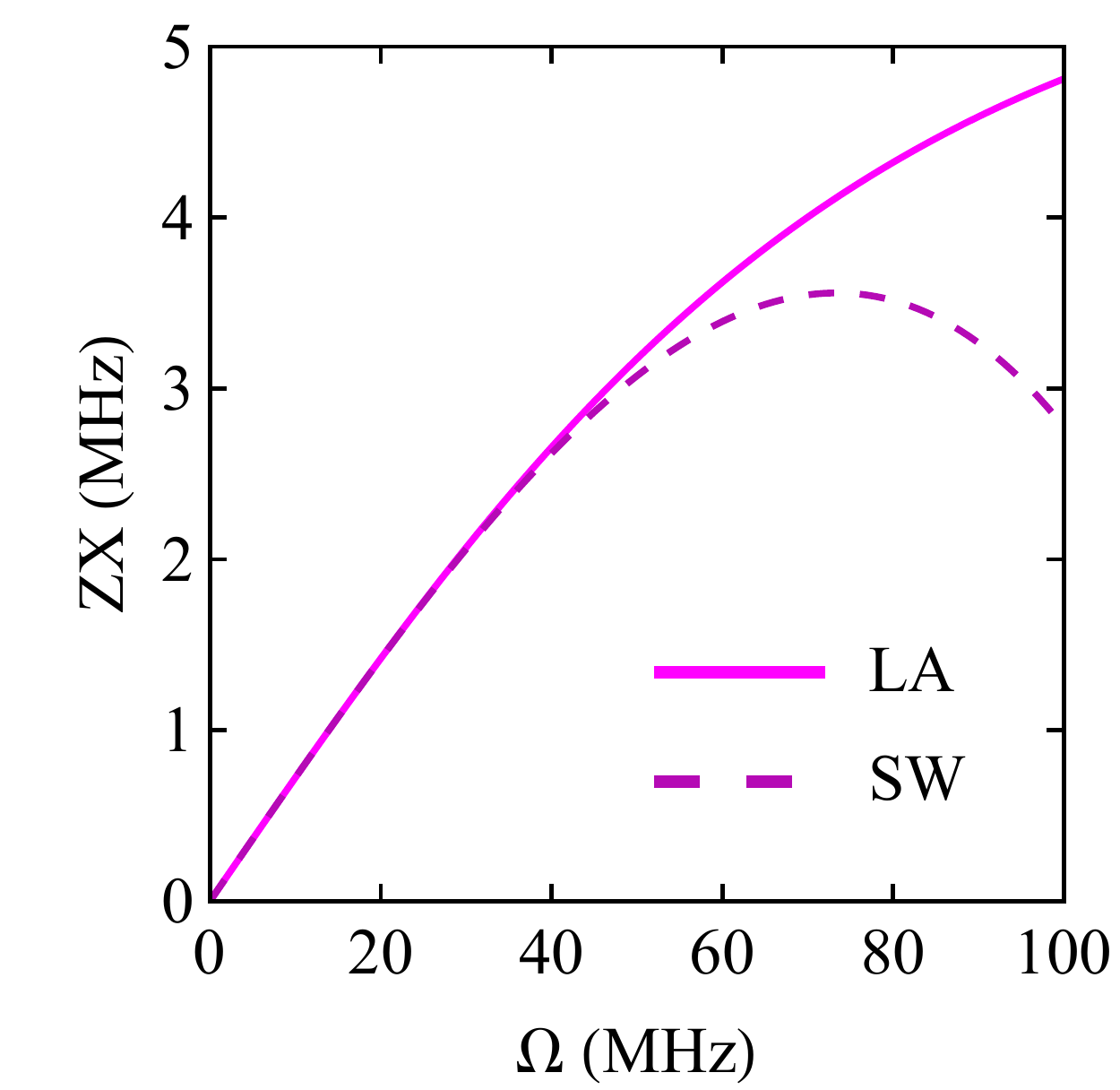}    \put(-120,112){(a)}
    %\put(-19,86){(b)}\\
   % \includegraphics[width=0.20\textwidth]{tr_st.pdf}
    \includegraphics[width=0.24\textwidth]{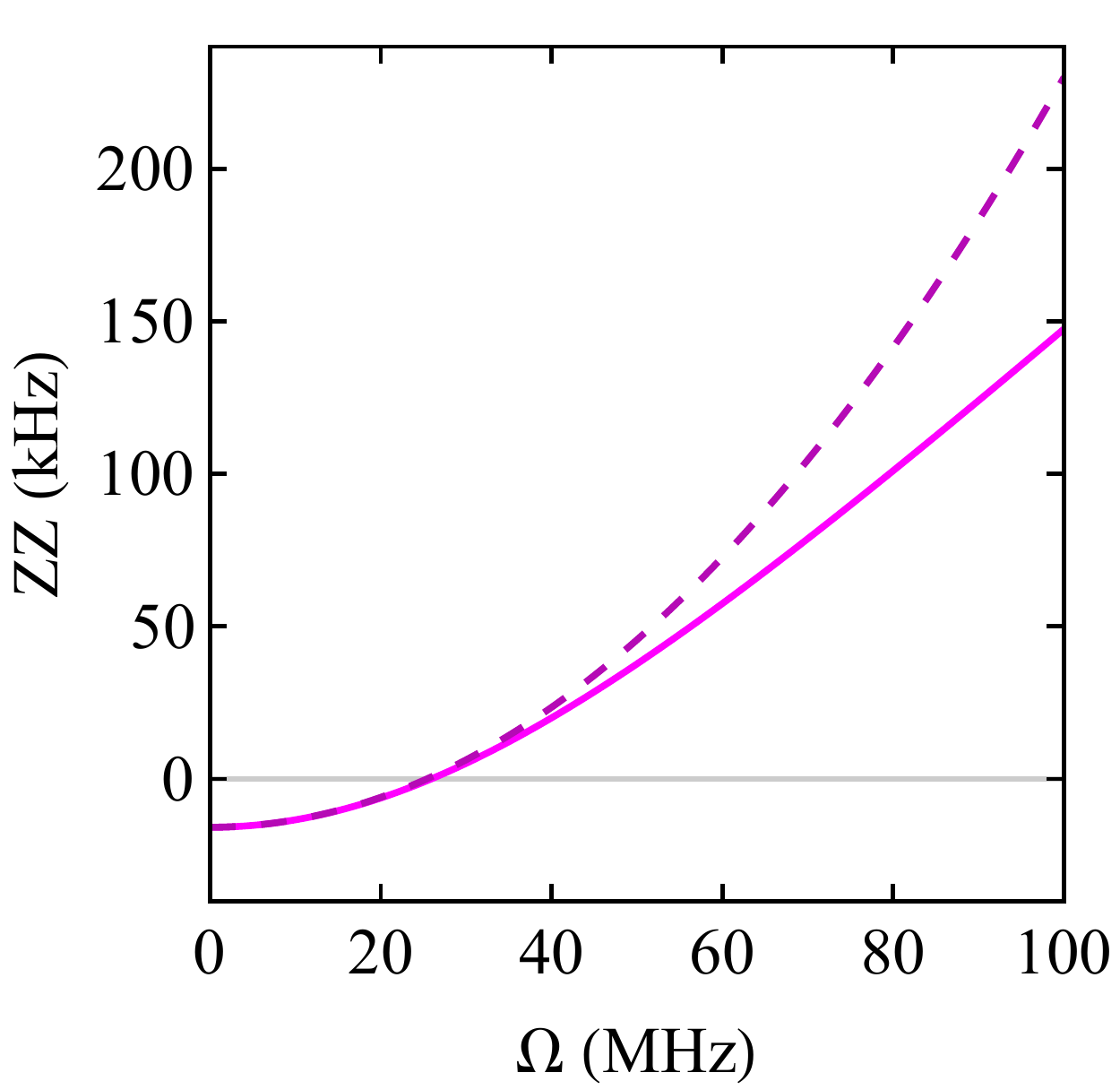}
    \put(-120,112){(b)} \vspace{-0.1in}
    \caption{ZZ and ZX coupling strengths in a CSFQ-transmon pair versus CR  amplitude $\Omega$, using Schrieffer-Wolff transformation (dashed) and Least Action transformation (solid).  $\Delta=$0.1GHz and the other parameters similar to Fig. \ref{figZZ}(b).} 
    \label{SW_LA}
    \end{center} 
\end{figure}

Although the same operator structure as of Eq. (\ref{eq.HCR})  can be found by  perturbative block diagonalization, however deviations from perturbation is visible within strong driving domain. We keep a record of the both sets of coupling strengths and compare them in the results taken in the rest of the paper.  Among all  terms that appear in Eq. (\ref{eq.HCR}) the only desired interaction is ZX as combining it with single qubit rotations provides two-qubit CNOT gate. Classical crosstalk effect, whose nature is still subject of research, can add up additional terms in the Hamiltonian Eq. (\ref{eq.HCR}) by changing the coupling constant of IX and IY terms. The ZY term can be eliminated by calibrating the global phase of CR pulse and this leaves us with its sibling ZX interaction.  Moreover applying an active cancellation pulse on target qubit with fine-tuned amplitude and phase can eliminate IY and IX terms \cite{Sheldon2016CRgate}. All these help to obtain gate ZX along with unwanted  ZZ term. Numerical results in a CSFQ-Transmon circuit has been plotted in Fig. \ref{SW_LA}. As one can see in weak amplitudes perturbation and nonperturbative results overlap, however as expected they grow differently by increasing the driving amplitude.

 \vspace{-0.1in}
 \vspace{-0.22in}

The ZZ interaction in a CR driven qubit-qubit circuit has two parts: 1) the static part due to computational level repulsion induced by non-computational levels, and 2) a dynamical ZZ part  induced by CR gate manipulation on the level repulsion. It is important to emphasize that since CR gate produces interactions other than ZZ coupling in the Hamiltonian (\ref{eq.HCR}), such as ZX, in presence of CR gate the ZZ coupling cannot be identified from Eq. (\ref{eq.ZZoriginal}) any more.  

Perturbation theory obtains that in the weak driving limit the dynamic ZZ component depends quadratically on CR amplitude $\Omega$ \cite{magesan2018effective}. Therefore the general structure of ZZ strength is
\beq 
\alpha_{\rm ZZ}=\zeta + \eta\Omega^2+O(\Omega^3),
\label{eq. ZZtot}  \vspace{-0.1in}
\eeq
with $\eta$ depending on qubit parameters such as anharmonicity and detuning frequency, as well as coupling strengths.  Using SW perturbation theory one can analytically determine it in weak driving  limit, see Appendix \ref{app eta}. 

\begin{figure}[t]
\begin{center}
  \includegraphics[width=0.235\textwidth]{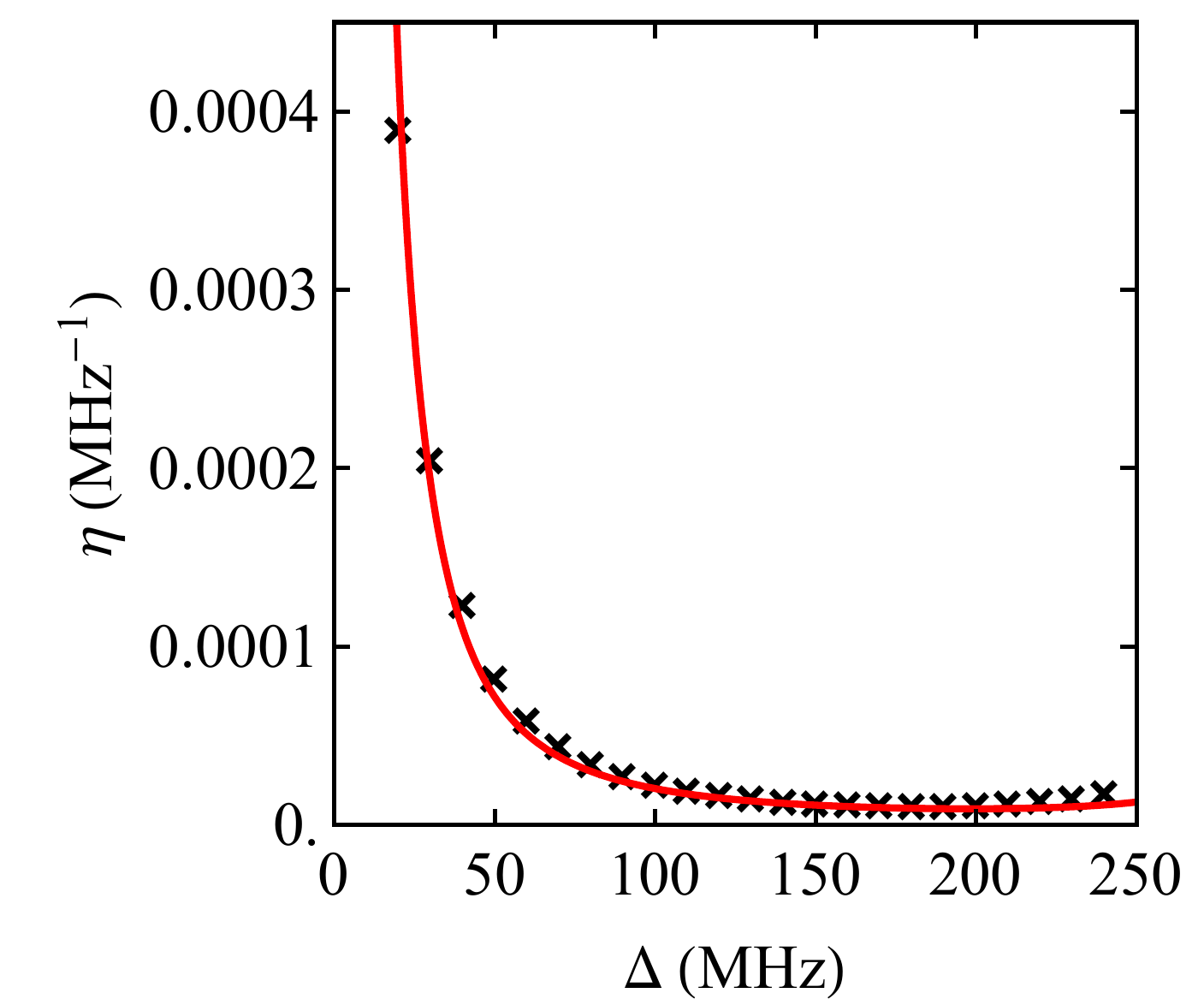}    
   \put(-118,92){(a)}
    %\put(-19,86){(b)}\\
   % \includegraphics[width=0.20\textwidth]{tr_st.pdf}
    \includegraphics[width=0.235\textwidth]{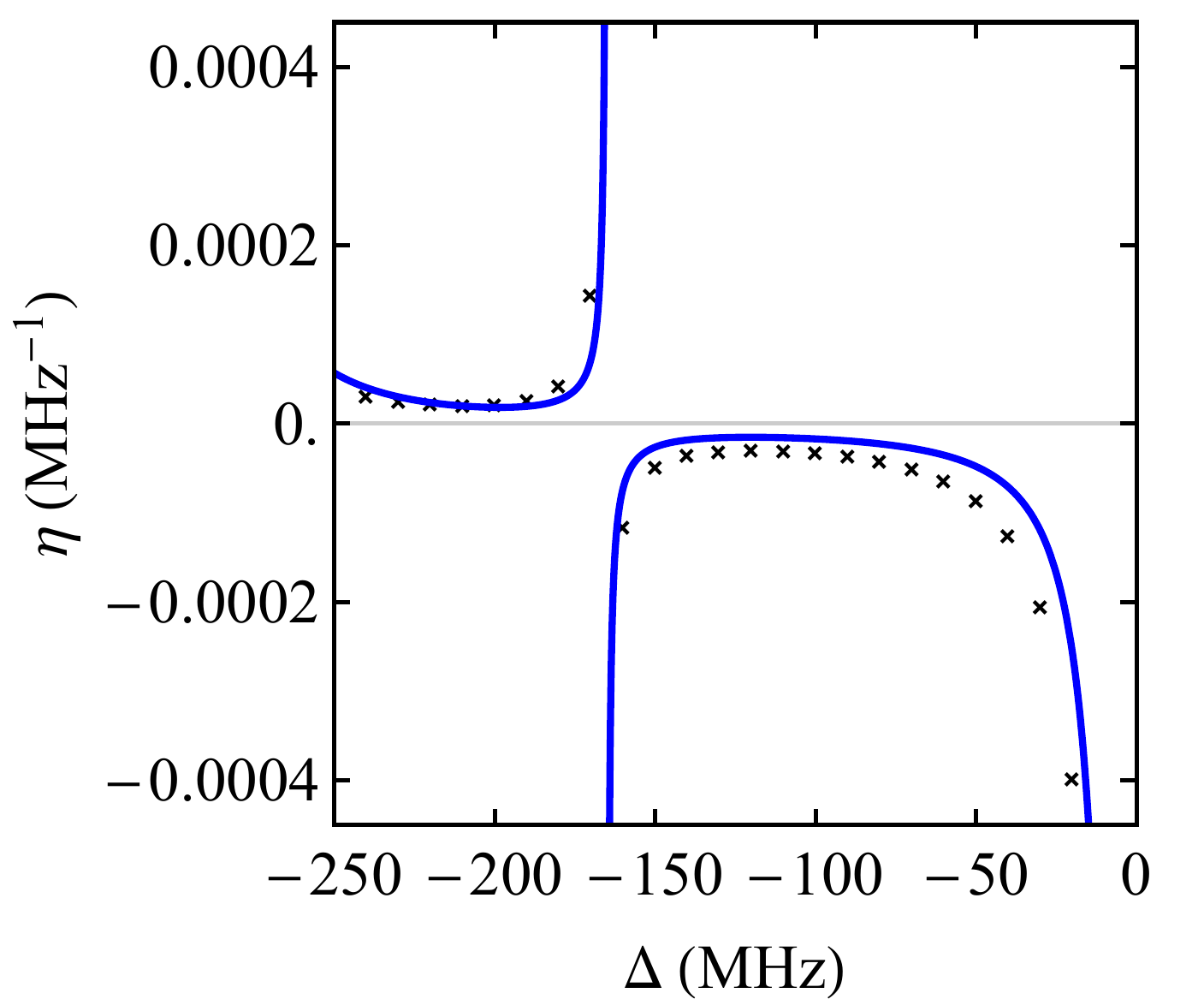}
    \put(-118,92){(b)} \vspace{-0.1in}
    \caption{$\eta$ as a function of qubit frequency detuning, (a)  CSFQ-transmon devices similar to  Fig. \ref{figZZ}(b), and  (b)  transmon-transmon devices similar to Fig. \ref{figZZ}(d) with $g_{12}=2.5$MHz. Cross points are numerical results from LA transformation and solid line from SW transformation.}  \vspace{-0.22in}
    \label{fig:eta}
    \end{center}
\end{figure}

The nonperturbative block diagonalization of LA transformation reproduces weak coupling results, however adds higher order corrections denoted by $O(\Omega^3)$ that contribute to deviations from perturbation theory in strong driving limit.  In Fig. \ref{fig:eta}  we plot perturbative  $\eta$ in solid lines and the LA nonperturbative $\eta$ in cross points, in (a) for a CSFQ-transmon and in (b) for a transmon-transmon circuit. One can see that CSFQ-transmon pair carries positive $\eta$, which makes CR gate to add up positive dynamic ZZ component on top of the static part. This may result in suppression of total ZZ strength if the static part is negative. In the transmon-transmon circuit $\eta$ is negative only at small detuning frequency. Perturbation theory shows divergence, however LA transformation finds that the divergence is unphysical and that ZZ strength remains finite.

% ------ section dyn ZZ
\vspace{-0.15in}
\section{Dynamical ZZ cancellation}
\label{sec. dynzz}

In previous section we studies the impact of CR gate on level repulsion that results in variation of total ZZ interaction from its value at idle qubits. Here we determine ZX and ZZ strengths for some CSFQ-transmon and transmon-trasnmon pairs at a large domain of CR pulse amplitudes. We show in examples how the dynamic ZZ interaction may or may not cancel the static one. This provides a unique opportunity to tune circuits parameters for obtaining opposite sign static and dynamic ZZ components. CR amplitude can control the magnitude of dynamic part and allows for  vanishing total ZZ strength.  In this section we show the dynamic ZZ freedom can take place in transmon-transmon as well as CSFQ-transmon pairs. Moreover the freedom is persistent as long as CR gate is active and this simultaneously improves the CR gate fidelity.

\begin{figure}[h]
\begin{center}
    \includegraphics[width=0.45\textwidth]{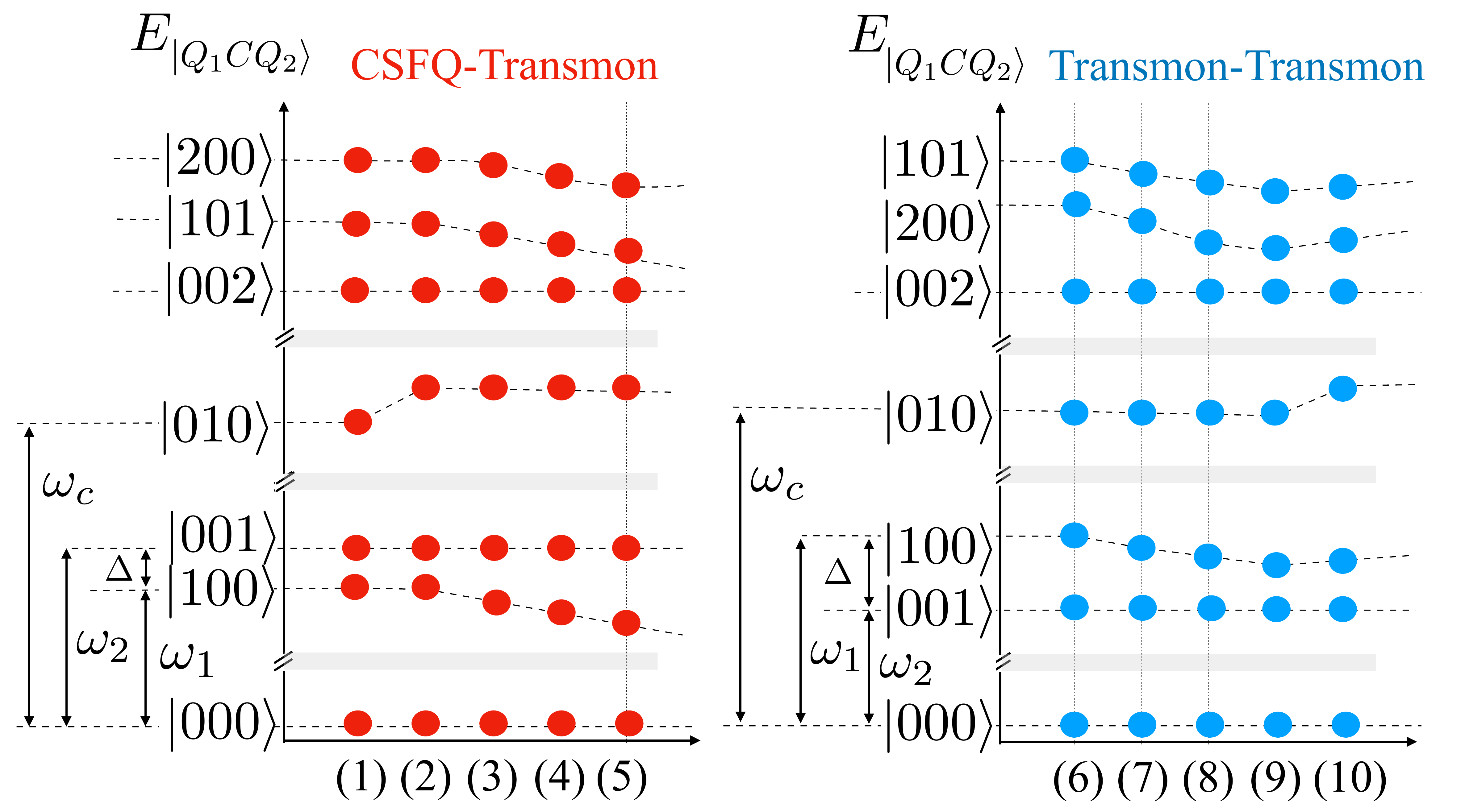}
    \put(-230,122){(a)}\\
   \hspace{3mm} \includegraphics[width=0.2\textwidth]{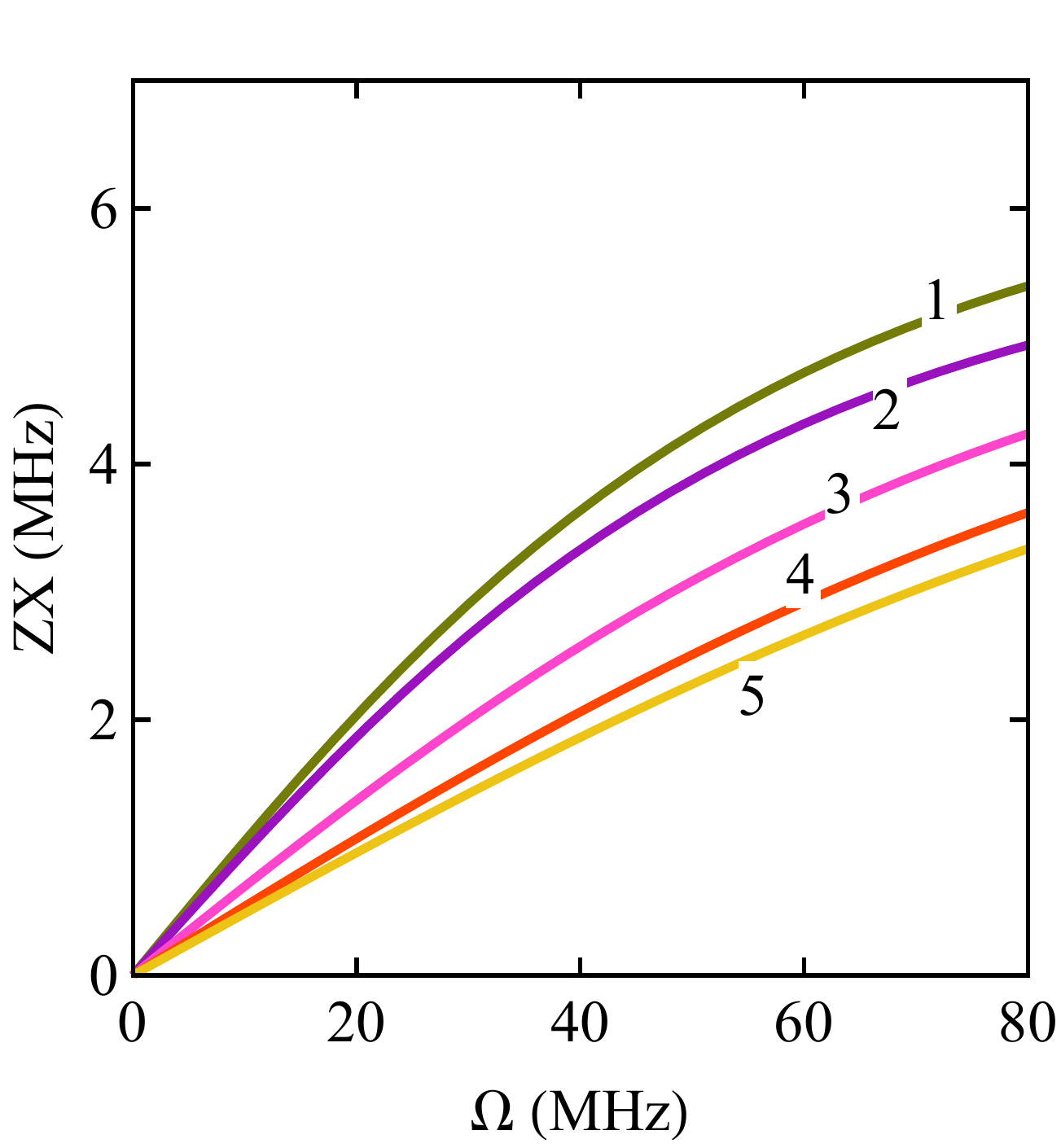}\hspace{5mm}
    \includegraphics[width=0.204\textwidth]{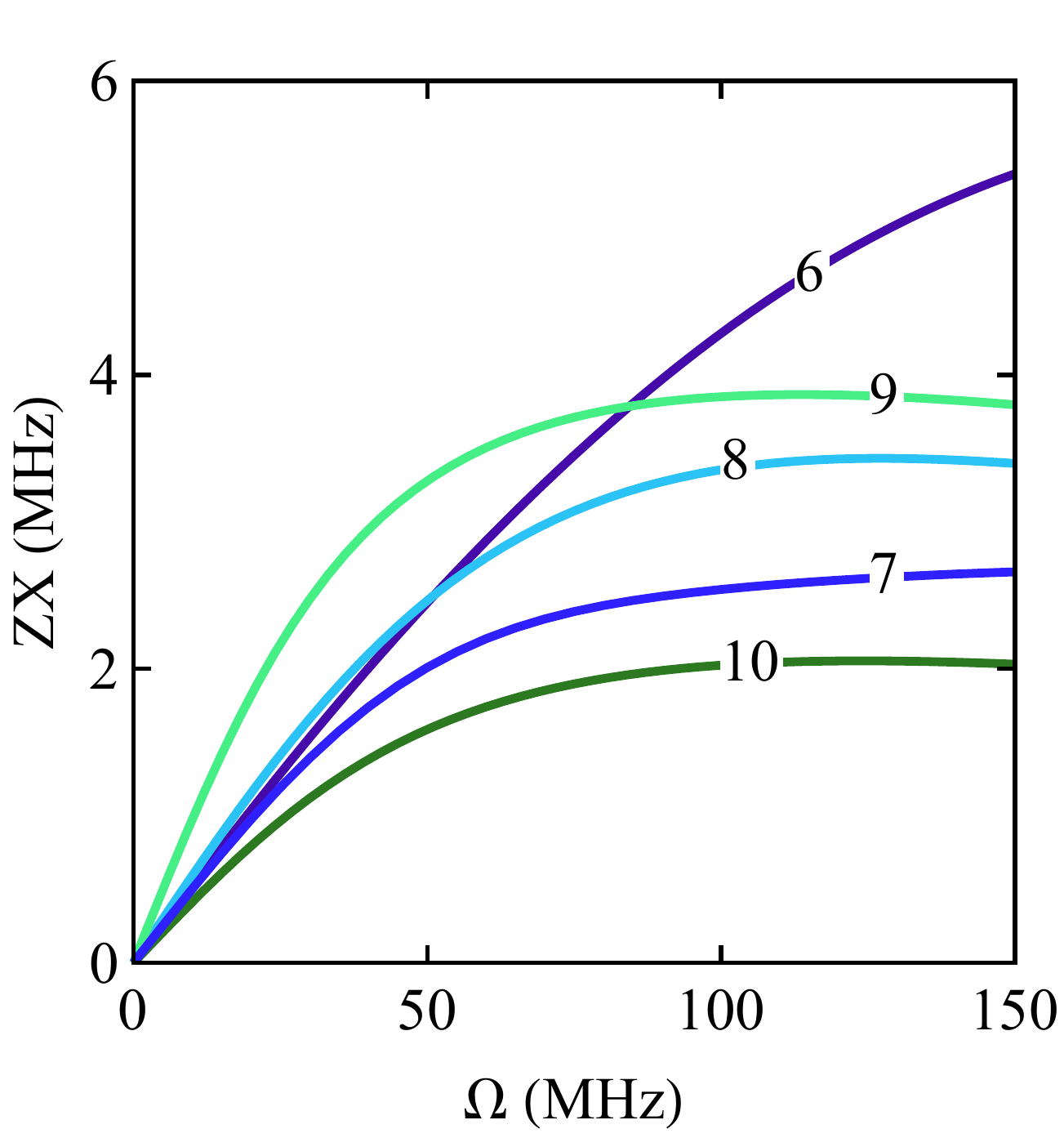}
    \put(-230,102){(b)}
    \put(-118,102){(c)}\\
     \includegraphics[width=0.22\textwidth]{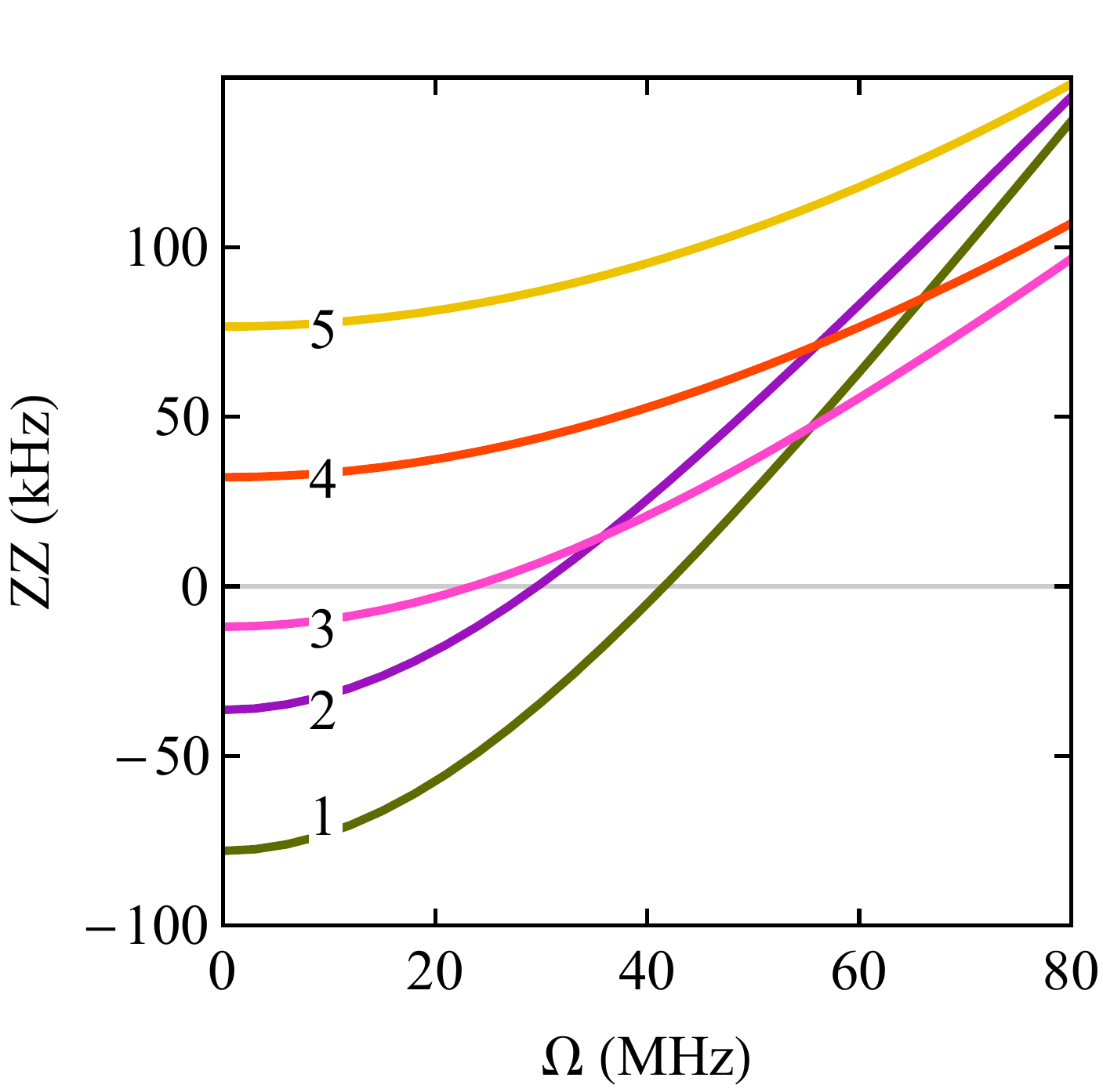}\hspace{2mm}
    \includegraphics[width=0.223\textwidth]{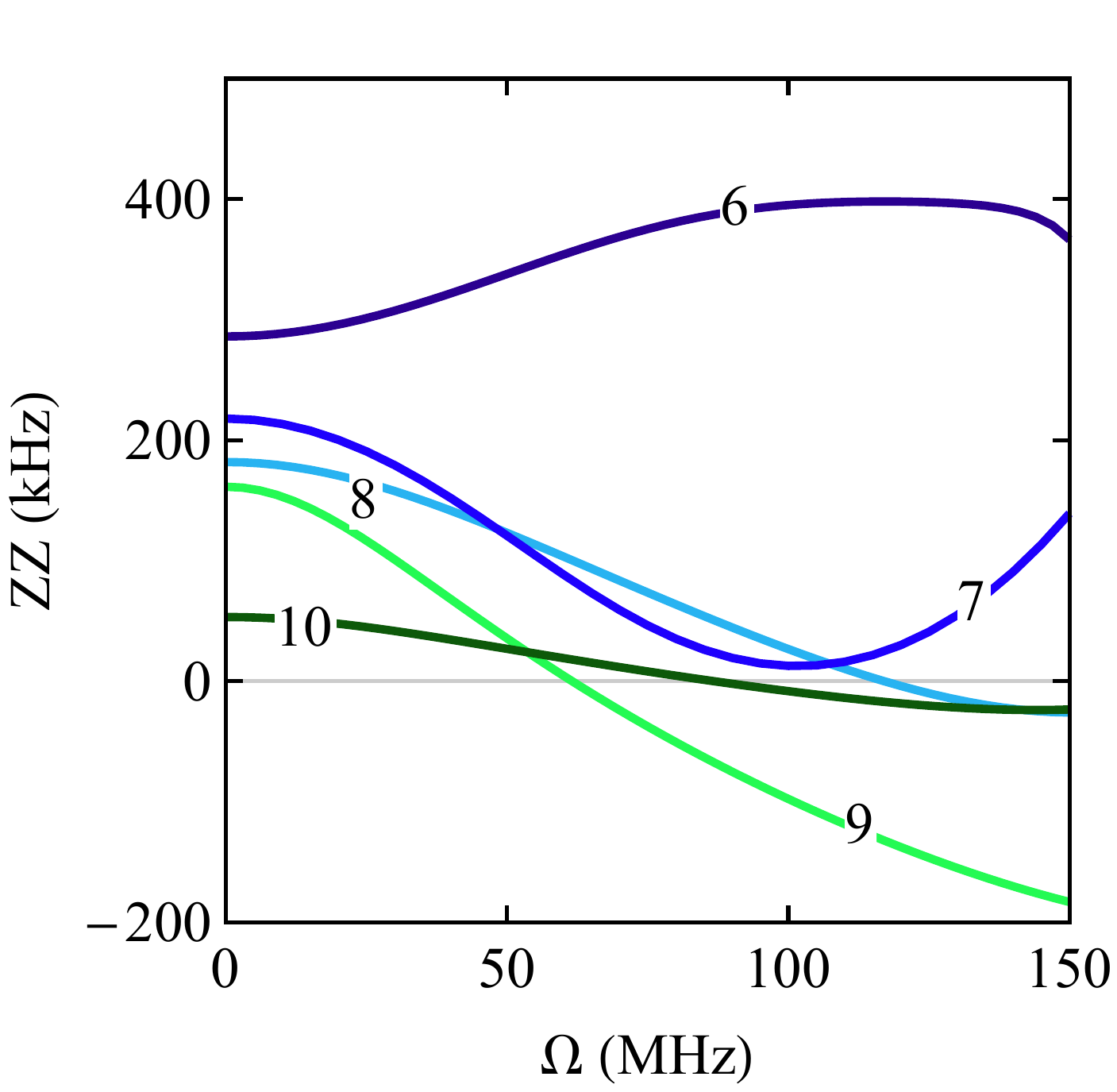}\hspace{5mm}
        \put(-230,102){(d)}
    \put(-118,102){(e)}  \vspace{-0.1in}
        \caption{ (a) Energy diagrams for  five CSFQ-transmon circuits (1-5) and five transmon-transmon circuits (6-10).  The coupling strength under CR gate using nonperturbative least action transformation for: (b) ZX interaction in devices 1-5,  (c) ZX interaction in 6-10, (d) ZZ interaction in 1-5,  and (e) ZZ interaction in 6-10. Common  parameters in 1-5 are $\delta_{1/2}=0.6,-0.33$ GHz, $g_{1c/2c}=80$ MHz, and $g_{12}=0$. In 1: $\Delta=70$ MHz and $\Delta_2=1.1$ GHz.  In 2-5: $\Delta=70, 105, 150, 180$ MHz  and $\Delta_2=1.2$ GHz. Common parameters in 6-10 are $\delta_{1/2}=-0.33$ GHz, $g_{1c/2c}=98, 83$ MHz, and $g_{12}=2.5$ MHz. In 6-9: $\Delta=-200, -150, -100, -50$ MHz and $\Delta_2=1.4$ GHz. In 10: $\Delta=-70$ MHz and $\Delta_2=2$ GHz.} \label{figdynamic}
    \end{center}  \vspace{-0.2in}
\end{figure}

In Fig. \ref{figdynamic} we study two types of qubit-qubit setups: five  CSFQ-transmon samples labelled from 1 to 5, and five transmon-transmon samples labelled from 6 to 10. The corresponding energy levels are depicted in Fig. \ref{figdynamic}(a). In the diagram the energy level of $|010\rangle$, $|100\rangle$, and $|001\rangle$ show their differences in the coupler and the two qubit frequencies. The noncomputational states $|002\rangle$ and $|200\rangle$ in transmon-transmon circuits are  both below $|101\rangle$ and in CSFQ-transmon circuits on its two sides.  Since in these examples the coupler frequency is far detuned from qubits, the repulsions between $|101\rangle$ and noncomputational levels  in transmon-transmon devices have the same sign and sum, and in CSFQ-transmon devices have different signs and subtract.

Applying CR pulse produces desired ZX entanglement between the two qubits. We determine interaction strength from nonperturbative LA transformation. In Fig. \ref{figdynamic}(b) and \ref{figdynamic}(c) the ZX strengths of CSFQ-transmon and transmon-transmon devices have been plotted, respectively. The strength of ZX coupling increases with the CR amplitude, however its growth starts to diminish as soon as $E_{11}$ comes near to other levels.  Figure \ref{figdynamic}(d) and \ref{figdynamic}(e) show total ZZ strength for CSFQ-transmon and transmon-trasmon devices. In samples 1-3 the static ZZ, i.e. at $\Omega=0$, are negative and in 4 and  5 positive. The static ZZ remains the same at all driving amplitudes $\Omega$, i.e. as it is always on, however adding the positive dynamic ZZ component the total suppresses in 1-3 and can go zero at certain amplitude, however in 4 and 5 the cancellation cannot take place.  Interestingly similar ZZ freedom takes place in transmon-transmon circuits as it can be seen in devices 8-10. Notice that device 6 is the IBM experimental device used in Ref. \cite{Sheldon2016CRgate} and we can see it does not show total ZZ freedom. 

Determining the CR amplitude at which total ZZ freedom may take place requires measuring the static ZZ coupling. This can take place by performing a Ramsey pulse sequence on Q1 at state $|0\rangle$ and repeating it at state $|1\rangle$. The difference in frequency between these experiments determines the static ZZ rate \cite{ku2020suppression}.  Once the static ZZ is determined, one can use Eq. (\ref{eq. ZZtot}) to determine how much of CR amplitude is required to set the total ZZ to zero. Equivalently one can also perform a quantum Hamiltonian tomography after applying the CR pulse, by measuring the target qubit state after projecting it on X, Y, and Z axis of the Bloch sphere \cite{Sheldon2016CRgate}. This determines the Pauli coefficients of Eq. (\ref{eq.HCR}), including ZZ term.  

We can use SW perturbation theory to determine the interaction strengths analytically and we expect sufficient accuracy of the results in weak driving limit. By solving the equation $\alpha_{\rm ZZ}=0$, the condition for dynamical ZZ freedom in the first order of $\Delta/\delta_2$ can be obtained at the particular CR amplitude (in the limit of $\Delta/\delta_2\ll 1$):
\beq
 \Omega^* =|\Delta|\sqrt{ \frac{2  (r+\gamma^2)}{r+\gamma(2+\gamma)}} \sqrt{ 1 - C\frac{\Delta}{\delta_2} },
\label{eq.dyncanc} 
\eeq
with $C\equiv \frac{ 1/2+2\gamma+ \gamma^2 + r^2  +  r \gamma (2+\gamma) + \gamma^2 (1+2 \gamma^2)/2r}{ (r +  \gamma^2) (r  + \gamma (2+\gamma) )}$ and  $ r\equiv \delta_1/\delta_2$.
Table \ref{tab} compares the CR amplitude $\Omega^*$ at which dynamical ZZ cancels out the static ZZ interaction. The amplitude $\Omega^*$ is determined using three different methods for devices 1-10. In the row labelled by LA we use nonperturbative least action method to determine total ZZ and find where it is zero. In $O(n)$ row we use the SW-evaluated  static ZZ coupling  $\zeta$ of Eq. (\ref{eq.zeta}) and the SW-evaluated $\eta$ in Appendix \ref{app eta} and  substitute them in Eq. (\ref{eq. ZZtot}) to obtain at what amplitude ZZ becomes zero. Below it we present the results from Eq. (\ref{eq.dyncanc}) and in the last row we evaluate the ratio of $\Delta/\delta_2$ in each device. One can see the results are better consistent in the limit of    $\Delta/\delta_2\ll 1$.

\begin{table}[h]
\label{tab.1}
\begin{tabular}{|c|c|c|c|c|c|c|c|c|c|c|}
\cline{2-11} \cline{3-11} \cline{4-11} \cline{5-11} \cline{6-11} \cline{7-11} \cline{8-11} \cline{9-11} \cline{10-11} \cline{11-11} 
\multicolumn{1}{c|}{} 
& (1) & (2) & (3) & (4) & (5) & (6) & (7) & (8) & (9) & (10)\tabularnewline
\hline 
LA & 42 &  30 & 24 & No &  No & No & No & 115  & 61 & 82 \tabularnewline
\hline 
$O(n)$ & 41 & 31 & 24 & No & No & No & 71 & 83 & 46 & 62\tabularnewline
\hline 
Eq(\ref{eq.dyncanc}) & 41 & 34 & 40 & 20 & No & 110 & 104 & 81 & 46 & 61\tabularnewline
\hline 
\hline
$\Delta/\delta_{2}$ & 0.11 & 0.11 & 0.17 & 0.25 & 0.3 & 0.61 & 0.45 & 0.3 & 0.15 & 0.21\tabularnewline
\hline 
\end{tabular}
\caption{$\Omega^*$ from different methods in devices 1-10 in MHz. `No' indicates devices with no dynamic ZZ freedom.}
\label{tab}
\end{table}

\begin{figure}[t]
\begin{center}
    \includegraphics[width=0.23\textwidth]{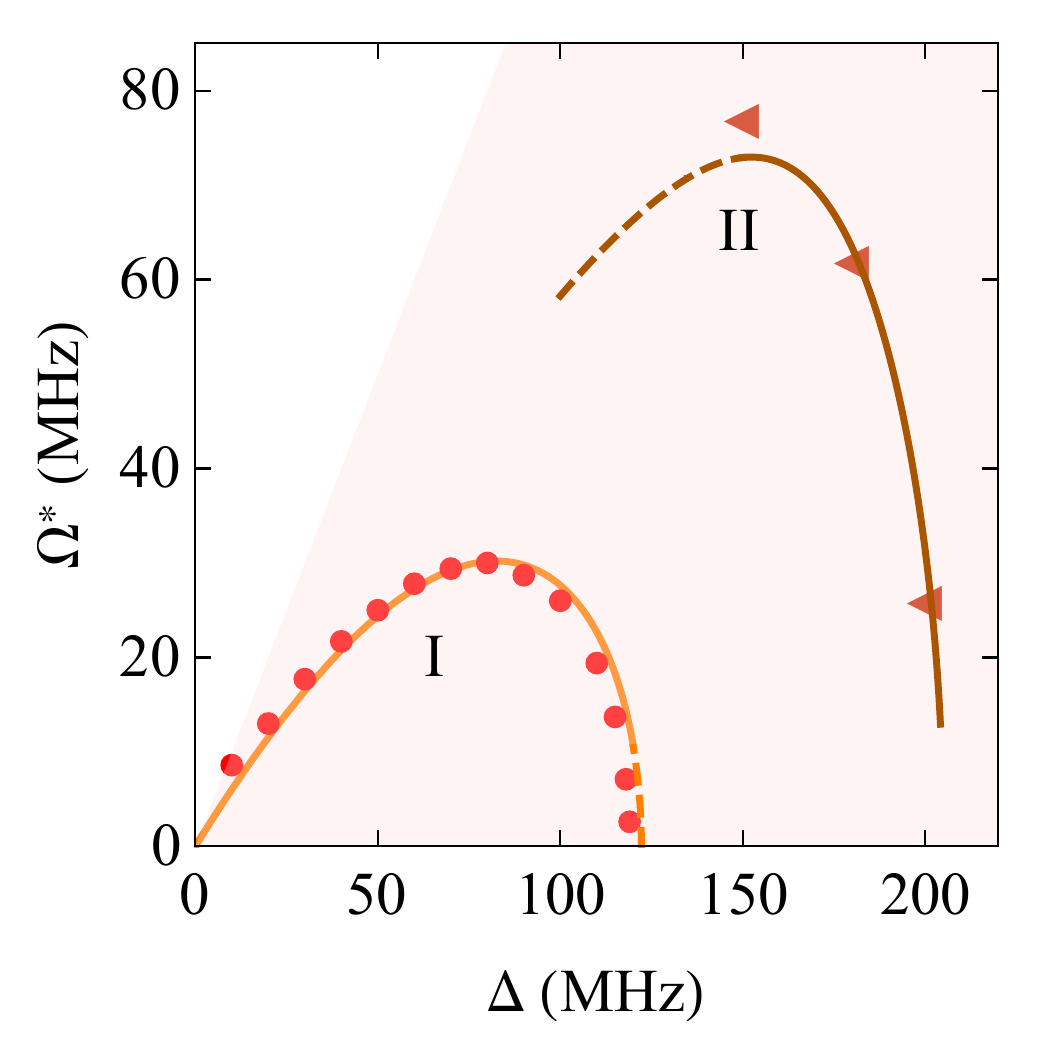}
    \put(-122,112){(a)}\hspace{2mm}
    %\put(-19,86){(b)}\\
   % \includegraphics[width=0.20\textwidth]{tr_st.pdf}
    \includegraphics[width=0.23\textwidth]{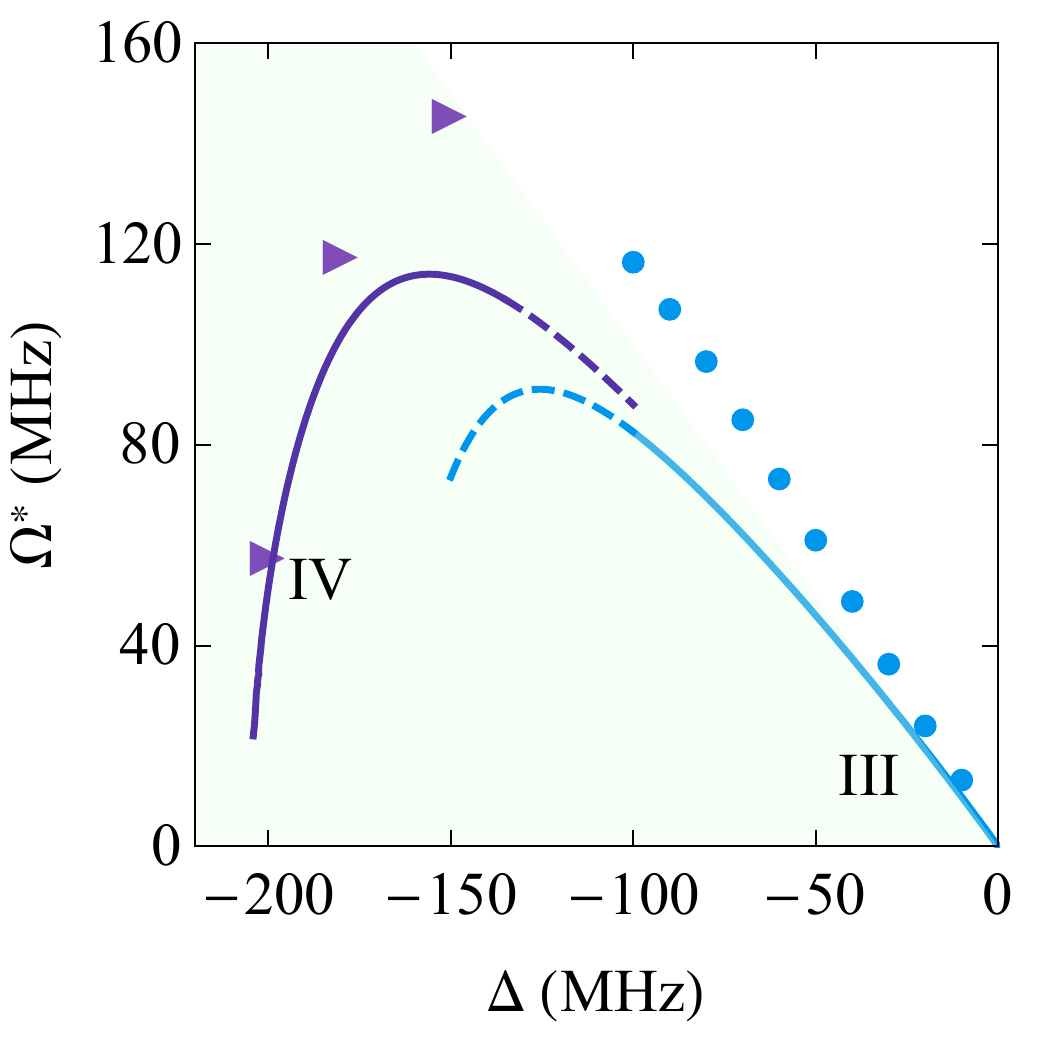}
   % \put(-121,86){(c)}
    \put(-122,112){(b)}
    \caption{The amplitude of dynamical ZZ freedom versus qubit detuning frequency, (a) for CSFQ-transmon with parameters similar to circuit 2-5 devices, \RN{1} with $\delta_{1/2}=0.6, -0.33$ GHz and \RN{2} with $\delta_{1/2}=0.41, -0.39$ GHz,  (b)  for transmon-transmon with parameters similar to 6-9 devices, \RN{3} with $\delta_{1/2}=-0.33,-0.33$ GHz and \RN{4} with $\delta_{1/2}=-0.41,-0.33$ GHz. Lines are perturbative results, dots and triangles are results taken from nonperurbative least action. Shaded area shows the validity domain of perturbation theory.}
    \label{figZZdy}
    \end{center}
\end{figure}

Let us further study the CR amplitudes at which dynamic freedom takes place. In what comes next we work with the parameters of  the devices 2-5  CSFQ-transmons and the devices 6-9  transmon-transmons, except that we keep detuning frequency $\Delta$ a variable and change the anharmonicity at large detuning.  We work out the CR amplitude at which dynamic ZZ freedom can take place within a large range of $\Delta$ in Fig. \ref{figZZdy}(a) and \ref{figZZdy}(b).  Dots and triangles shows results taken using nonperturbative least action transformation and lines are SW perturbative results.  Shaded area shows the validity domain of perturbation theory in the parameter $\Omega/\Delta$, these results show that dynamical cancellation is not limited by the qubit-qubit detuning.  One can see perturbation theory is a crude approximation for transmon-transmon devices, while it works better for CSFQ-Transmon pairs. This is mainly because the static ZZ strength in transmon-transmon pairs is usually large and then cancelling it requires strong driving amplitudes whose accuracy falls outside of perturbation techniques.

One of the noticeable characteristics of dynamic ZZ freedom in CSFQ-transmon pairs as seen in Fig. \ref{figZZdy}(a) is that by increasing detuning frequency  first cancellation amplitude increases, and at larger detuning frequency the amplitude squeezes. In transmon-transmon pairs of Fig. \ref{figZZdy}(b) the amplitude monotonically changes.  These behaviours are consistent with what we found above for the way how $\zeta$ and $\eta$ scale with detuning $\Delta$.

\section{CR gate error}
\label{sec error}

 A pair idle qubits carry a Hamiltonian that contains ZI, IZ and ZZ Pauli coefficient.  CR gate changes some of these coefficients and also as shown in Eq. (\ref{eq.HCR}) introduces some unwanted interactions. Instead of a single CR pulse one can use an echoed CR pulse on the control qubit, which a CR pulse for duration $t$, a $\pi$ rotations about X axis, a $\pi$-shifted CR pulse for the same duration, and another $\pi$ rotation about X axis. Applying an echoed CR gate on the control qubit along with an active cancellation pulse with fine-tuned phase and amplitude on target qubit eliminate all unwanted interactions and leave us with the following couplings only: ZX and ZZ. For details see Ref. \cite{Corcoles2013process, sundaresan2020reducing}.  Applying the echoed CR gate results in an oscillation in the target qubit with the  frequency $f_{\rm eCR} = 2 \sqrt{\alpha_{\rm ZX}^2+\alpha_{\rm ZZ}^2/4}\approx 2 \alpha_{\rm ZX}$.

To quantify the performance of echoed-CR gate on dynamic ZZ free qubits, we numerically simulate the CR gate for several devices in Fig.\ref{figdynamic}. Here we consider the gate length includes two rounded square CR pulses with ignorable rise and fall times, and each of which is followed by a 40ns long $\pi$ pulse. As described in Ref.  \cite{ku2020suppression}, the two-qubit gate will be of the form ${\rm exp}[-i\theta \rm ZX/2]$ with $\theta=2\pi f_{eCR}\tau$ and $\tau$ being the flat-top length of each CR tone. Setting $\theta=\pi/2$ makes a unitary gate that entangles the two qubits. When performing a CR gate, the flap top of the single CR tone $\tau$ satisfies $\tau=1/8\alpha_{\rm ZX}$ when CR driving amplitude is small, and the total gate length $t_g=(2\tau+80)$ ns.

\begin{figure}
     \centering
     \includegraphics[width=.38\textwidth]{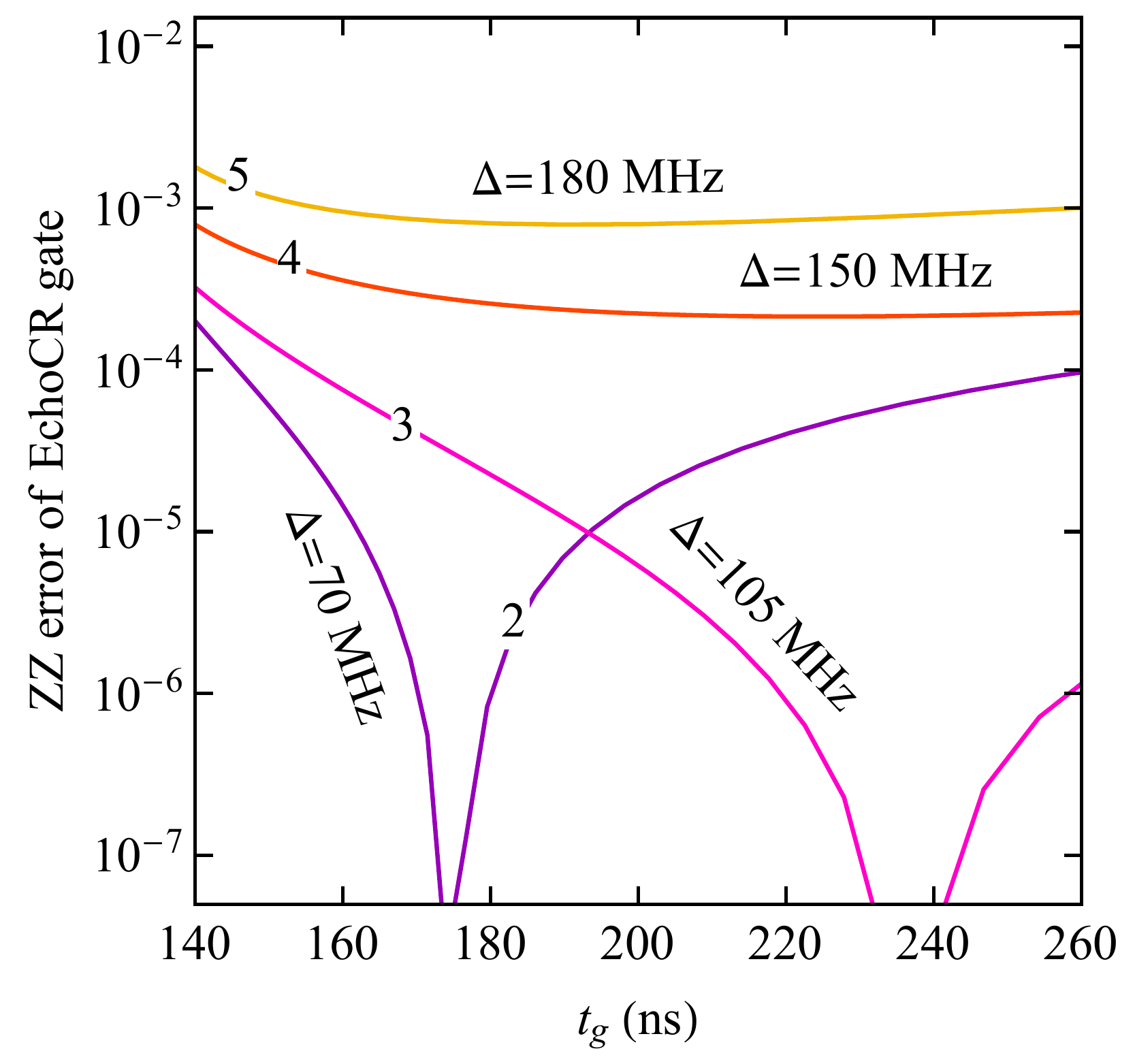}
     %\put(-190,148){(a)}\\
     \put(-202,175){(a)}\\
     \includegraphics[width=.38\textwidth]{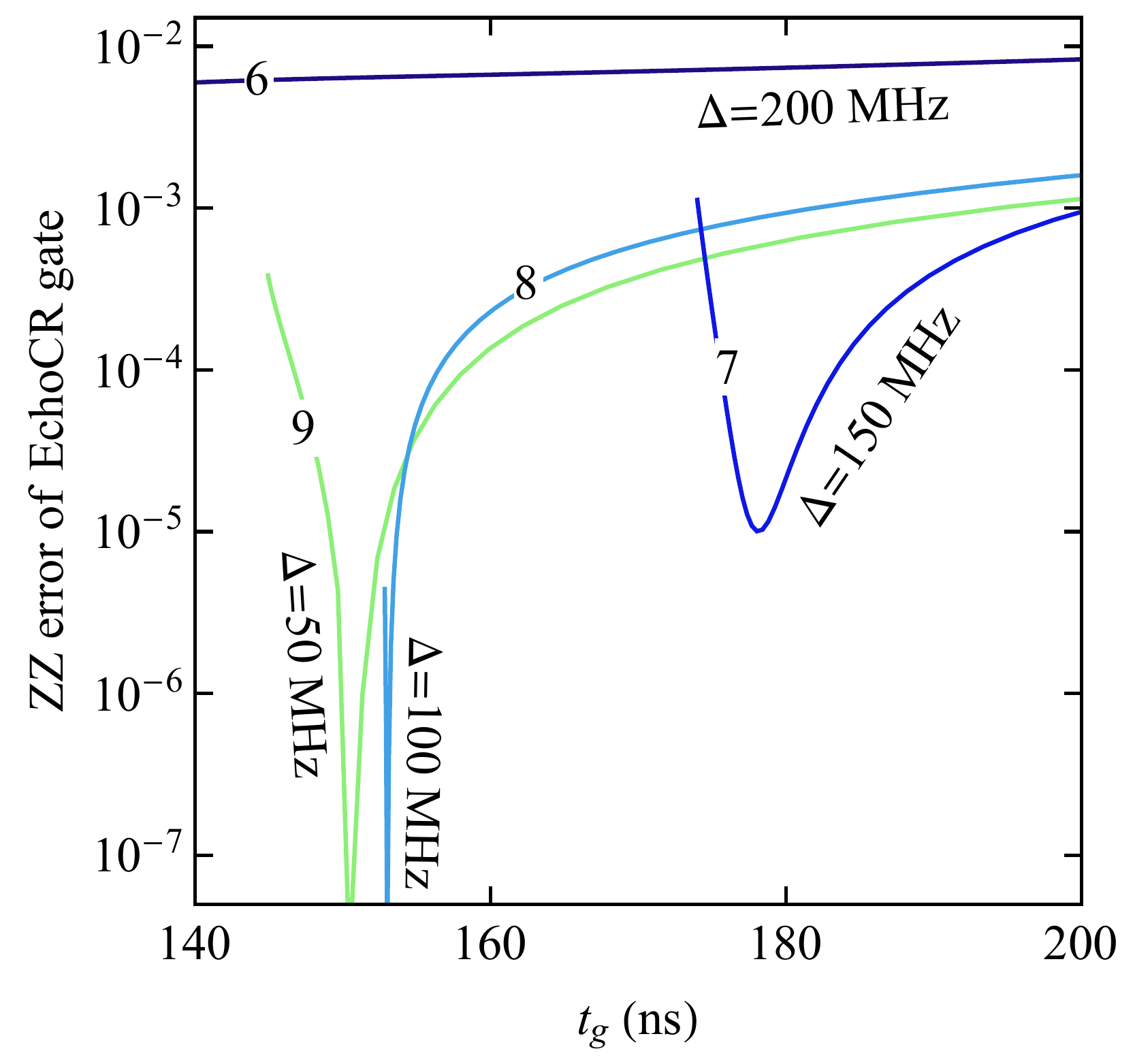}
     \put(-202,175){(b)}\vspace{-0.1in}
     %\put(-190,148){(b)}
     \caption{ZZ error of the echo CR gate as a function of gate length and two-qubit detuning in (a) CSFQ-transmon devices (b) transmon-transmon devices.}
     \label{fidelity}
\end{figure}

We simulate an echoed CR pulse sequence for implementing a ZX($\pi/2$) gate. 
We compute the two-qubit error per gate by evaluating how the unitary evolution of the echoed CR gate evolves an initial state. The presence of ZZ interaction determines a state-dependent phase error in the desired state. We evaluate the infidelity of final state. Figure \ref{fidelity} shows the CR gate error caused by ZZ interaction as a function of gate length and qubit-qubit detuning with infinite coherence time. In these plots we ignore the decoherence effect on the gate as assume that qubits can have desirably long coherence times $T_1$ and $T_2$. In CSFQ-transmon devices 2 and 3 where total ZZ can be dynamically set to zero, we can get the ZX rotation free of parasitic ZZ interaction and therefore the gate error drops at certain gate times.  For device 2  Fig. \ref{figdynamic}(a) and \ref{figdynamic}(d) show that where the dynamic freedom takes place is at $\alpha_{\rm ZX}\sim 2.7$ MHz.  Such a frequency requires $\tau\sim 46$ns for each CR pulse to perform $\pi/2$ ZX rotation. Considering the second CR pulse and the additional $\pi$ rotation will sum the total echoed-CR pulse length  to ~172 ns. In Fig. \ref{fidelity}(a) one can see that device 2 performs a perfect gate with no error at this gate length. In device 3 the cancellation takes place at  a $\alpha_{\rm ZX}$
that is smaller by a factor of 1/1.7 and this causes the prolongation of the gate to become ~235 ns. While the gate error in devices 4 and 5 decreases  as the gate becomes longer due to reduction of  total ZZ interaction, however the gate error in absence of decoherence stays can be in the scale of $10^{-3}$. 

In transmon-transmon devices almost similar behavior is expected and one can find errorless ZX($\pi/2$) in dynamical ZZ free trasmons. The perfect gate time in devices 8 and 9 are shorter compared to CSFQ-transmon pairs. The reason for such improvement is that in these transmon-transmon devices the dynamic cancellation takes place at an amplitude that causes faster ZX rotation. However, one can see in Fig. \ref{fidelity}(b) that some transmon-transmon devices such as 7, 8, and 9 show some cutoff in their minimum gate length for ZX($\pi/2$) rotation. Figure \ref{figdynamic}(c) shows that $\alpha_{\rm ZX}$ rate starts to saturate after some amplitudes and cannot increase anymore. This saturation puts limitation on flap top length $\tau$ such that it cannot become shorter than a minimum, i.e. $\tau_{\rm min} = 1/8\alpha_{\rm ZX}^{\rm max}$. This will introduce a cutoff  on echo-CR gate length to become limited to longer than $t_g^{\rm min}=(1/4\alpha_{\rm ZX}^{\rm max}+80)$ ns. For instance device 7 in Fig. \ref{figdynamic}(c) reaches to a saturation at $\sim 2.5$MHz and this introduces a gate time cutoff below $\sim 180$ ns as shown in Fig. \ref{fidelity}(b). Moreover the device 7 shows some dynamic ZZ suppression to a minimum of $\sim$20kHz and our analysis show that at this gate time the error although cannot be eliminated but it can be suppressed to $10^{-5}$ for transmons without decoherence error. 
 
 Overall, to perform an ideal cross-resonance gate, the following two conditions should be satisfied: 1) static ZZ interaction is zero when two qubits are in the idle state. 2) Maximum ZX component and zero total ZZ interaction should be realized synchronously when the two qubits are under driving. Although it is more convenient to implement dynamical ZZ cancellation on a CR gate, this approach can still be generalized to suppress the build-in ZZ interaction in any gates. In large quantum processors we expect to have a huge number of pairs of qubits that have non-zero ZZ interaction contributing to the gate error, if we apply the well-designed CR-type pulses on a pair of qubits and make a $2\pi$ rotation by ZX operator, leaving qubits with no ZX rotation. Such a system will not suffer from parasitic ZZ interaction at all while it maintains its original state.

\vspace{-0.2in}
\section{Discussion}

In this paper, we studied the parasite ZZ interaction and the possibility of freeing qubits from it. Zeroing the ZZ strength can take place in two ways: 1) in idle qubits circuit parameters can be found such that the static ZZ strength becomes zero.  The static ZZ freedom can take place in qubits with opposite sign anharmonicity. 2) in driven qubits with a microwave pulse a new ZZ component adds on top of the static part  making it possible to cancel it out, making total ZZ strength zero. This dynamic ZZ freedom takes place at a driving amplitude that could be determined from circuit QED. 

Eliminating the static ZZ will make idle qubits to no longer suffer from accumulating state dependent phase error across the circuit. Moreover, when a two-qubit gate is active, eliminating  all of the parasitic ZZ interaction as long as the gate is active results in a large increase in the gate fidelity.

Our derivation was restricted to two interacting qubits.   The ZZ free qubits can have many advantages; for instance a complete understanding of pairwise suppression of ZZ interaction in a large circuit with many qubits can result in quantum computation on the circuit with less error.  Future research is needed to study clusters of interacting three-body and higher many-body qubits. Although the general expectation is that the coupling strength of these terms to be weaker than pairwise interactions, however they can generate additional gate error as well as crosstalk error across the circuit. Elimination of parasitic many body interactions such as ZZZ couplings depends on qubit connectivity, i.e. whether qubits are connected in a loop or are in linear connection. This paves the road toward optimal operation for reducing quantum computational error to below the threshold of error-correction.

\section{Acknowledgement}
We thank David DiVincenzo and Britton Plourde for insightful discussions. We also thank the OpenSuperQ project (820363) of the EU Flagship on Quantum Technology, H2020-FETFLAG-2018-03, for support.

 \appendix

\section{CSFQ Hamiltonian}
\label{app.CSFQ}
Figure \ref{fig. diag}(a) helps to write the CSFQ Lagrangian $\mathscr{L}=\frac{1}{2} (\Phi_0/2\pi)^2 \{  (C_{Sh}+\alpha C_J)(\dot{\phi}_1-\dot{\phi}_3)^2 + C_J (\dot{\phi}_1-\dot{\phi}_2)^2 + C_J (\dot{\phi}_2-\dot{\phi}_3)^2 \} +  E_{J} \{\cos(\phi_1-\phi_2) +\cos(\phi_2-\phi_3) + \alpha  \cos (2\pi f -  \phi_1+\phi_3) \}$ with external flux number $f\equiv \Phi_{\textup{ext}}/\Phi_0 $ and  the external magnetic flux $\Phi_{\textup{ext}}$ penetrating the loop. Defining new phase variables $\phi=\phi_1- \phi_3$ and $\phi'=\phi_1-2\phi_2 + \phi_3$ can help to uncouple degrees of freedom. %by which the qubit Hamiltonian becomes $H_{Q2}=(1/2)\{E_{Cm}n_m^2   +  E_{Cp} n_p^2\} -  E_{J} \{2\cos(\phi_m/2) \cos(\phi_p/2) - \alpha \cos (2\pi \Phi/\Phi_0 -  \phi_m)\}$ 
Considering the charging energy of a typical capacitance $C$ being $E_{C}\equiv e^2/2C$, the effective capacitance associated with the modes $\phi$ and $\phi'$ are $C\equiv C_{Sh}+\alpha C_J+C_J/2 $ and $C'=C_J/2$, respectively.  The key to C-shunt flux qubit is the large shunting capacitance $C_{Sh} \gg C_J$ which indicates that the mode $\phi'$ oscillates with an order of magnitude larger frequency compared to the oscillation of $\phi$ mode. Therefore the large charging energy of $\phi'$ mode makes its  contribution to qubit potential negligible. This makes qubit potential separable and eventually the $\phi'$ mode can be safely discarded from dynamics. 
This helps to write down the CSFQ qubit Hamiltonian 
 $H=4 E_{C} n^2  - 2E_{J} \cos\left(\phi/2\right)  - \alpha E_{J} \cos\left(2\pi f -  \phi \right)$. The 1D potential has a single minimum for  $\alpha<1/2$, which is the domain of our interest, otherwise indicates double minima. 
Bringing the qubit to the bottom of its potential minimum, namely ``Sweet Spot'' or (SS), will produce the longest $T_2$ coherence for the qubit. 
 
\begin{figure}
     \centering
     \includegraphics[width=.25\textwidth]{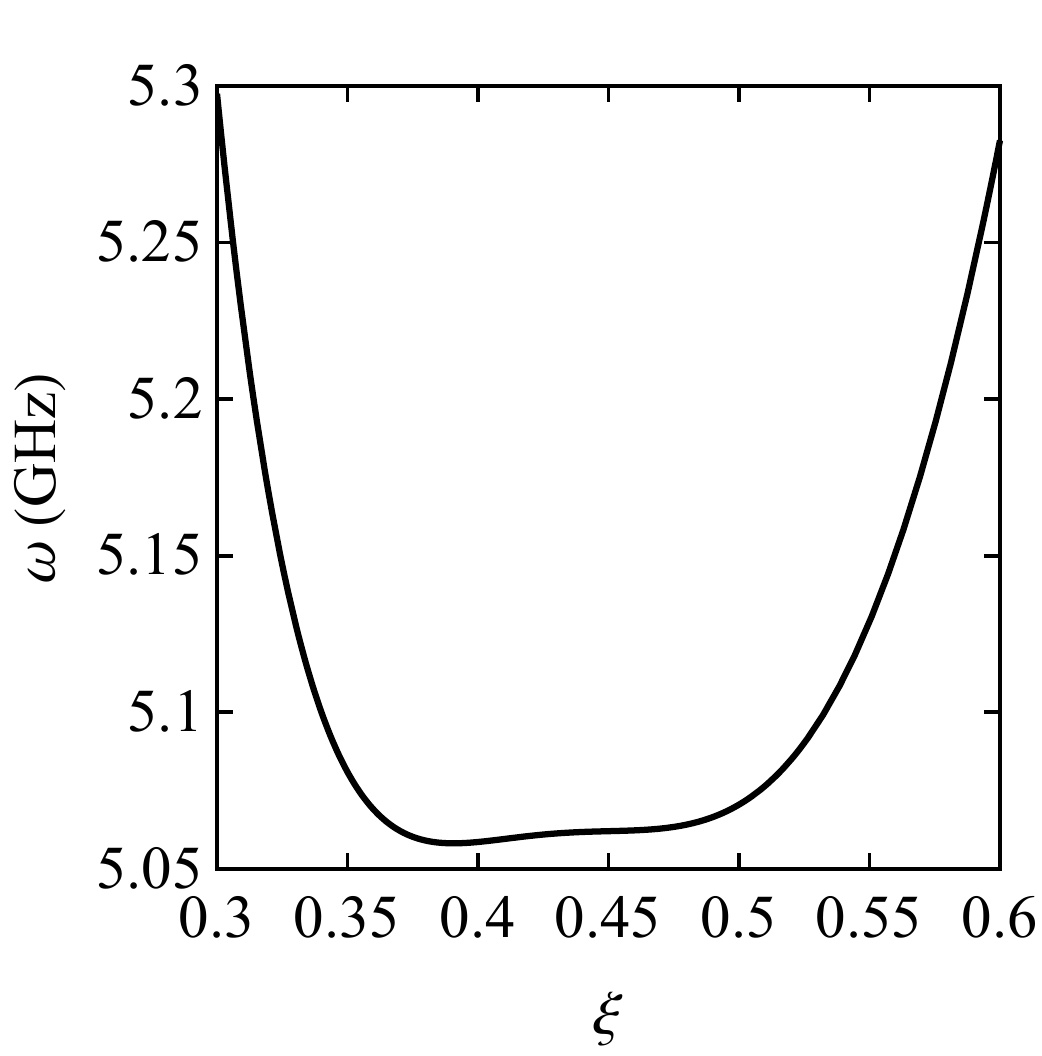}
     \caption{Frequency as a function of $\xi$ at the sweet spot. Simulation parameters $E_C=0.292$ GHz, $E_J=108.9$ GHz, $\alpha=0.43$ and $f=0.5$.}
     \label{sfig1}
     \vspace{-0.1in}
\end{figure}

\begin{figure}
     \centering
     \includegraphics[width=.225\textwidth]{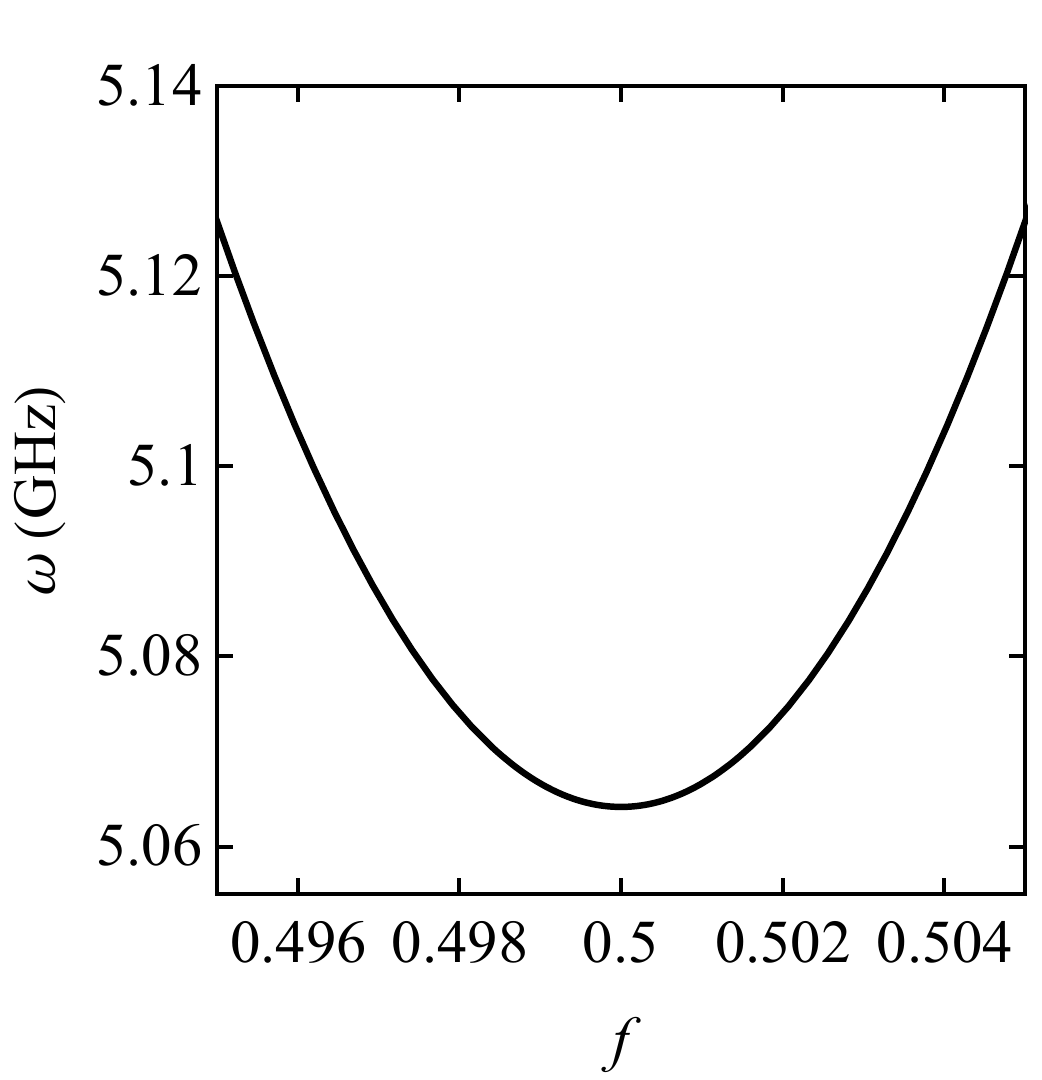}\quad
     \includegraphics[width=.225\textwidth]{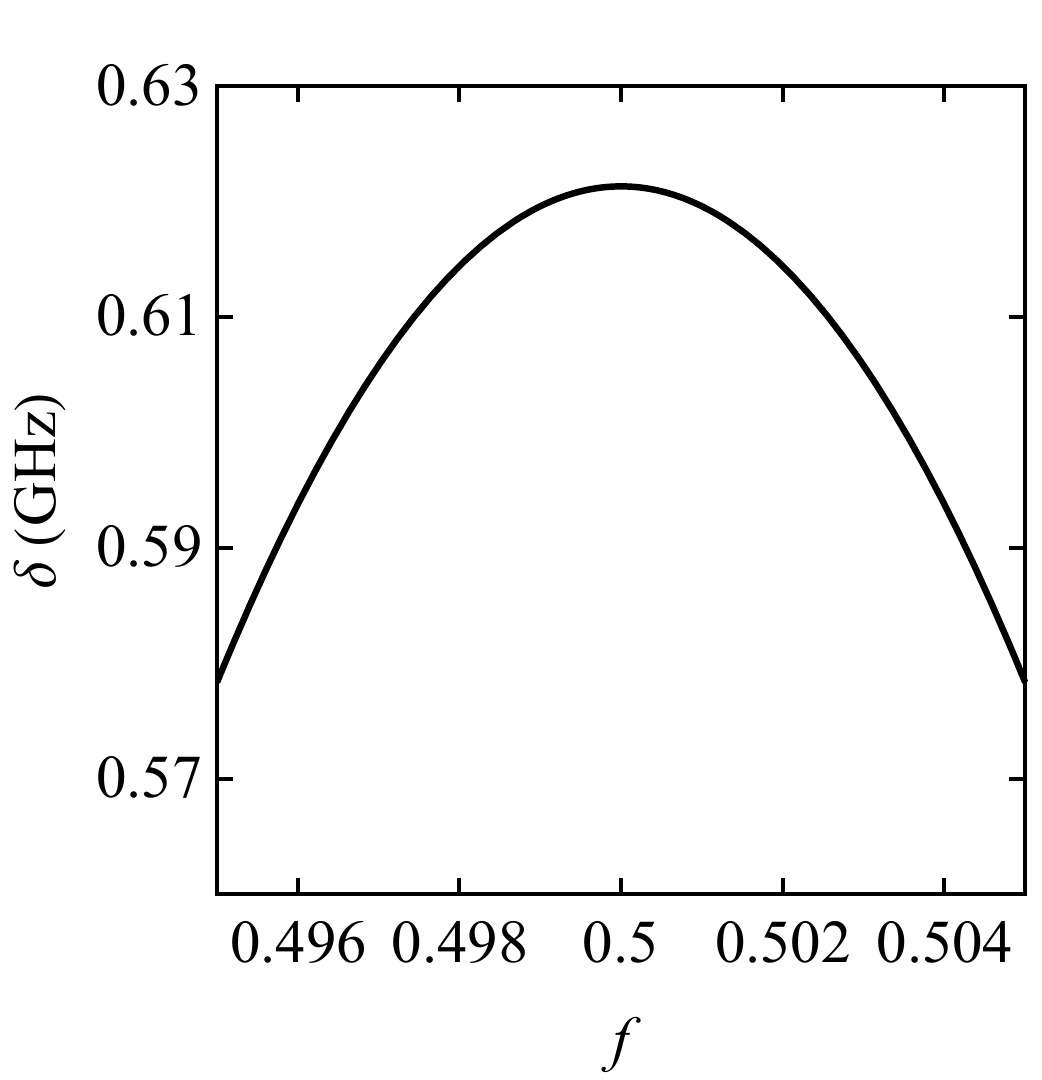}
     \put(-243,110){(a)}
     \put(-123,110){(b)}
     \caption{Frequency (a) and anharmonicity (b) as a function of external flux $f$ using perturbation theory.}
     \label{sfig2}
\end{figure}

Compare to a transmon, higher order terms (>4) in the expansion of the CSFQ potential also contributes to the eigenvalues, which will change the zero point fluctuation to an unknown number. To be more precise, the Hamiltonian can be quantized in terms of field operators, e.g. $n=i(\hat{a}-\hat{a}^{\dagger})/2\xi$ and  $\phi=\xi(\hat{a}+\hat{a}^{\dagger})$ with $\xi$ being the expansion parameter which will minimize the total energy, the normal ordered Hamiltonian then can be written as 
 \beqr \nonumber 
H &=& -\frac{E_c}{\xi^2} \left(\hat{a}^{\dagger}-\hat{a}\right)^{2}+ \sum_{u=0}^{\infty}\xi^{2u+2} \sum_{v=0}^{u} \frac{U_{2u+2}(\phi_0)}{2^{u-v}\left(u-v\right)!} \\ && \times \nonumber  \sum_{w=-\left(v+1\right)}^{v+1}  \frac{\left(\hat{a}^{\dagger}\right)^{v+1+w}\left(\hat{a}\right)^{v+1-w}}{\left(v+1+w\right)! \left(v+1-w\right)!}  \\ 
&+& \sum_{u=0}^{\infty}\xi^{2u+1} \sum_{v=0}^{u} \frac{U_{2u+1}(\phi_0)}{2^{u-v}\left(u-v\right)!} \\ && \times \nonumber  \sum_{w=-\left(v+1\right)}^{v} \frac{\left(\hat{a}^{\dagger}\right)^{v+1+w}\left(\hat{a}\right)^{v-w}}{\left(v+1+w\right)!\left(v-w\right)!}
%&\approx& \omega(f) \hat{a}^\dagger \hat{a} +  \frac{\delta(f)}{12} (\hat{a} + \hat{a}^\dagger)^4, \ \  \delta>0,
\eeqr
with $U_{n}(\phi_0)\equiv{\partial^{n}U(\phi_0)}/{\partial\phi_0^{n}}$ and $U(\phi_0)\equiv -2E_{J}\cos\phi_0/2-\alpha E_{J}\cos\left(2\pi f-\phi_0\right)$ and $\phi_0$ being the phase of minimum $U(\phi_0)$, i.e. $\phi_0=-2 \pi \alpha (\delta f)/(1/2-\alpha)$, which vanishes at sweet spot.  %The leading order of CSFQ qubit frequency at sweet spot is $\hbar \omega(f=1/2)= E_c/2 \phiz^2+ 5E_J \phiz^2 (1-2\alpha )/4$, by substituting zero pint fluctuations the qubit frequency becomes $\omega= (\sqrt{8 E_J E_c}/\hbar) (27-52 \alpha)/8$. Anharmonicity can also be found  will $\delta(f=1/2)=E_2-2\hbar\omega=...$ STOP END (1).]
 By solving Schr\"{o}dinger equation, the eigenenergies $E_n$ can be obtained using perturbation theory. Unperturbed eigenvalues $E_n^{(0)}$ and first three order corrections are given by
% \beqr \nonumber 
%E_{n}^{(0)}&=&\left(\sum_{l=1}^{L}\frac{U^{(2l)}\xi^{2l}}{\left(2l-2\right)!!}+\frac{2E_{Cm}}{\xi^{2}}\right)n+\sum_{k=2}^{n}\sum_{l=2}^{L}\frac{U^{(2l)}\xi^{2l}}{\left(2l-2k\right)!!}\frac{n!}{\left(n-k\right)!}\\ \nonumber
%\delta E_{n}^{(1)}&=&0, \ \ 
%\delta E_{n}^{(2)}=\sum_{k\neq n}\frac{V_{nk}^{2}}{E_{n}^{(0)}-E_{k}^{(0)}}\\ \nonumber
%\delta E_{n}^{(3)}&=&\sum_{k\neq n}\sum_{m\neq n}\frac{V_{nm}V_{mk}V_{kn}}{\left(E_{n}^{(0)}-E_{k}^{(0)}\right)\left(E_{n}^{(0)}-E_{m}^{(0)}\right)}
%\eeqr
\begin{equation}
\begin{split}
E_{n}^{(0)}&=\left(\sum_{l=1}^{L}\frac{U^{(2l)}\xi^{2l}}{\left(2l-2\right)!!}+\frac{2E_{C}}{\xi^{2}}\right)n\\
&+\sum_{k=2}^{L}\sum_{l=2}^{L}\frac{U^{(2l)}\xi^{2l}}{\left(2l-2k\right)!!}\frac{n!}{\left(n-k\right)!}\\
E_{n}^{(1)}&=0\\
E_{n}^{(2)}&=\sum_{k\neq n}^{n+L}\frac{V_{nk}^{2}}{E_{n}^{(0)}-E_{k}^{(0)}}\\
E_{n}^{(3)}&=\sum_{k\neq n}^{n+L}\sum_{m\neq n}^{n+L}\frac{V_{nm}V_{mk}V_{kn}}{\left(E_{n}^{(0)}-E_{k}^{(0)}\right)\left(E_{n}^{(0)}-E_{m}^{(0)}\right)}
\label{E}
\end{split}
\end{equation}
with
\begin{equation}
\begin{split}
V_{nm}&=x_{|m-n|,|m-n|}\sqrt{\frac{\max(m,n)!}{\min(m,n)!}}\\
&+\sum_{s=0}^{\min(m,n)}x_{\max(m,n)-s,\min(m,n)-s}\sqrt{\frac{n!m!}{s!}}\\
x_{a,b}&=\delta_{ab}\sum_{k=a}^L\sum_{u=0}\Theta(k-a-2u)\frac{U^{(k)}\xi^k}{(k-a)!!a!}\\
&+(1-\delta_{ab})\sum_{k=a}^{L}\sum_{u=0}\Theta(k-a-2u)\frac{U^{(k+b)}\xi^{k+b}}{b(k-a)!!a!}\\
&-\delta_{a2}\delta_{b2}\frac{E_{C}}{\xi^{2}}\nonumber
\end{split}
\end{equation}
Sum up unperturbed energy and all three orders corrections, one can obtain the analytical formula of CSFQ eigenvalues, then frequency $\omega_{01}=(E_1-E_0)/\hbar$ and anharmonicity $\delta=(E_2-2E_1+E_0)/\hbar$ can be evaluated. Here is an example to illustrate how to find these parameters. Consider a CSFQ with $E_C=0.292$ GHz, $E_J=108.9$ GHz, $\alpha=0.43$ and $f=0.5$. Firstly, expand the potential to 20th order (L=10) at the sweet spot and plot the frequency $f_{01}$ as a function of $\xi$. 
Determine $\xi$ by finding the minimum as shown in Fig. \ref{sfig1} (approximately $\xi\approx\phiz/\sqrt{2}$), then substitute $\xi$ back to the Eq. (\ref{E}), the corresponding frequency and anharmonicity spectra are shown in Fig. \ref{sfig2}.

 \section{Full Hamiltonian}
 \label{app fullH}
 
To understand how the interaction terms change the bare basis, we apply the full Hamiltonian (\ref{eq. ham})  on the states $\left|n_1,n_c,n_2\right>$. Consider the first order of $g_{ij}$ ($i\neq j$ and $i,j=1,2,c$), the result is written as
\beqr \nonumber
&& \hat{H}{|n_1,n_{c},n_2\rangle} =    \\ \nonumber &&  [\omega_1(n_1)+\omega_{c}(n_{c}) +\omega_2(n_2) ] \  {|n_1,n_{c},n_2\rangle}  \\ \nonumber &&   + g_{12}  [  \sqrt{n_1 n_2}\ |n_1-1,n_{c},n_2-1\rangle     \\ \nonumber &&  \ \ \ \ \ \ \  +  \sqrt{n_1 (n_2+1)} \ |n_1-1,n_{c},n_2+1\rangle    \\ \nonumber &&   \ \ \ \ \ \ \ + \sqrt{(n_1+1) n_2}\ |n_1+1,n_{c},n_2-1\rangle 
  \\ \nonumber &&   \ \ \ \ \ \ \ + \sqrt{(n_1+1) (n_2+1)}\ |n_1+1,n_{c},n_2+1\rangle ]
  \\ \nonumber &&   + g_{1c}  [  \sqrt{n_1 n_{c}}\ |n_1-1,n_{c}-1,n_2\rangle     \\ \nonumber &&  \ \ \ \ \ \ \  +  \sqrt{n_1 (n_{c}+1)} \ |n_1-1,n_{c}+1,n_2\rangle    \\ \nonumber &&   \ \ \ \ \ \ \ + \sqrt{(n_1+1) n_{c}}\ |n_1+1,n_{c}-1,n_2\rangle 
  \\ \nonumber &&   \ \ \ \ \ \ \ + \sqrt{(n_1+1) (n_{c}+1)}\ |n_1+1,n_{c}+1,n_2\rangle ]
  \\ \nonumber &&   + g_{2c}  [  \sqrt{n_2 n_{c}}\ |n_1,n_{c}-1,n_2-1\rangle     \\ \nonumber &&  \ \ \ \ \ \ \  +  \sqrt{n_2 (n_{c}+1)} \ |n_1,n_{c}+1,n_2-1\rangle    \\ \nonumber &&   \ \ \ \ \ \ \ + \sqrt{(n_2+1) n_{c}}\ |n_1,n_{c}-1,n_2+1\rangle 
  \\ \nonumber &&   \ \ \ \ \ \ \ + \sqrt{(n_2+1) (n_{c}+1)}\ |n_1,n_{c}+1,n_2+1\rangle]
 \eeqr
The equation above shows that when interactions turn off, $\left|n_1,n_c,n_2\right>$ is the eigenstate of the non-interacting Hamiltonian, when we turn on the interactions, level crossing takes place with the coupling strength scaling with the first order of $g_{ij}$ as shown in Fig. \ref{fig. diag}(c).
 
 \section{Principle of Least Action}
 \label{app LA}
Consider a general $d\times d$ Hermitian Hamiltonian H, which can be transformed into a block diagonal matrix $\mathscr{H}$ using Eq. (\ref{eq. leastaction})  with  two blocks $\mathscr{H}_{nn}$ and $\mathscr{H}_{mm}$, where $n$ and $m$ are the dimensions of the blocks and satisfy $d=n+m$ and $n\leq m$. The corresponding matrix of all eigenvectors $S$ can also be divided into 4 blocks, namely
\beq
S =
\begin{pmatrix}
S_{nn} &S_{nm}\\
S_{mn}& S_{mm}
\end{pmatrix}
\eeq
As described in Ref. \cite{Cederbaum_1989}, the unitary transformation $T$ can be simplified as $T=U(U^{\dagger}U)^{-1/2}$ where
\beq
U=
\begin{pmatrix}
1 &X\\
-X^{\dagger}& 1
\end{pmatrix},  U^{\dagger}U=\begin{pmatrix}
1 +X X^{\dagger}&0\\
0& 1+X^{\dagger}X
\end{pmatrix}\\
\eeq

with  $X=-(S_{mn}S_{nn}^{-1})^{\dagger}=S_{nm}S_{mm}^{-1}$. Therefore, only part of the eigenvector matrix $S$ is needed, for example, the first $n$ eigenvectors. To obtain the blocks in the computational subspace in a large matrix, we first decouple the computational subspace from higher levels using the principle of least action above, and then repeat this approach in the $4\times4$ matrix.

\section{Dynamical Quadratic Factor}
\label{app eta}
{\it Transmon-Transmon pair:} the ratio $\gamma$ in Eq. (\ref{eq.Jratio}) is around 1, so the perturbative dynamical quadratic factor $\eta$ can be simplified as 
\beq
\begin{split}
\eta&=\frac{J_{01}^2}{2\delta\Delta^2(\delta-2\Delta)(\delta-\Delta)^3(\delta+\Delta)^2}\left[8\delta^6-15\delta^5\Delta\right.\\
&\left.-18\delta^4\Delta^2+38\delta^3\Delta^3+6\delta^2\Delta^4-\delta\Delta^5+2\Delta^6\right]
\end{split}
\eeq
with $\delta_1=\delta_2=\delta$.

{\it Transmon-CSFQ pair:} the ratio $\gamma$ in Eq. (\ref{eq.Jratio}) is a slow changing parameter, approximately we can assume $\eta=\eta({\Delta=0})\approx 3/2$, the perturbative dynamical quadratic factor $\eta$ can be simplified as 
\beq
\begin{split}
\eta&=\frac{J_{01}^2}{16\delta\Delta^2(\delta+\Delta)(2\delta+\Delta)^3}\left[8\Delta^5-208\delta^5\right.\\
&\left.-472\delta^4\Delta-304\delta^3\Delta^2+57\delta^2\Delta^3+97\delta\Delta^4\right]
\end{split}
\eeq
with $\delta_1\approx-2\delta_2=-2\delta$.

\bibliography{theorypaper}

\begin{thebibliography}{41}%
\makeatletter
\providecommand \@ifxundefined [1]{%
 \@ifx{#1\undefined}
}%
\providecommand \@ifnum [1]{%
 \ifnum #1\expandafter \@firstoftwo
 \else \expandafter \@secondoftwo
 \fi
}%
\providecommand \@ifx [1]{%
 \ifx #1\expandafter \@firstoftwo
 \else \expandafter \@secondoftwo
 \fi
}%
\providecommand \natexlab [1]{#1}%
\providecommand \enquote  [1]{``#1''}%
\providecommand \bibnamefont  [1]{#1}%
\providecommand \bibfnamefont [1]{#1}%
\providecommand \citenamefont [1]{#1}%
\providecommand \href@noop [0]{\@secondoftwo}%
\providecommand \href [0]{\begingroup \@sanitize@url \@href}%
\providecommand \@href[1]{\@@startlink{#1}\@@href}%
\providecommand \@@href[1]{\endgroup#1\@@endlink}%
\providecommand \@sanitize@url [0]{\catcode `\\12\catcode `\$12\catcode
  `\&12\catcode `\#12\catcode `\^12\catcode `\_12\catcode `\%12\relax}%
\providecommand \@@startlink[1]{}%
\providecommand \@@endlink[0]{}%
\providecommand \url  [0]{\begingroup\@sanitize@url \@url }%
\providecommand \@url [1]{\endgroup\@href {#1}{\urlprefix }}%
\providecommand \urlprefix  [0]{URL }%
\providecommand \Eprint [0]{\href }%
\providecommand \doibase [0]{https://doi.org/}%
\providecommand \selectlanguage [0]{\@gobble}%
\providecommand \bibinfo  [0]{\@secondoftwo}%
\providecommand \bibfield  [0]{\@secondoftwo}%
\providecommand \translation [1]{[#1]}%
\providecommand \BibitemOpen [0]{}%
\providecommand \bibitemStop [0]{}%
\providecommand \bibitemNoStop [0]{.\EOS\space}%
\providecommand \EOS [0]{\spacefactor3000\relax}%
\providecommand \BibitemShut  [1]{\csname bibitem#1\endcsname}%
\let\auto@bib@innerbib\@empty
%</preamble>
\bibitem [{\citenamefont {Ayanzadeh}\ \emph {et~al.}(2020)\citenamefont
  {Ayanzadeh}, \citenamefont {Halem},\ and\ \citenamefont
  {Finin}}]{ayanzadeh2020reinforcement}%
  \BibitemOpen
  \bibfield  {author} {\bibinfo {author} {\bibfnamefont {R.}~\bibnamefont
  {Ayanzadeh}}, \bibinfo {author} {\bibfnamefont {M.}~\bibnamefont {Halem}},\
  and\ \bibinfo {author} {\bibfnamefont {T.}~\bibnamefont {Finin}},\ }\href
  {https://doi.org/10.1038/s41598-020-64078-1} {\bibfield  {journal} {\bibinfo
  {journal} {Scientific Reports}\ }\textbf {\bibinfo {volume} {10}},\ \bibinfo
  {pages} {1} (\bibinfo {year} {2020})}\BibitemShut {NoStop}%
\bibitem [{\citenamefont {Brydges}\ \emph {et~al.}(2019)\citenamefont
  {Brydges}, \citenamefont {Elben}, \citenamefont {Jurcevic}, \citenamefont
  {Vermersch}, \citenamefont {Maier}, \citenamefont {Lanyon}, \citenamefont
  {Zoller}, \citenamefont {Blatt},\ and\ \citenamefont {Roos}}]{Brydges_2019}%
  \BibitemOpen
  \bibfield  {author} {\bibinfo {author} {\bibfnamefont {T.}~\bibnamefont
  {Brydges}}, \bibinfo {author} {\bibfnamefont {A.}~\bibnamefont {Elben}},
  \bibinfo {author} {\bibfnamefont {P.}~\bibnamefont {Jurcevic}}, \bibinfo
  {author} {\bibfnamefont {B.}~\bibnamefont {Vermersch}}, \bibinfo {author}
  {\bibfnamefont {C.}~\bibnamefont {Maier}}, \bibinfo {author} {\bibfnamefont
  {B.~P.}\ \bibnamefont {Lanyon}}, \bibinfo {author} {\bibfnamefont
  {P.}~\bibnamefont {Zoller}}, \bibinfo {author} {\bibfnamefont
  {R.}~\bibnamefont {Blatt}},\ and\ \bibinfo {author} {\bibfnamefont {C.~F.}\
  \bibnamefont {Roos}},\ }\href {https://doi.org/10.1126/science.aau4963}
  {\bibfield  {journal} {\bibinfo  {journal} {Science}\ }\textbf {\bibinfo
  {volume} {364}},\ \bibinfo {pages} {260} (\bibinfo {year}
  {2019})}\BibitemShut {NoStop}%
\bibitem [{\citenamefont {Foxen}\ \emph {et~al.}(2020)\citenamefont {Foxen},
  \citenamefont {Neill}, \citenamefont {Dunsworth}, \citenamefont {Roushan},
  \citenamefont {Chiaro}, \citenamefont {Megrant}, \citenamefont {Kelly},
  \citenamefont {Chen}, \citenamefont {Satzinger}, \citenamefont {Barends},
  \citenamefont {Arute}, \citenamefont {Arya}, \citenamefont {Babbush},
  \citenamefont {Bacon}, \citenamefont {Bardin}, \citenamefont {Boixo},
  \citenamefont {Buell}, \citenamefont {Burkett}, \citenamefont {Chen},
  \citenamefont {Collins}, \citenamefont {Farhi}, \citenamefont {Fowler},
  \citenamefont {Gidney}, \citenamefont {Giustina}, \citenamefont {Graff},
  \citenamefont {Harrigan}, \citenamefont {Huang}, \citenamefont {Isakov},
  \citenamefont {Jeffrey}, \citenamefont {Jiang}, \citenamefont {Kafri},
  \citenamefont {Kechedzhi}, \citenamefont {Klimov}, \citenamefont {Korotkov},
  \citenamefont {Kostritsa}, \citenamefont {Landhuis}, \citenamefont {Lucero},
  \citenamefont {McClean}, \citenamefont {McEwen}, \citenamefont {Mi},
  \citenamefont {Mohseni}, \citenamefont {Mutus}, \citenamefont {Naaman},
  \citenamefont {Neeley}, \citenamefont {Niu}, \citenamefont {Petukhov},
  \citenamefont {Quintana}, \citenamefont {Rubin}, \citenamefont {Sank},
  \citenamefont {Smelyanskiy}, \citenamefont {Vainsencher}, \citenamefont
  {White}, \citenamefont {Yao}, \citenamefont {Yeh}, \citenamefont {Zalcman},
  \citenamefont {Neven},\ and\ \citenamefont
  {Martinis}}]{foxen2020demonstrating}%
  \BibitemOpen
  \bibfield  {author} {\bibinfo {author} {\bibfnamefont {B.}~\bibnamefont
  {Foxen}}, \bibinfo {author} {\bibfnamefont {C.}~\bibnamefont {Neill}},
  \bibinfo {author} {\bibfnamefont {A.}~\bibnamefont {Dunsworth}}, \bibinfo
  {author} {\bibfnamefont {P.}~\bibnamefont {Roushan}}, \bibinfo {author}
  {\bibfnamefont {B.}~\bibnamefont {Chiaro}}, \bibinfo {author} {\bibfnamefont
  {A.}~\bibnamefont {Megrant}}, \bibinfo {author} {\bibfnamefont
  {J.}~\bibnamefont {Kelly}}, \bibinfo {author} {\bibfnamefont
  {Z.}~\bibnamefont {Chen}}, \bibinfo {author} {\bibfnamefont {K.}~\bibnamefont
  {Satzinger}}, \bibinfo {author} {\bibfnamefont {R.}~\bibnamefont {Barends}},
  \bibinfo {author} {\bibfnamefont {F.}~\bibnamefont {Arute}}, \bibinfo
  {author} {\bibfnamefont {K.}~\bibnamefont {Arya}}, \bibinfo {author}
  {\bibfnamefont {R.}~\bibnamefont {Babbush}}, \bibinfo {author} {\bibfnamefont
  {D.}~\bibnamefont {Bacon}}, \bibinfo {author} {\bibfnamefont {J.~C.}\
  \bibnamefont {Bardin}}, \bibinfo {author} {\bibfnamefont {S.}~\bibnamefont
  {Boixo}}, \bibinfo {author} {\bibfnamefont {D.}~\bibnamefont {Buell}},
  \bibinfo {author} {\bibfnamefont {B.}~\bibnamefont {Burkett}}, \bibinfo
  {author} {\bibfnamefont {Y.}~\bibnamefont {Chen}}, \bibinfo {author}
  {\bibfnamefont {R.}~\bibnamefont {Collins}}, \bibinfo {author} {\bibfnamefont
  {E.}~\bibnamefont {Farhi}}, \bibinfo {author} {\bibfnamefont
  {A.}~\bibnamefont {Fowler}}, \bibinfo {author} {\bibfnamefont
  {C.}~\bibnamefont {Gidney}}, \bibinfo {author} {\bibfnamefont
  {M.}~\bibnamefont {Giustina}}, \bibinfo {author} {\bibfnamefont
  {R.}~\bibnamefont {Graff}}, \bibinfo {author} {\bibfnamefont
  {M.}~\bibnamefont {Harrigan}}, \bibinfo {author} {\bibfnamefont
  {T.}~\bibnamefont {Huang}}, \bibinfo {author} {\bibfnamefont {S.~V.}\
  \bibnamefont {Isakov}}, \bibinfo {author} {\bibfnamefont {E.}~\bibnamefont
  {Jeffrey}}, \bibinfo {author} {\bibfnamefont {Z.}~\bibnamefont {Jiang}},
  \bibinfo {author} {\bibfnamefont {D.}~\bibnamefont {Kafri}}, \bibinfo
  {author} {\bibfnamefont {K.}~\bibnamefont {Kechedzhi}}, \bibinfo {author}
  {\bibfnamefont {P.}~\bibnamefont {Klimov}}, \bibinfo {author} {\bibfnamefont
  {A.}~\bibnamefont {Korotkov}}, \bibinfo {author} {\bibfnamefont
  {F.}~\bibnamefont {Kostritsa}}, \bibinfo {author} {\bibfnamefont
  {D.}~\bibnamefont {Landhuis}}, \bibinfo {author} {\bibfnamefont
  {E.}~\bibnamefont {Lucero}}, \bibinfo {author} {\bibfnamefont
  {J.}~\bibnamefont {McClean}}, \bibinfo {author} {\bibfnamefont
  {M.}~\bibnamefont {McEwen}}, \bibinfo {author} {\bibfnamefont
  {X.}~\bibnamefont {Mi}}, \bibinfo {author} {\bibfnamefont {M.}~\bibnamefont
  {Mohseni}}, \bibinfo {author} {\bibfnamefont {J.~Y.}\ \bibnamefont {Mutus}},
  \bibinfo {author} {\bibfnamefont {O.}~\bibnamefont {Naaman}}, \bibinfo
  {author} {\bibfnamefont {M.}~\bibnamefont {Neeley}}, \bibinfo {author}
  {\bibfnamefont {M.}~\bibnamefont {Niu}}, \bibinfo {author} {\bibfnamefont
  {A.}~\bibnamefont {Petukhov}}, \bibinfo {author} {\bibfnamefont
  {C.}~\bibnamefont {Quintana}}, \bibinfo {author} {\bibfnamefont
  {N.}~\bibnamefont {Rubin}}, \bibinfo {author} {\bibfnamefont
  {D.}~\bibnamefont {Sank}}, \bibinfo {author} {\bibfnamefont {V.}~\bibnamefont
  {Smelyanskiy}}, \bibinfo {author} {\bibfnamefont {A.}~\bibnamefont
  {Vainsencher}}, \bibinfo {author} {\bibfnamefont {T.~C.}\ \bibnamefont
  {White}}, \bibinfo {author} {\bibfnamefont {Z.}~\bibnamefont {Yao}}, \bibinfo
  {author} {\bibfnamefont {P.}~\bibnamefont {Yeh}}, \bibinfo {author}
  {\bibfnamefont {A.}~\bibnamefont {Zalcman}}, \bibinfo {author} {\bibfnamefont
  {H.}~\bibnamefont {Neven}},\ and\ \bibinfo {author} {\bibfnamefont {J.~M.}\
  \bibnamefont {Martinis}} (\bibinfo {collaboration} {Google AI Quantum}),\
  }\href {https://doi.org/10.1103/PhysRevLett.125.120504} {\bibfield  {journal}
  {\bibinfo  {journal} {Phys. Rev. Lett.}\ }\textbf {\bibinfo {volume} {125}},\
  \bibinfo {pages} {120504} (\bibinfo {year} {2020})}\BibitemShut {NoStop}%
\bibitem [{\citenamefont {Arute}\ \emph {et~al.}(2019)\citenamefont {Arute},
  \citenamefont {Arya}, \citenamefont {Babbush}, \citenamefont {Bacon},
  \citenamefont {Bardin}, \citenamefont {Barends}, \citenamefont {Biswas},
  \citenamefont {Boixo}, \citenamefont {Brandao}, \citenamefont {Buell},
  \citenamefont {Burkett}, \citenamefont {Chen}, \citenamefont {Chen},
  \citenamefont {Chiaro}, \citenamefont {Collins}, \citenamefont {Courtney},
  \citenamefont {Dunsworth}, \citenamefont {Farhi}, \citenamefont {Foxen},
  \citenamefont {Fowler}, \citenamefont {Gidney}, \citenamefont {Giustina},
  \citenamefont {Graff}, \citenamefont {Guerin}, \citenamefont {Habegger},
  \citenamefont {Harrigan}, \citenamefont {Hartmann}, \citenamefont {Ho},
  \citenamefont {Hoffmann}, \citenamefont {Huang}, \citenamefont {Humble},
  \citenamefont {Isakov}, \citenamefont {Jeffrey}, \citenamefont {Jiang},
  \citenamefont {Kafri}, \citenamefont {Kechedzhi}, \citenamefont {Kelly},
  \citenamefont {Klimov}, \citenamefont {Knysh}, \citenamefont {Korotkov},
  \citenamefont {Kostritsa}, \citenamefont {Landhuis}, \citenamefont
  {Lindmark}, \citenamefont {Lucero}, \citenamefont {Lyakh}, \citenamefont
  {Mandr{\`a}}, \citenamefont {McClean}, \citenamefont {McEwen}, \citenamefont
  {Megrant}, \citenamefont {Mi}, \citenamefont {Michielsen}, \citenamefont
  {Mohseni}, \citenamefont {Mutus}, \citenamefont {Naaman}, \citenamefont
  {Neeley}, \citenamefont {Neill}, \citenamefont {Niu}, \citenamefont {Ostby},
  \citenamefont {Petukhov}, \citenamefont {Platt}, \citenamefont {Quintana},
  \citenamefont {Rieffel}, \citenamefont {Roushan}, \citenamefont {Rubin},
  \citenamefont {Sank}, \citenamefont {Satzinger}, \citenamefont {Smelyanskiy},
  \citenamefont {Sung}, \citenamefont {Trevithick}, \citenamefont
  {Vainsencher}, \citenamefont {Villalonga}, \citenamefont {White},
  \citenamefont {Yao}, \citenamefont {Yeh}, \citenamefont {Zalcman},
  \citenamefont {Neven},\ and\ \citenamefont {Martinis}}]{Arute:2019aa}%
  \BibitemOpen
  \bibfield  {author} {\bibinfo {author} {\bibfnamefont {F.}~\bibnamefont
  {Arute}}, \bibinfo {author} {\bibfnamefont {K.}~\bibnamefont {Arya}},
  \bibinfo {author} {\bibfnamefont {R.}~\bibnamefont {Babbush}}, \bibinfo
  {author} {\bibfnamefont {D.}~\bibnamefont {Bacon}}, \bibinfo {author}
  {\bibfnamefont {J.~C.}\ \bibnamefont {Bardin}}, \bibinfo {author}
  {\bibfnamefont {R.}~\bibnamefont {Barends}}, \bibinfo {author} {\bibfnamefont
  {R.}~\bibnamefont {Biswas}}, \bibinfo {author} {\bibfnamefont
  {S.}~\bibnamefont {Boixo}}, \bibinfo {author} {\bibfnamefont {F.~G. S.~L.}\
  \bibnamefont {Brandao}}, \bibinfo {author} {\bibfnamefont {D.~A.}\
  \bibnamefont {Buell}}, \bibinfo {author} {\bibfnamefont {B.}~\bibnamefont
  {Burkett}}, \bibinfo {author} {\bibfnamefont {Y.}~\bibnamefont {Chen}},
  \bibinfo {author} {\bibfnamefont {Z.}~\bibnamefont {Chen}}, \bibinfo {author}
  {\bibfnamefont {B.}~\bibnamefont {Chiaro}}, \bibinfo {author} {\bibfnamefont
  {R.}~\bibnamefont {Collins}}, \bibinfo {author} {\bibfnamefont
  {W.}~\bibnamefont {Courtney}}, \bibinfo {author} {\bibfnamefont
  {A.}~\bibnamefont {Dunsworth}}, \bibinfo {author} {\bibfnamefont
  {E.}~\bibnamefont {Farhi}}, \bibinfo {author} {\bibfnamefont
  {B.}~\bibnamefont {Foxen}}, \bibinfo {author} {\bibfnamefont
  {A.}~\bibnamefont {Fowler}}, \bibinfo {author} {\bibfnamefont
  {C.}~\bibnamefont {Gidney}}, \bibinfo {author} {\bibfnamefont
  {M.}~\bibnamefont {Giustina}}, \bibinfo {author} {\bibfnamefont
  {R.}~\bibnamefont {Graff}}, \bibinfo {author} {\bibfnamefont
  {K.}~\bibnamefont {Guerin}}, \bibinfo {author} {\bibfnamefont
  {S.}~\bibnamefont {Habegger}}, \bibinfo {author} {\bibfnamefont {M.~P.}\
  \bibnamefont {Harrigan}}, \bibinfo {author} {\bibfnamefont {M.~J.}\
  \bibnamefont {Hartmann}}, \bibinfo {author} {\bibfnamefont {A.}~\bibnamefont
  {Ho}}, \bibinfo {author} {\bibfnamefont {M.}~\bibnamefont {Hoffmann}},
  \bibinfo {author} {\bibfnamefont {T.}~\bibnamefont {Huang}}, \bibinfo
  {author} {\bibfnamefont {T.~S.}\ \bibnamefont {Humble}}, \bibinfo {author}
  {\bibfnamefont {S.~V.}\ \bibnamefont {Isakov}}, \bibinfo {author}
  {\bibfnamefont {E.}~\bibnamefont {Jeffrey}}, \bibinfo {author} {\bibfnamefont
  {Z.}~\bibnamefont {Jiang}}, \bibinfo {author} {\bibfnamefont
  {D.}~\bibnamefont {Kafri}}, \bibinfo {author} {\bibfnamefont
  {K.}~\bibnamefont {Kechedzhi}}, \bibinfo {author} {\bibfnamefont
  {J.}~\bibnamefont {Kelly}}, \bibinfo {author} {\bibfnamefont {P.~V.}\
  \bibnamefont {Klimov}}, \bibinfo {author} {\bibfnamefont {S.}~\bibnamefont
  {Knysh}}, \bibinfo {author} {\bibfnamefont {A.}~\bibnamefont {Korotkov}},
  \bibinfo {author} {\bibfnamefont {F.}~\bibnamefont {Kostritsa}}, \bibinfo
  {author} {\bibfnamefont {D.}~\bibnamefont {Landhuis}}, \bibinfo {author}
  {\bibfnamefont {M.}~\bibnamefont {Lindmark}}, \bibinfo {author}
  {\bibfnamefont {E.}~\bibnamefont {Lucero}}, \bibinfo {author} {\bibfnamefont
  {D.}~\bibnamefont {Lyakh}}, \bibinfo {author} {\bibfnamefont
  {S.}~\bibnamefont {Mandr{\`a}}}, \bibinfo {author} {\bibfnamefont {J.~R.}\
  \bibnamefont {McClean}}, \bibinfo {author} {\bibfnamefont {M.}~\bibnamefont
  {McEwen}}, \bibinfo {author} {\bibfnamefont {A.}~\bibnamefont {Megrant}},
  \bibinfo {author} {\bibfnamefont {X.}~\bibnamefont {Mi}}, \bibinfo {author}
  {\bibfnamefont {K.}~\bibnamefont {Michielsen}}, \bibinfo {author}
  {\bibfnamefont {M.}~\bibnamefont {Mohseni}}, \bibinfo {author} {\bibfnamefont
  {J.}~\bibnamefont {Mutus}}, \bibinfo {author} {\bibfnamefont
  {O.}~\bibnamefont {Naaman}}, \bibinfo {author} {\bibfnamefont
  {M.}~\bibnamefont {Neeley}}, \bibinfo {author} {\bibfnamefont
  {C.}~\bibnamefont {Neill}}, \bibinfo {author} {\bibfnamefont {M.~Y.}\
  \bibnamefont {Niu}}, \bibinfo {author} {\bibfnamefont {E.}~\bibnamefont
  {Ostby}}, \bibinfo {author} {\bibfnamefont {A.}~\bibnamefont {Petukhov}},
  \bibinfo {author} {\bibfnamefont {J.~C.}\ \bibnamefont {Platt}}, \bibinfo
  {author} {\bibfnamefont {C.}~\bibnamefont {Quintana}}, \bibinfo {author}
  {\bibfnamefont {E.~G.}\ \bibnamefont {Rieffel}}, \bibinfo {author}
  {\bibfnamefont {P.}~\bibnamefont {Roushan}}, \bibinfo {author} {\bibfnamefont
  {N.~C.}\ \bibnamefont {Rubin}}, \bibinfo {author} {\bibfnamefont
  {D.}~\bibnamefont {Sank}}, \bibinfo {author} {\bibfnamefont {K.~J.}\
  \bibnamefont {Satzinger}}, \bibinfo {author} {\bibfnamefont {V.}~\bibnamefont
  {Smelyanskiy}}, \bibinfo {author} {\bibfnamefont {K.~J.}\ \bibnamefont
  {Sung}}, \bibinfo {author} {\bibfnamefont {M.~D.}\ \bibnamefont
  {Trevithick}}, \bibinfo {author} {\bibfnamefont {A.}~\bibnamefont
  {Vainsencher}}, \bibinfo {author} {\bibfnamefont {B.}~\bibnamefont
  {Villalonga}}, \bibinfo {author} {\bibfnamefont {T.}~\bibnamefont {White}},
  \bibinfo {author} {\bibfnamefont {Z.~J.}\ \bibnamefont {Yao}}, \bibinfo
  {author} {\bibfnamefont {P.}~\bibnamefont {Yeh}}, \bibinfo {author}
  {\bibfnamefont {A.}~\bibnamefont {Zalcman}}, \bibinfo {author} {\bibfnamefont
  {H.}~\bibnamefont {Neven}},\ and\ \bibinfo {author} {\bibfnamefont {J.~M.}\
  \bibnamefont {Martinis}},\ }\href {https://doi.org/10.1038/s41586-019-1666-5}
  {\bibfield  {journal} {\bibinfo  {journal} {Nature}\ }\textbf {\bibinfo
  {volume} {574}},\ \bibinfo {pages} {505} (\bibinfo {year}
  {2019})}\BibitemShut {NoStop}%
\bibitem [{\citenamefont {Kjaergaard}\ \emph {et~al.}(2020)\citenamefont
  {Kjaergaard}, \citenamefont {Schwartz}, \citenamefont {Braum{\"u}ller},
  \citenamefont {Krantz}, \citenamefont {Wang}, \citenamefont {Gustavsson},\
  and\ \citenamefont {Oliver}}]{kjaergaard2020superconducting}%
  \BibitemOpen
  \bibfield  {author} {\bibinfo {author} {\bibfnamefont {M.}~\bibnamefont
  {Kjaergaard}}, \bibinfo {author} {\bibfnamefont {M.~E.}\ \bibnamefont
  {Schwartz}}, \bibinfo {author} {\bibfnamefont {J.}~\bibnamefont
  {Braum{\"u}ller}}, \bibinfo {author} {\bibfnamefont {P.}~\bibnamefont
  {Krantz}}, \bibinfo {author} {\bibfnamefont {J.~I.-J.}\ \bibnamefont {Wang}},
  \bibinfo {author} {\bibfnamefont {S.}~\bibnamefont {Gustavsson}},\ and\
  \bibinfo {author} {\bibfnamefont {W.~D.}\ \bibnamefont {Oliver}},\ }\href
  {https://doi.org/10.1146/annurev-conmatphys-031119-050605} {\bibfield
  {journal} {\bibinfo  {journal} {Annual Review of Condensed Matter Physics}\
  }\textbf {\bibinfo {volume} {11}} (\bibinfo {year} {2020})}\BibitemShut
  {NoStop}%
\bibitem [{\citenamefont {Martinis}\ \emph {et~al.}(2020)\citenamefont
  {Martinis}, \citenamefont {Devoret},\ and\ \citenamefont
  {Clarke}}]{martinis2020quantum}%
  \BibitemOpen
  \bibfield  {author} {\bibinfo {author} {\bibfnamefont {J.~M.}\ \bibnamefont
  {Martinis}}, \bibinfo {author} {\bibfnamefont {M.~H.}\ \bibnamefont
  {Devoret}},\ and\ \bibinfo {author} {\bibfnamefont {J.}~\bibnamefont
  {Clarke}},\ }\href {https://doi.org/10.1038/s41567-020-0829-5} {\bibfield
  {journal} {\bibinfo  {journal} {Nature Physics}\ }\textbf {\bibinfo {volume}
  {16}},\ \bibinfo {pages} {234} (\bibinfo {year} {2020})}\BibitemShut
  {NoStop}%
\bibitem [{\citenamefont {Bialczak}\ \emph {et~al.}(2011)\citenamefont
  {Bialczak}, \citenamefont {Ansmann}, \citenamefont {Hofheinz}, \citenamefont
  {Lenander}, \citenamefont {Lucero}, \citenamefont {Neeley}, \citenamefont
  {O'Connell}, \citenamefont {Sank}, \citenamefont {Wang}, \citenamefont
  {Weides}, \citenamefont {Wenner}, \citenamefont {Yamamoto}, \citenamefont
  {Cleland},\ and\ \citenamefont {Martinis}}]{Bialcza12Fast}%
  \BibitemOpen
  \bibfield  {author} {\bibinfo {author} {\bibfnamefont {R.~C.}\ \bibnamefont
  {Bialczak}}, \bibinfo {author} {\bibfnamefont {M.}~\bibnamefont {Ansmann}},
  \bibinfo {author} {\bibfnamefont {M.}~\bibnamefont {Hofheinz}}, \bibinfo
  {author} {\bibfnamefont {M.}~\bibnamefont {Lenander}}, \bibinfo {author}
  {\bibfnamefont {E.}~\bibnamefont {Lucero}}, \bibinfo {author} {\bibfnamefont
  {M.}~\bibnamefont {Neeley}}, \bibinfo {author} {\bibfnamefont {A.~D.}\
  \bibnamefont {O'Connell}}, \bibinfo {author} {\bibfnamefont {D.}~\bibnamefont
  {Sank}}, \bibinfo {author} {\bibfnamefont {H.}~\bibnamefont {Wang}}, \bibinfo
  {author} {\bibfnamefont {M.}~\bibnamefont {Weides}}, \bibinfo {author}
  {\bibfnamefont {J.}~\bibnamefont {Wenner}}, \bibinfo {author} {\bibfnamefont
  {T.}~\bibnamefont {Yamamoto}}, \bibinfo {author} {\bibfnamefont {A.~N.}\
  \bibnamefont {Cleland}},\ and\ \bibinfo {author} {\bibfnamefont {J.~M.}\
  \bibnamefont {Martinis}},\ }\href
  {https://doi.org/10.1103/PhysRevLett.106.060501} {\bibfield  {journal}
  {\bibinfo  {journal} {Phys. Rev. Lett.}\ }\textbf {\bibinfo {volume} {106}},\
  \bibinfo {pages} {060501} (\bibinfo {year} {2011})}\BibitemShut {NoStop}%
\bibitem [{\citenamefont {Gustavsson}\ \emph {et~al.}(2016)\citenamefont
  {Gustavsson}, \citenamefont {Yan}, \citenamefont {Catelani}, \citenamefont
  {Bylander}, \citenamefont {Kamal}, \citenamefont {Birenbaum}, \citenamefont
  {Hover}, \citenamefont {Rosenberg}, \citenamefont {Samach}, \citenamefont
  {Sears}, \citenamefont {Weber}, \citenamefont {Yoder}, \citenamefont
  {Clarke}, \citenamefont {Kerman}, \citenamefont {Yoshihara}, \citenamefont
  {Nakamura}, \citenamefont {Orlando},\ and\ \citenamefont
  {Oliver}}]{gustavsson2016suppressing}%
  \BibitemOpen
  \bibfield  {author} {\bibinfo {author} {\bibfnamefont {S.}~\bibnamefont
  {Gustavsson}}, \bibinfo {author} {\bibfnamefont {F.}~\bibnamefont {Yan}},
  \bibinfo {author} {\bibfnamefont {G.}~\bibnamefont {Catelani}}, \bibinfo
  {author} {\bibfnamefont {J.}~\bibnamefont {Bylander}}, \bibinfo {author}
  {\bibfnamefont {A.}~\bibnamefont {Kamal}}, \bibinfo {author} {\bibfnamefont
  {J.}~\bibnamefont {Birenbaum}}, \bibinfo {author} {\bibfnamefont
  {D.}~\bibnamefont {Hover}}, \bibinfo {author} {\bibfnamefont
  {D.}~\bibnamefont {Rosenberg}}, \bibinfo {author} {\bibfnamefont
  {G.}~\bibnamefont {Samach}}, \bibinfo {author} {\bibfnamefont {A.~P.}\
  \bibnamefont {Sears}}, \bibinfo {author} {\bibfnamefont {S.~J.}\ \bibnamefont
  {Weber}}, \bibinfo {author} {\bibfnamefont {J.~L.}\ \bibnamefont {Yoder}},
  \bibinfo {author} {\bibfnamefont {J.}~\bibnamefont {Clarke}}, \bibinfo
  {author} {\bibfnamefont {A.~J.}\ \bibnamefont {Kerman}}, \bibinfo {author}
  {\bibfnamefont {F.}~\bibnamefont {Yoshihara}}, \bibinfo {author}
  {\bibfnamefont {Y.}~\bibnamefont {Nakamura}}, \bibinfo {author}
  {\bibfnamefont {T.~P.}\ \bibnamefont {Orlando}},\ and\ \bibinfo {author}
  {\bibfnamefont {W.~D.}\ \bibnamefont {Oliver}},\ }\href
  {https://doi.org/10.1126/science.aah5844} {\bibfield  {journal} {\bibinfo
  {journal} {Science}\ }\textbf {\bibinfo {volume} {354}},\ \bibinfo {pages}
  {1573} (\bibinfo {year} {2016})}\BibitemShut {NoStop}%
\bibitem [{\citenamefont {Ansari}(2015)}]{ansari2015rate}%
  \BibitemOpen
  \bibfield  {author} {\bibinfo {author} {\bibfnamefont {M.~H.}\ \bibnamefont
  {Ansari}},\ }\href {https://doi.org/10.1088/0953-2048/28/4/045005} {\bibfield
   {journal} {\bibinfo  {journal} {Superconductor Science and Technology}\
  }\textbf {\bibinfo {volume} {28}},\ \bibinfo {pages} {045005} (\bibinfo
  {year} {2015})}\BibitemShut {NoStop}%
\bibitem [{\citenamefont {Ansari}\ \emph {et~al.}(2013)\citenamefont {Ansari},
  \citenamefont {Wilhelm}, \citenamefont {Sinha},\ and\ \citenamefont
  {Sinha}}]{ansari2013effect}%
  \BibitemOpen
  \bibfield  {author} {\bibinfo {author} {\bibfnamefont {M.~H.}\ \bibnamefont
  {Ansari}}, \bibinfo {author} {\bibfnamefont {F.~K.}\ \bibnamefont {Wilhelm}},
  \bibinfo {author} {\bibfnamefont {U.}~\bibnamefont {Sinha}},\ and\ \bibinfo
  {author} {\bibfnamefont {A.}~\bibnamefont {Sinha}},\ }\href
  {https://doi.org/10.1088/0953-2048/26/12/125013} {\bibfield  {journal}
  {\bibinfo  {journal} {Superconductor Science and Technology}\ }\textbf
  {\bibinfo {volume} {26}},\ \bibinfo {pages} {125013} (\bibinfo {year}
  {2013})}\BibitemShut {NoStop}%
\bibitem [{\citenamefont {Serniak}\ \emph {et~al.}(2018)\citenamefont
  {Serniak}, \citenamefont {Hays}, \citenamefont {de~Lange}, \citenamefont
  {Diamond}, \citenamefont {Shankar}, \citenamefont {Burkhart}, \citenamefont
  {Frunzio}, \citenamefont {Houzet},\ and\ \citenamefont
  {Devoret}}]{serniak2018hot}%
  \BibitemOpen
  \bibfield  {author} {\bibinfo {author} {\bibfnamefont {K.}~\bibnamefont
  {Serniak}}, \bibinfo {author} {\bibfnamefont {M.}~\bibnamefont {Hays}},
  \bibinfo {author} {\bibfnamefont {G.}~\bibnamefont {de~Lange}}, \bibinfo
  {author} {\bibfnamefont {S.}~\bibnamefont {Diamond}}, \bibinfo {author}
  {\bibfnamefont {S.}~\bibnamefont {Shankar}}, \bibinfo {author} {\bibfnamefont
  {L.~D.}\ \bibnamefont {Burkhart}}, \bibinfo {author} {\bibfnamefont
  {L.}~\bibnamefont {Frunzio}}, \bibinfo {author} {\bibfnamefont
  {M.}~\bibnamefont {Houzet}},\ and\ \bibinfo {author} {\bibfnamefont {M.~H.}\
  \bibnamefont {Devoret}},\ }\href
  {https://doi.org/10.1103/PhysRevLett.121.157701} {\bibfield  {journal}
  {\bibinfo  {journal} {Phys. Rev. Lett.}\ }\textbf {\bibinfo {volume} {121}},\
  \bibinfo {pages} {157701} (\bibinfo {year} {2018})}\BibitemShut {NoStop}%
\bibitem [{\citenamefont {Ansari}\ and\ \citenamefont
  {Wilhelm}(2011)}]{ansari2011noise}%
  \BibitemOpen
  \bibfield  {author} {\bibinfo {author} {\bibfnamefont {M.~H.}\ \bibnamefont
  {Ansari}}\ and\ \bibinfo {author} {\bibfnamefont {F.~K.}\ \bibnamefont
  {Wilhelm}},\ }\href {https://doi.org/10.1103/PhysRevB.84.235102} {\bibfield
  {journal} {\bibinfo  {journal} {Phys. Rev. B}\ }\textbf {\bibinfo {volume}
  {84}},\ \bibinfo {pages} {235102} (\bibinfo {year} {2011})}\BibitemShut
  {NoStop}%
\bibitem [{\citenamefont {Bal}\ \emph {et~al.}(2015)\citenamefont {Bal},
  \citenamefont {Ansari}, \citenamefont {Orgiazzi}, \citenamefont {Lutchyn},\
  and\ \citenamefont {Lupascu}}]{bal2015dynamics}%
  \BibitemOpen
  \bibfield  {author} {\bibinfo {author} {\bibfnamefont {M.}~\bibnamefont
  {Bal}}, \bibinfo {author} {\bibfnamefont {M.~H.}\ \bibnamefont {Ansari}},
  \bibinfo {author} {\bibfnamefont {J.-L.}\ \bibnamefont {Orgiazzi}}, \bibinfo
  {author} {\bibfnamefont {R.~M.}\ \bibnamefont {Lutchyn}},\ and\ \bibinfo
  {author} {\bibfnamefont {A.}~\bibnamefont {Lupascu}},\ }\href
  {https://doi.org/10.1103/PhysRevB.91.195434} {\bibfield  {journal} {\bibinfo
  {journal} {Phys. Rev. B}\ }\textbf {\bibinfo {volume} {91}},\ \bibinfo
  {pages} {195434} (\bibinfo {year} {2015})}\BibitemShut {NoStop}%
\bibitem [{\citenamefont {Krantz}\ \emph {et~al.}(2019)\citenamefont {Krantz},
  \citenamefont {Kjaergaard}, \citenamefont {Yan}, \citenamefont {Orlando},
  \citenamefont {Gustavsson},\ and\ \citenamefont
  {Oliver}}]{krantz2019quantum}%
  \BibitemOpen
  \bibfield  {author} {\bibinfo {author} {\bibfnamefont {P.}~\bibnamefont
  {Krantz}}, \bibinfo {author} {\bibfnamefont {M.}~\bibnamefont {Kjaergaard}},
  \bibinfo {author} {\bibfnamefont {F.}~\bibnamefont {Yan}}, \bibinfo {author}
  {\bibfnamefont {T.~P.}\ \bibnamefont {Orlando}}, \bibinfo {author}
  {\bibfnamefont {S.}~\bibnamefont {Gustavsson}},\ and\ \bibinfo {author}
  {\bibfnamefont {W.~D.}\ \bibnamefont {Oliver}},\ }\href
  {https://doi.org/10.1063/1.5089550} {\bibfield  {journal} {\bibinfo
  {journal} {Applied Physics Reviews}\ }\textbf {\bibinfo {volume} {6}},\
  \bibinfo {pages} {021318} (\bibinfo {year} {2019})}\BibitemShut {NoStop}%
\bibitem [{\citenamefont {Caldwell}\ \emph {et~al.}(2018)\citenamefont
  {Caldwell}, \citenamefont {Didier}, \citenamefont {Ryan}, \citenamefont
  {Sete}, \citenamefont {Hudson}, \citenamefont {Karalekas}, \citenamefont
  {Manenti}, \citenamefont {da~Silva}, \citenamefont {Sinclair}, \citenamefont
  {Acala}, \citenamefont {Alidoust}, \citenamefont {Angeles}, \citenamefont
  {Bestwick}, \citenamefont {Block}, \citenamefont {Bloom}, \citenamefont
  {Bradley}, \citenamefont {Bui}, \citenamefont {Capelluto}, \citenamefont
  {Chilcott}, \citenamefont {Cordova}, \citenamefont {Crossman}, \citenamefont
  {Curtis}, \citenamefont {Deshpande}, \citenamefont {Bouayadi}, \citenamefont
  {Girshovich}, \citenamefont {Hong}, \citenamefont {Kuang}, \citenamefont
  {Lenihan}, \citenamefont {Manning}, \citenamefont {Marchenkov}, \citenamefont
  {Marshall}, \citenamefont {Maydra}, \citenamefont {Mohan}, \citenamefont
  {O'Brien}, \citenamefont {Osborn}, \citenamefont {Otterbach}, \citenamefont
  {Papageorge}, \citenamefont {Paquette}, \citenamefont {Pelstring},
  \citenamefont {Polloreno}, \citenamefont {Prawiroatmodjo}, \citenamefont
  {Rawat}, \citenamefont {Reagor}, \citenamefont {Renzas}, \citenamefont
  {Rubin}, \citenamefont {Russell}, \citenamefont {Rust}, \citenamefont
  {Scarabelli}, \citenamefont {Scheer}, \citenamefont {Selvanayagam},
  \citenamefont {Smith}, \citenamefont {Staley}, \citenamefont {Suska},
  \citenamefont {Tezak}, \citenamefont {Thompson}, \citenamefont {To},
  \citenamefont {Vahidpour}, \citenamefont {Vodrahalli}, \citenamefont
  {Whyland}, \citenamefont {Yadav}, \citenamefont {Zeng},\ and\ \citenamefont
  {Rigetti}}]{Caldwell2018para}%
  \BibitemOpen
  \bibfield  {author} {\bibinfo {author} {\bibfnamefont {S.~A.}\ \bibnamefont
  {Caldwell}}, \bibinfo {author} {\bibfnamefont {N.}~\bibnamefont {Didier}},
  \bibinfo {author} {\bibfnamefont {C.~A.}\ \bibnamefont {Ryan}}, \bibinfo
  {author} {\bibfnamefont {E.~A.}\ \bibnamefont {Sete}}, \bibinfo {author}
  {\bibfnamefont {A.}~\bibnamefont {Hudson}}, \bibinfo {author} {\bibfnamefont
  {P.}~\bibnamefont {Karalekas}}, \bibinfo {author} {\bibfnamefont
  {R.}~\bibnamefont {Manenti}}, \bibinfo {author} {\bibfnamefont {M.~P.}\
  \bibnamefont {da~Silva}}, \bibinfo {author} {\bibfnamefont {R.}~\bibnamefont
  {Sinclair}}, \bibinfo {author} {\bibfnamefont {E.}~\bibnamefont {Acala}},
  \bibinfo {author} {\bibfnamefont {N.}~\bibnamefont {Alidoust}}, \bibinfo
  {author} {\bibfnamefont {J.}~\bibnamefont {Angeles}}, \bibinfo {author}
  {\bibfnamefont {A.}~\bibnamefont {Bestwick}}, \bibinfo {author}
  {\bibfnamefont {M.}~\bibnamefont {Block}}, \bibinfo {author} {\bibfnamefont
  {B.}~\bibnamefont {Bloom}}, \bibinfo {author} {\bibfnamefont
  {A.}~\bibnamefont {Bradley}}, \bibinfo {author} {\bibfnamefont
  {C.}~\bibnamefont {Bui}}, \bibinfo {author} {\bibfnamefont {L.}~\bibnamefont
  {Capelluto}}, \bibinfo {author} {\bibfnamefont {R.}~\bibnamefont {Chilcott}},
  \bibinfo {author} {\bibfnamefont {J.}~\bibnamefont {Cordova}}, \bibinfo
  {author} {\bibfnamefont {G.}~\bibnamefont {Crossman}}, \bibinfo {author}
  {\bibfnamefont {M.}~\bibnamefont {Curtis}}, \bibinfo {author} {\bibfnamefont
  {S.}~\bibnamefont {Deshpande}}, \bibinfo {author} {\bibfnamefont {T.~E.}\
  \bibnamefont {Bouayadi}}, \bibinfo {author} {\bibfnamefont {D.}~\bibnamefont
  {Girshovich}}, \bibinfo {author} {\bibfnamefont {S.}~\bibnamefont {Hong}},
  \bibinfo {author} {\bibfnamefont {K.}~\bibnamefont {Kuang}}, \bibinfo
  {author} {\bibfnamefont {M.}~\bibnamefont {Lenihan}}, \bibinfo {author}
  {\bibfnamefont {T.}~\bibnamefont {Manning}}, \bibinfo {author} {\bibfnamefont
  {A.}~\bibnamefont {Marchenkov}}, \bibinfo {author} {\bibfnamefont
  {J.}~\bibnamefont {Marshall}}, \bibinfo {author} {\bibfnamefont
  {R.}~\bibnamefont {Maydra}}, \bibinfo {author} {\bibfnamefont
  {Y.}~\bibnamefont {Mohan}}, \bibinfo {author} {\bibfnamefont
  {W.}~\bibnamefont {O'Brien}}, \bibinfo {author} {\bibfnamefont
  {C.}~\bibnamefont {Osborn}}, \bibinfo {author} {\bibfnamefont
  {J.}~\bibnamefont {Otterbach}}, \bibinfo {author} {\bibfnamefont
  {A.}~\bibnamefont {Papageorge}}, \bibinfo {author} {\bibfnamefont {J.-P.}\
  \bibnamefont {Paquette}}, \bibinfo {author} {\bibfnamefont {M.}~\bibnamefont
  {Pelstring}}, \bibinfo {author} {\bibfnamefont {A.}~\bibnamefont
  {Polloreno}}, \bibinfo {author} {\bibfnamefont {G.}~\bibnamefont
  {Prawiroatmodjo}}, \bibinfo {author} {\bibfnamefont {V.}~\bibnamefont
  {Rawat}}, \bibinfo {author} {\bibfnamefont {M.}~\bibnamefont {Reagor}},
  \bibinfo {author} {\bibfnamefont {R.}~\bibnamefont {Renzas}}, \bibinfo
  {author} {\bibfnamefont {N.}~\bibnamefont {Rubin}}, \bibinfo {author}
  {\bibfnamefont {D.}~\bibnamefont {Russell}}, \bibinfo {author} {\bibfnamefont
  {M.}~\bibnamefont {Rust}}, \bibinfo {author} {\bibfnamefont {D.}~\bibnamefont
  {Scarabelli}}, \bibinfo {author} {\bibfnamefont {M.}~\bibnamefont {Scheer}},
  \bibinfo {author} {\bibfnamefont {M.}~\bibnamefont {Selvanayagam}}, \bibinfo
  {author} {\bibfnamefont {R.}~\bibnamefont {Smith}}, \bibinfo {author}
  {\bibfnamefont {A.}~\bibnamefont {Staley}}, \bibinfo {author} {\bibfnamefont
  {M.}~\bibnamefont {Suska}}, \bibinfo {author} {\bibfnamefont
  {N.}~\bibnamefont {Tezak}}, \bibinfo {author} {\bibfnamefont {D.~C.}\
  \bibnamefont {Thompson}}, \bibinfo {author} {\bibfnamefont {T.-W.}\
  \bibnamefont {To}}, \bibinfo {author} {\bibfnamefont {M.}~\bibnamefont
  {Vahidpour}}, \bibinfo {author} {\bibfnamefont {N.}~\bibnamefont
  {Vodrahalli}}, \bibinfo {author} {\bibfnamefont {T.}~\bibnamefont {Whyland}},
  \bibinfo {author} {\bibfnamefont {K.}~\bibnamefont {Yadav}}, \bibinfo
  {author} {\bibfnamefont {W.}~\bibnamefont {Zeng}},\ and\ \bibinfo {author}
  {\bibfnamefont {C.}~\bibnamefont {Rigetti}},\ }\href
  {https://doi.org/10.1103/PhysRevApplied.10.034050} {\bibfield  {journal}
  {\bibinfo  {journal} {Phys. Rev. Applied}\ }\textbf {\bibinfo {volume}
  {10}},\ \bibinfo {pages} {034050} (\bibinfo {year} {2018})}\BibitemShut
  {NoStop}%
\bibitem [{\citenamefont {McKay}\ \emph {et~al.}(2016)\citenamefont {McKay},
  \citenamefont {Filipp}, \citenamefont {Mezzacapo}, \citenamefont {Magesan},
  \citenamefont {Chow},\ and\ \citenamefont {Gambetta}}]{McKay2016universal}%
  \BibitemOpen
  \bibfield  {author} {\bibinfo {author} {\bibfnamefont {D.~C.}\ \bibnamefont
  {McKay}}, \bibinfo {author} {\bibfnamefont {S.}~\bibnamefont {Filipp}},
  \bibinfo {author} {\bibfnamefont {A.}~\bibnamefont {Mezzacapo}}, \bibinfo
  {author} {\bibfnamefont {E.}~\bibnamefont {Magesan}}, \bibinfo {author}
  {\bibfnamefont {J.~M.}\ \bibnamefont {Chow}},\ and\ \bibinfo {author}
  {\bibfnamefont {J.~M.}\ \bibnamefont {Gambetta}},\ }\href
  {https://doi.org/10.1103/PhysRevApplied.6.064007} {\bibfield  {journal}
  {\bibinfo  {journal} {Phys. Rev. Applied}\ }\textbf {\bibinfo {volume} {6}},\
  \bibinfo {pages} {064007} (\bibinfo {year} {2016})}\BibitemShut {NoStop}%
\bibitem [{\citenamefont {Walter}\ \emph {et~al.}(2017)\citenamefont {Walter},
  \citenamefont {Kurpiers}, \citenamefont {Gasparinetti}, \citenamefont
  {Magnard}, \citenamefont {Poto\ifmmode~\check{c}\else \v{c}\fi{}nik},
  \citenamefont {Salath\'e}, \citenamefont {Pechal}, \citenamefont {Mondal},
  \citenamefont {Oppliger}, \citenamefont {Eichler},\ and\ \citenamefont
  {Wallraff}}]{Walter2017rapid}%
  \BibitemOpen
  \bibfield  {author} {\bibinfo {author} {\bibfnamefont {T.}~\bibnamefont
  {Walter}}, \bibinfo {author} {\bibfnamefont {P.}~\bibnamefont {Kurpiers}},
  \bibinfo {author} {\bibfnamefont {S.}~\bibnamefont {Gasparinetti}}, \bibinfo
  {author} {\bibfnamefont {P.}~\bibnamefont {Magnard}}, \bibinfo {author}
  {\bibfnamefont {A.}~\bibnamefont {Poto\ifmmode~\check{c}\else
  \v{c}\fi{}nik}}, \bibinfo {author} {\bibfnamefont {Y.}~\bibnamefont
  {Salath\'e}}, \bibinfo {author} {\bibfnamefont {M.}~\bibnamefont {Pechal}},
  \bibinfo {author} {\bibfnamefont {M.}~\bibnamefont {Mondal}}, \bibinfo
  {author} {\bibfnamefont {M.}~\bibnamefont {Oppliger}}, \bibinfo {author}
  {\bibfnamefont {C.}~\bibnamefont {Eichler}},\ and\ \bibinfo {author}
  {\bibfnamefont {A.}~\bibnamefont {Wallraff}},\ }\href
  {https://doi.org/10.1103/PhysRevApplied.7.054020} {\bibfield  {journal}
  {\bibinfo  {journal} {Phys. Rev. Applied}\ }\textbf {\bibinfo {volume} {7}},\
  \bibinfo {pages} {054020} (\bibinfo {year} {2017})}\BibitemShut {NoStop}%
\bibitem [{\citenamefont {Blais}\ \emph {et~al.}()\citenamefont {Blais},
  \citenamefont {Grimsmo}, \citenamefont {Girvin},\ and\ \citenamefont
  {Wallraff}}]{blais2020circuit}%
  \BibitemOpen
  \bibfield  {author} {\bibinfo {author} {\bibfnamefont {A.}~\bibnamefont
  {Blais}}, \bibinfo {author} {\bibfnamefont {A.~L.}\ \bibnamefont {Grimsmo}},
  \bibinfo {author} {\bibfnamefont {S.}~\bibnamefont {Girvin}},\ and\ \bibinfo
  {author} {\bibfnamefont {A.}~\bibnamefont {Wallraff}},\ }\href@noop {} {\
  }\Eprint {https://arxiv.org/abs/2005.12667} {arXiv:2005.12667} \BibitemShut
  {NoStop}%
\bibitem [{\citenamefont {Ku}\ \emph {et~al.}(2020)\citenamefont {Ku},
  \citenamefont {Xu}, \citenamefont {Brink}, \citenamefont {McKay},
  \citenamefont {Hertzberg}, \citenamefont {Ansari},\ and\ \citenamefont
  {Plourde}}]{ku2020suppression}%
  \BibitemOpen
  \bibfield  {author} {\bibinfo {author} {\bibfnamefont {J.}~\bibnamefont
  {Ku}}, \bibinfo {author} {\bibfnamefont {X.}~\bibnamefont {Xu}}, \bibinfo
  {author} {\bibfnamefont {M.}~\bibnamefont {Brink}}, \bibinfo {author}
  {\bibfnamefont {D.~C.}\ \bibnamefont {McKay}}, \bibinfo {author}
  {\bibfnamefont {J.~B.}\ \bibnamefont {Hertzberg}}, \bibinfo {author}
  {\bibfnamefont {M.~H.}\ \bibnamefont {Ansari}},\ and\ \bibinfo {author}
  {\bibfnamefont {B.~L.~T.}\ \bibnamefont {Plourde}},\ }\href
  {https://doi.org/10.1103/PhysRevLett.125.200504} {\bibfield  {journal}
  {\bibinfo  {journal} {Phys. Rev. Lett.}\ }\textbf {\bibinfo {volume} {125}},\
  \bibinfo {pages} {200504} (\bibinfo {year} {2020})}\BibitemShut {NoStop}%
\bibitem [{\citenamefont {Mundada}\ \emph {et~al.}(2019)\citenamefont
  {Mundada}, \citenamefont {Zhang}, \citenamefont {Hazard},\ and\ \citenamefont
  {Houck}}]{mundada2019suppression}%
  \BibitemOpen
  \bibfield  {author} {\bibinfo {author} {\bibfnamefont {P.}~\bibnamefont
  {Mundada}}, \bibinfo {author} {\bibfnamefont {G.}~\bibnamefont {Zhang}},
  \bibinfo {author} {\bibfnamefont {T.}~\bibnamefont {Hazard}},\ and\ \bibinfo
  {author} {\bibfnamefont {A.}~\bibnamefont {Houck}},\ }\href
  {https://doi.org/10.1103/PhysRevApplied.12.054023} {\bibfield  {journal}
  {\bibinfo  {journal} {Phys. Rev. Applied}\ }\textbf {\bibinfo {volume}
  {12}},\ \bibinfo {pages} {054023} (\bibinfo {year} {2019})}\BibitemShut
  {NoStop}%
\bibitem [{\citenamefont {McKay}\ \emph {et~al.}(2019)\citenamefont {McKay},
  \citenamefont {Sheldon}, \citenamefont {Smolin}, \citenamefont {Chow},\ and\
  \citenamefont {Gambetta}}]{mckay2019three}%
  \BibitemOpen
  \bibfield  {author} {\bibinfo {author} {\bibfnamefont {D.~C.}\ \bibnamefont
  {McKay}}, \bibinfo {author} {\bibfnamefont {S.}~\bibnamefont {Sheldon}},
  \bibinfo {author} {\bibfnamefont {J.~A.}\ \bibnamefont {Smolin}}, \bibinfo
  {author} {\bibfnamefont {J.~M.}\ \bibnamefont {Chow}},\ and\ \bibinfo
  {author} {\bibfnamefont {J.~M.}\ \bibnamefont {Gambetta}},\ }\href
  {https://doi.org/10.1103/PhysRevLett.122.200502} {\bibfield  {journal}
  {\bibinfo  {journal} {Phys. Rev. Lett.}\ }\textbf {\bibinfo {volume} {122}},\
  \bibinfo {pages} {200502} (\bibinfo {year} {2019})}\BibitemShut {NoStop}%
\bibitem [{\citenamefont {Krinner}\ \emph
  {et~al.}(2020{\natexlab{a}})\citenamefont {Krinner}, \citenamefont {Lazar},
  \citenamefont {Remm}, \citenamefont {Andersen}, \citenamefont {Lacroix},
  \citenamefont {Norris}, \citenamefont {Hellings}, \citenamefont {Gabureac},
  \citenamefont {Eichler},\ and\ \citenamefont
  {Wallraff}}]{PhysRevApplied.14.024042}%
  \BibitemOpen
  \bibfield  {author} {\bibinfo {author} {\bibfnamefont {S.}~\bibnamefont
  {Krinner}}, \bibinfo {author} {\bibfnamefont {S.}~\bibnamefont {Lazar}},
  \bibinfo {author} {\bibfnamefont {A.}~\bibnamefont {Remm}}, \bibinfo {author}
  {\bibfnamefont {C.~K.}\ \bibnamefont {Andersen}}, \bibinfo {author}
  {\bibfnamefont {N.}~\bibnamefont {Lacroix}}, \bibinfo {author} {\bibfnamefont
  {G.~J.}\ \bibnamefont {Norris}}, \bibinfo {author} {\bibfnamefont
  {C.}~\bibnamefont {Hellings}}, \bibinfo {author} {\bibfnamefont
  {M.}~\bibnamefont {Gabureac}}, \bibinfo {author} {\bibfnamefont
  {C.}~\bibnamefont {Eichler}},\ and\ \bibinfo {author} {\bibfnamefont
  {A.}~\bibnamefont {Wallraff}},\ }\href
  {https://doi.org/10.1103/PhysRevApplied.14.024042} {\bibfield  {journal}
  {\bibinfo  {journal} {Phys. Rev. Applied}\ }\textbf {\bibinfo {volume}
  {14}},\ \bibinfo {pages} {024042} (\bibinfo {year}
  {2020}{\natexlab{a}})}\BibitemShut {NoStop}%
\bibitem [{\citenamefont {Zhao}\ \emph {et~al.}(2020)\citenamefont {Zhao},
  \citenamefont {Xu}, \citenamefont {Lan}, \citenamefont {Chu}, \citenamefont
  {Tan}, \citenamefont {Yu},\ and\ \citenamefont {Yu}}]{zhao2020high}%
  \BibitemOpen
  \bibfield  {author} {\bibinfo {author} {\bibfnamefont {P.}~\bibnamefont
  {Zhao}}, \bibinfo {author} {\bibfnamefont {P.}~\bibnamefont {Xu}}, \bibinfo
  {author} {\bibfnamefont {D.}~\bibnamefont {Lan}}, \bibinfo {author}
  {\bibfnamefont {J.}~\bibnamefont {Chu}}, \bibinfo {author} {\bibfnamefont
  {X.}~\bibnamefont {Tan}}, \bibinfo {author} {\bibfnamefont {H.}~\bibnamefont
  {Yu}},\ and\ \bibinfo {author} {\bibfnamefont {Y.}~\bibnamefont {Yu}},\
  }\href {https://doi.org/10.1103/PhysRevLett.125.200503} {\bibfield  {journal}
  {\bibinfo  {journal} {Phys. Rev. Lett.}\ }\textbf {\bibinfo {volume} {125}},\
  \bibinfo {pages} {200503} (\bibinfo {year} {2020})}\BibitemShut {NoStop}%
\bibitem [{\citenamefont {Li}\ \emph {et~al.}(2020)\citenamefont {Li},
  \citenamefont {Cai}, \citenamefont {Yan}, \citenamefont {Wang}, \citenamefont
  {Pan}, \citenamefont {Ma}, \citenamefont {Cai}, \citenamefont {Han},
  \citenamefont {Hua}, \citenamefont {Han}, \citenamefont {Wu}, \citenamefont
  {Zhang}, \citenamefont {Wang}, \citenamefont {Song}, \citenamefont {Duan},\
  and\ \citenamefont {Sun}}]{li2019tunable}%
  \BibitemOpen
  \bibfield  {author} {\bibinfo {author} {\bibfnamefont {X.}~\bibnamefont
  {Li}}, \bibinfo {author} {\bibfnamefont {T.}~\bibnamefont {Cai}}, \bibinfo
  {author} {\bibfnamefont {H.}~\bibnamefont {Yan}}, \bibinfo {author}
  {\bibfnamefont {Z.}~\bibnamefont {Wang}}, \bibinfo {author} {\bibfnamefont
  {X.}~\bibnamefont {Pan}}, \bibinfo {author} {\bibfnamefont {Y.}~\bibnamefont
  {Ma}}, \bibinfo {author} {\bibfnamefont {W.}~\bibnamefont {Cai}}, \bibinfo
  {author} {\bibfnamefont {J.}~\bibnamefont {Han}}, \bibinfo {author}
  {\bibfnamefont {Z.}~\bibnamefont {Hua}}, \bibinfo {author} {\bibfnamefont
  {X.}~\bibnamefont {Han}}, \bibinfo {author} {\bibfnamefont {Y.}~\bibnamefont
  {Wu}}, \bibinfo {author} {\bibfnamefont {H.}~\bibnamefont {Zhang}}, \bibinfo
  {author} {\bibfnamefont {H.}~\bibnamefont {Wang}}, \bibinfo {author}
  {\bibfnamefont {Y.}~\bibnamefont {Song}}, \bibinfo {author} {\bibfnamefont
  {L.}~\bibnamefont {Duan}},\ and\ \bibinfo {author} {\bibfnamefont
  {L.}~\bibnamefont {Sun}},\ }\href
  {https://doi.org/10.1103/PhysRevApplied.14.024070} {\bibfield  {journal}
  {\bibinfo  {journal} {Phys. Rev. Applied}\ }\textbf {\bibinfo {volume}
  {14}},\ \bibinfo {pages} {024070} (\bibinfo {year} {2020})}\BibitemShut
  {NoStop}%
\bibitem [{\citenamefont {Malekakhlagh}\ \emph {et~al.}(2020)\citenamefont
  {Malekakhlagh}, \citenamefont {Magesan},\ and\ \citenamefont
  {McKay}}]{malekakhlagh2020first}%
  \BibitemOpen
  \bibfield  {author} {\bibinfo {author} {\bibfnamefont {M.}~\bibnamefont
  {Malekakhlagh}}, \bibinfo {author} {\bibfnamefont {E.}~\bibnamefont
  {Magesan}},\ and\ \bibinfo {author} {\bibfnamefont {D.~C.}\ \bibnamefont
  {McKay}},\ }\href {https://doi.org/10.1103/PhysRevA.102.042605} {\bibfield
  {journal} {\bibinfo  {journal} {Phys. Rev. A}\ }\textbf {\bibinfo {volume}
  {102}},\ \bibinfo {pages} {042605} (\bibinfo {year} {2020})}\BibitemShut
  {NoStop}%
\bibitem [{\citenamefont {Rigetti}\ and\ \citenamefont
  {Devoret}(2010)}]{Rigetti2010CR}%
  \BibitemOpen
  \bibfield  {author} {\bibinfo {author} {\bibfnamefont {C.}~\bibnamefont
  {Rigetti}}\ and\ \bibinfo {author} {\bibfnamefont {M.}~\bibnamefont
  {Devoret}},\ }\href {https://doi.org/10.1103/PhysRevB.81.134507} {\bibfield
  {journal} {\bibinfo  {journal} {Phys. Rev. B}\ }\textbf {\bibinfo {volume}
  {81}},\ \bibinfo {pages} {134507} (\bibinfo {year} {2010})}\BibitemShut
  {NoStop}%
\bibitem [{\citenamefont {Steffen}\ \emph {et~al.}(2010)\citenamefont
  {Steffen}, \citenamefont {Kumar}, \citenamefont {DiVincenzo}, \citenamefont
  {Rozen}, \citenamefont {Keefe}, \citenamefont {Rothwell},\ and\ \citenamefont
  {Ketchen}}]{Steffen2010high}%
  \BibitemOpen
  \bibfield  {author} {\bibinfo {author} {\bibfnamefont {M.}~\bibnamefont
  {Steffen}}, \bibinfo {author} {\bibfnamefont {S.}~\bibnamefont {Kumar}},
  \bibinfo {author} {\bibfnamefont {D.~P.}\ \bibnamefont {DiVincenzo}},
  \bibinfo {author} {\bibfnamefont {J.~R.}\ \bibnamefont {Rozen}}, \bibinfo
  {author} {\bibfnamefont {G.~A.}\ \bibnamefont {Keefe}}, \bibinfo {author}
  {\bibfnamefont {M.~B.}\ \bibnamefont {Rothwell}},\ and\ \bibinfo {author}
  {\bibfnamefont {M.~B.}\ \bibnamefont {Ketchen}},\ }\href
  {https://doi.org/10.1103/PhysRevLett.105.100502} {\bibfield  {journal}
  {\bibinfo  {journal} {Phys. Rev. Lett.}\ }\textbf {\bibinfo {volume} {105}},\
  \bibinfo {pages} {100502} (\bibinfo {year} {2010})}\BibitemShut {NoStop}%
\bibitem [{\citenamefont {Didier}\ \emph {et~al.}(2018)\citenamefont {Didier},
  \citenamefont {Sete}, \citenamefont {da~Silva},\ and\ \citenamefont
  {Rigetti}}]{Didier2018Analytical}%
  \BibitemOpen
  \bibfield  {author} {\bibinfo {author} {\bibfnamefont {N.}~\bibnamefont
  {Didier}}, \bibinfo {author} {\bibfnamefont {E.~A.}\ \bibnamefont {Sete}},
  \bibinfo {author} {\bibfnamefont {M.~P.}\ \bibnamefont {da~Silva}},\ and\
  \bibinfo {author} {\bibfnamefont {C.}~\bibnamefont {Rigetti}},\ }\href
  {https://doi.org/10.1103/PhysRevA.97.022330} {\bibfield  {journal} {\bibinfo
  {journal} {Phys. Rev. A}\ }\textbf {\bibinfo {volume} {97}},\ \bibinfo
  {pages} {022330} (\bibinfo {year} {2018})}\BibitemShut {NoStop}%
\bibitem [{\citenamefont {Reed}\ \emph {et~al.}(2012)\citenamefont {Reed},
  \citenamefont {DiCarlo}, \citenamefont {Nigg}, \citenamefont {Sun},
  \citenamefont {Frunzio}, \citenamefont {Girvin},\ and\ \citenamefont
  {Schoelkopf}}]{Reed:2012ab}%
  \BibitemOpen
  \bibfield  {author} {\bibinfo {author} {\bibfnamefont {M.~D.}\ \bibnamefont
  {Reed}}, \bibinfo {author} {\bibfnamefont {L.}~\bibnamefont {DiCarlo}},
  \bibinfo {author} {\bibfnamefont {S.~E.}\ \bibnamefont {Nigg}}, \bibinfo
  {author} {\bibfnamefont {L.}~\bibnamefont {Sun}}, \bibinfo {author}
  {\bibfnamefont {L.}~\bibnamefont {Frunzio}}, \bibinfo {author} {\bibfnamefont
  {S.~M.}\ \bibnamefont {Girvin}},\ and\ \bibinfo {author} {\bibfnamefont
  {R.~J.}\ \bibnamefont {Schoelkopf}},\ }\href
  {https://doi.org/10.1038/nature10786} {\bibfield  {journal} {\bibinfo
  {journal} {Nature}\ }\textbf {\bibinfo {volume} {482}},\ \bibinfo {pages}
  {382} (\bibinfo {year} {2012})}\BibitemShut {NoStop}%
\bibitem [{\citenamefont {Koch}\ \emph {et~al.}(2007)\citenamefont {Koch},
  \citenamefont {Yu}, \citenamefont {Gambetta}, \citenamefont {Houck},
  \citenamefont {Schuster}, \citenamefont {Majer}, \citenamefont {Blais},
  \citenamefont {Devoret}, \citenamefont {Girvin},\ and\ \citenamefont
  {Schoelkopf}}]{koch2007charge}%
  \BibitemOpen
  \bibfield  {author} {\bibinfo {author} {\bibfnamefont {J.}~\bibnamefont
  {Koch}}, \bibinfo {author} {\bibfnamefont {T.~M.}\ \bibnamefont {Yu}},
  \bibinfo {author} {\bibfnamefont {J.}~\bibnamefont {Gambetta}}, \bibinfo
  {author} {\bibfnamefont {A.~A.}\ \bibnamefont {Houck}}, \bibinfo {author}
  {\bibfnamefont {D.~I.}\ \bibnamefont {Schuster}}, \bibinfo {author}
  {\bibfnamefont {J.}~\bibnamefont {Majer}}, \bibinfo {author} {\bibfnamefont
  {A.}~\bibnamefont {Blais}}, \bibinfo {author} {\bibfnamefont {M.~H.}\
  \bibnamefont {Devoret}}, \bibinfo {author} {\bibfnamefont {S.~M.}\
  \bibnamefont {Girvin}},\ and\ \bibinfo {author} {\bibfnamefont {R.~J.}\
  \bibnamefont {Schoelkopf}},\ }\href
  {https://doi.org/10.1103/PhysRevA.76.042319} {\bibfield  {journal} {\bibinfo
  {journal} {Phys. Rev. A}\ }\textbf {\bibinfo {volume} {76}},\ \bibinfo
  {pages} {042319} (\bibinfo {year} {2007})}\BibitemShut {NoStop}%
\bibitem [{\citenamefont {Bravyi}\ \emph {et~al.}(2011)\citenamefont {Bravyi},
  \citenamefont {DiVincenzo},\ and\ \citenamefont {Loss}}]{BRAVYI20112793}%
  \BibitemOpen
  \bibfield  {author} {\bibinfo {author} {\bibfnamefont {S.}~\bibnamefont
  {Bravyi}}, \bibinfo {author} {\bibfnamefont {D.~P.}\ \bibnamefont
  {DiVincenzo}},\ and\ \bibinfo {author} {\bibfnamefont {D.}~\bibnamefont
  {Loss}},\ }\href {https://doi.org/https://doi.org/10.1016/j.aop.2011.06.004}
  {\bibfield  {journal} {\bibinfo  {journal} {Annals of Physics}\ }\textbf
  {\bibinfo {volume} {326}},\ \bibinfo {pages} {2793 } (\bibinfo {year}
  {2011})}\BibitemShut {NoStop}%
\bibitem [{\citenamefont {Ansari}(2019)}]{ansari2019superconducting}%
  \BibitemOpen
  \bibfield  {author} {\bibinfo {author} {\bibfnamefont {M.~H.}\ \bibnamefont
  {Ansari}},\ }\href {https://doi.org/10.1103/PhysRevB.100.024509} {\bibfield
  {journal} {\bibinfo  {journal} {Physical Review B}\ }\textbf {\bibinfo
  {volume} {100}},\ \bibinfo {pages} {024509} (\bibinfo {year}
  {2019})}\BibitemShut {NoStop}%
\bibitem [{\citenamefont {Cederbaum}\ \emph {et~al.}(1989)\citenamefont
  {Cederbaum}, \citenamefont {Schirmer},\ and\ \citenamefont
  {Meyer}}]{Cederbaum_1989}%
  \BibitemOpen
  \bibfield  {author} {\bibinfo {author} {\bibfnamefont {L.~S.}\ \bibnamefont
  {Cederbaum}}, \bibinfo {author} {\bibfnamefont {J.}~\bibnamefont
  {Schirmer}},\ and\ \bibinfo {author} {\bibfnamefont {H.~D.}\ \bibnamefont
  {Meyer}},\ }\href {https://doi.org/10.1088/0305-4470/22/13/035} {\bibfield
  {journal} {\bibinfo  {journal} {Journal of Physics A: Mathematical and
  General}\ }\textbf {\bibinfo {volume} {22}},\ \bibinfo {pages} {2427}
  (\bibinfo {year} {1989})}\BibitemShut {NoStop}%
\bibitem [{\citenamefont {Magesan}\ and\ \citenamefont
  {Gambetta}(2020)}]{magesan2018effective}%
  \BibitemOpen
  \bibfield  {author} {\bibinfo {author} {\bibfnamefont {E.}~\bibnamefont
  {Magesan}}\ and\ \bibinfo {author} {\bibfnamefont {J.~M.}\ \bibnamefont
  {Gambetta}},\ }\href {https://doi.org/10.1103/PhysRevA.101.052308} {\bibfield
   {journal} {\bibinfo  {journal} {Phys. Rev. A}\ }\textbf {\bibinfo {volume}
  {101}},\ \bibinfo {pages} {052308} (\bibinfo {year} {2020})}\BibitemShut
  {NoStop}%
\bibitem [{\citenamefont {Goerz}\ \emph {et~al.}(2017)\citenamefont {Goerz},
  \citenamefont {Motzoi}, \citenamefont {Whaley},\ and\ \citenamefont
  {Koch}}]{goerz2017charting}%
  \BibitemOpen
  \bibfield  {author} {\bibinfo {author} {\bibfnamefont {M.~H.}\ \bibnamefont
  {Goerz}}, \bibinfo {author} {\bibfnamefont {F.}~\bibnamefont {Motzoi}},
  \bibinfo {author} {\bibfnamefont {K.~B.}\ \bibnamefont {Whaley}},\ and\
  \bibinfo {author} {\bibfnamefont {C.~P.}\ \bibnamefont {Koch}},\ }\href
  {https://doi.org/10.1038/s41534-017-0036-0} {\bibfield  {journal} {\bibinfo
  {journal} {npj Quantum Information}\ }\textbf {\bibinfo {volume} {3}},\
  \bibinfo {pages} {37} (\bibinfo {year} {2017})}\BibitemShut {NoStop}%
\bibitem [{\citenamefont {Krinner}\ \emph
  {et~al.}(2020{\natexlab{b}})\citenamefont {Krinner}, \citenamefont
  {Kurpiers}, \citenamefont {Royer}, \citenamefont {Magnard}, \citenamefont
  {Tsitsilin}, \citenamefont {Besse}, \citenamefont {Remm}, \citenamefont
  {Blais},\ and\ \citenamefont {Wallraff}}]{krinner2020demonstration}%
  \BibitemOpen
  \bibfield  {author} {\bibinfo {author} {\bibfnamefont {S.}~\bibnamefont
  {Krinner}}, \bibinfo {author} {\bibfnamefont {P.}~\bibnamefont {Kurpiers}},
  \bibinfo {author} {\bibfnamefont {B.}~\bibnamefont {Royer}}, \bibinfo
  {author} {\bibfnamefont {P.}~\bibnamefont {Magnard}}, \bibinfo {author}
  {\bibfnamefont {I.}~\bibnamefont {Tsitsilin}}, \bibinfo {author}
  {\bibfnamefont {J.-C.}\ \bibnamefont {Besse}}, \bibinfo {author}
  {\bibfnamefont {A.}~\bibnamefont {Remm}}, \bibinfo {author} {\bibfnamefont
  {A.}~\bibnamefont {Blais}},\ and\ \bibinfo {author} {\bibfnamefont
  {A.}~\bibnamefont {Wallraff}},\ }\href
  {https://doi.org/10.1103/PhysRevApplied.14.044039} {\bibfield  {journal}
  {\bibinfo  {journal} {Phys. Rev. Applied}\ }\textbf {\bibinfo {volume}
  {14}},\ \bibinfo {pages} {044039} (\bibinfo {year}
  {2020}{\natexlab{b}})}\BibitemShut {NoStop}%
\bibitem [{\citenamefont {Collodo}\ \emph {et~al.}(2020)\citenamefont
  {Collodo}, \citenamefont {Herrmann}, \citenamefont {Lacroix}, \citenamefont
  {Andersen}, \citenamefont {Remm}, \citenamefont {Lazar}, \citenamefont
  {Besse}, \citenamefont {Walter}, \citenamefont {Wallraff},\ and\
  \citenamefont {Eichler}}]{collodo2020implementation}%
  \BibitemOpen
  \bibfield  {author} {\bibinfo {author} {\bibfnamefont {M.~C.}\ \bibnamefont
  {Collodo}}, \bibinfo {author} {\bibfnamefont {J.}~\bibnamefont {Herrmann}},
  \bibinfo {author} {\bibfnamefont {N.}~\bibnamefont {Lacroix}}, \bibinfo
  {author} {\bibfnamefont {C.~K.}\ \bibnamefont {Andersen}}, \bibinfo {author}
  {\bibfnamefont {A.}~\bibnamefont {Remm}}, \bibinfo {author} {\bibfnamefont
  {S.}~\bibnamefont {Lazar}}, \bibinfo {author} {\bibfnamefont {J.-C.}\
  \bibnamefont {Besse}}, \bibinfo {author} {\bibfnamefont {T.}~\bibnamefont
  {Walter}}, \bibinfo {author} {\bibfnamefont {A.}~\bibnamefont {Wallraff}},\
  and\ \bibinfo {author} {\bibfnamefont {C.}~\bibnamefont {Eichler}},\ }\href
  {https://doi.org/10.1103/PhysRevLett.125.240502} {\bibfield  {journal}
  {\bibinfo  {journal} {Phys. Rev. Lett.}\ }\textbf {\bibinfo {volume} {125}},\
  \bibinfo {pages} {240502} (\bibinfo {year} {2020})}\BibitemShut {NoStop}%
\bibitem [{\citenamefont {Xu}\ \emph {et~al.}(2020)\citenamefont {Xu},
  \citenamefont {Chu}, \citenamefont {Yuan}, \citenamefont {Qiu}, \citenamefont
  {Zhou}, \citenamefont {Zhang}, \citenamefont {Tan}, \citenamefont {Yu},
  \citenamefont {Liu}, \citenamefont {Li}, \citenamefont {Yan},\ and\
  \citenamefont {Yu}}]{xu2020highfidelity}%
  \BibitemOpen
  \bibfield  {author} {\bibinfo {author} {\bibfnamefont {Y.}~\bibnamefont
  {Xu}}, \bibinfo {author} {\bibfnamefont {J.}~\bibnamefont {Chu}}, \bibinfo
  {author} {\bibfnamefont {J.}~\bibnamefont {Yuan}}, \bibinfo {author}
  {\bibfnamefont {J.}~\bibnamefont {Qiu}}, \bibinfo {author} {\bibfnamefont
  {Y.}~\bibnamefont {Zhou}}, \bibinfo {author} {\bibfnamefont {L.}~\bibnamefont
  {Zhang}}, \bibinfo {author} {\bibfnamefont {X.}~\bibnamefont {Tan}}, \bibinfo
  {author} {\bibfnamefont {Y.}~\bibnamefont {Yu}}, \bibinfo {author}
  {\bibfnamefont {S.}~\bibnamefont {Liu}}, \bibinfo {author} {\bibfnamefont
  {J.}~\bibnamefont {Li}}, \bibinfo {author} {\bibfnamefont {F.}~\bibnamefont
  {Yan}},\ and\ \bibinfo {author} {\bibfnamefont {D.}~\bibnamefont {Yu}},\
  }\href {https://doi.org/10.1103/PhysRevLett.125.240503} {\bibfield  {journal}
  {\bibinfo  {journal} {Phys. Rev. Lett.}\ }\textbf {\bibinfo {volume} {125}},\
  \bibinfo {pages} {240503} (\bibinfo {year} {2020})}\BibitemShut {NoStop}%
\bibitem [{\citenamefont {Sheldon}\ \emph {et~al.}(2016)\citenamefont
  {Sheldon}, \citenamefont {Magesan}, \citenamefont {Chow},\ and\ \citenamefont
  {Gambetta}}]{Sheldon2016CRgate}%
  \BibitemOpen
  \bibfield  {author} {\bibinfo {author} {\bibfnamefont {S.}~\bibnamefont
  {Sheldon}}, \bibinfo {author} {\bibfnamefont {E.}~\bibnamefont {Magesan}},
  \bibinfo {author} {\bibfnamefont {J.~M.}\ \bibnamefont {Chow}},\ and\
  \bibinfo {author} {\bibfnamefont {J.~M.}\ \bibnamefont {Gambetta}},\ }\href
  {https://doi.org/10.1103/PhysRevA.93.060302} {\bibfield  {journal} {\bibinfo
  {journal} {Phys. Rev. A}\ }\textbf {\bibinfo {volume} {93}},\ \bibinfo
  {pages} {060302(R)} (\bibinfo {year} {2016})}\BibitemShut {NoStop}%
\bibitem [{\citenamefont {C\'orcoles}\ \emph {et~al.}(2013)\citenamefont
  {C\'orcoles}, \citenamefont {Gambetta}, \citenamefont {Chow}, \citenamefont
  {Smolin}, \citenamefont {Ware}, \citenamefont {Strand}, \citenamefont
  {Plourde},\ and\ \citenamefont {Steffen}}]{Corcoles2013process}%
  \BibitemOpen
  \bibfield  {author} {\bibinfo {author} {\bibfnamefont {A.~D.}\ \bibnamefont
  {C\'orcoles}}, \bibinfo {author} {\bibfnamefont {J.~M.}\ \bibnamefont
  {Gambetta}}, \bibinfo {author} {\bibfnamefont {J.~M.}\ \bibnamefont {Chow}},
  \bibinfo {author} {\bibfnamefont {J.~A.}\ \bibnamefont {Smolin}}, \bibinfo
  {author} {\bibfnamefont {M.}~\bibnamefont {Ware}}, \bibinfo {author}
  {\bibfnamefont {J.}~\bibnamefont {Strand}}, \bibinfo {author} {\bibfnamefont
  {B.~L.~T.}\ \bibnamefont {Plourde}},\ and\ \bibinfo {author} {\bibfnamefont
  {M.}~\bibnamefont {Steffen}},\ }\href
  {https://doi.org/10.1103/PhysRevA.87.030301} {\bibfield  {journal} {\bibinfo
  {journal} {Phys. Rev. A}\ }\textbf {\bibinfo {volume} {87}},\ \bibinfo
  {pages} {030301(R)} (\bibinfo {year} {2013})}\BibitemShut {NoStop}%
\bibitem [{\citenamefont {Sundaresan}\ \emph {et~al.}(2020)\citenamefont
  {Sundaresan}, \citenamefont {Lauer}, \citenamefont {Pritchett}, \citenamefont
  {Magesan}, \citenamefont {Jurcevic},\ and\ \citenamefont
  {Gambetta}}]{sundaresan2020reducing}%
  \BibitemOpen
  \bibfield  {author} {\bibinfo {author} {\bibfnamefont {N.}~\bibnamefont
  {Sundaresan}}, \bibinfo {author} {\bibfnamefont {I.}~\bibnamefont {Lauer}},
  \bibinfo {author} {\bibfnamefont {E.}~\bibnamefont {Pritchett}}, \bibinfo
  {author} {\bibfnamefont {E.}~\bibnamefont {Magesan}}, \bibinfo {author}
  {\bibfnamefont {P.}~\bibnamefont {Jurcevic}},\ and\ \bibinfo {author}
  {\bibfnamefont {J.~M.}\ \bibnamefont {Gambetta}},\ }\href
  {https://doi.org/10.1103/PRXQuantum.1.020318} {\bibfield  {journal} {\bibinfo
   {journal} {PRX Quantum}\ }\textbf {\bibinfo {volume} {1}},\ \bibinfo {pages}
  {020318} (\bibinfo {year} {2020})}\BibitemShut {NoStop}%
\end{thebibliography}%
\end{document}